\documentclass[journal,10pt,doublecolumn]{IEEEtran}

\usepackage{xcolor}
\usepackage[utf8]{inputenc} 
\usepackage[T1]{fontenc}
\usepackage{url}
\usepackage{ifthen}
\usepackage{booktabs}
\usepackage{tabularray}
\usepackage{cite}
\usepackage{comment}
\usepackage[cmex10]{amsmath} 
\usepackage{graphicx}
\usepackage{subcaption}
\usepackage{cellspace}
\usepackage{adjustbox} 
\usepackage{makecell} 

\usepackage{amsthm}
\usepackage{amsmath}
\usepackage{amsmath,amssymb,amsfonts}
\usepackage{algorithm}
\usepackage{algpseudocode}

\newtheorem{theorem}{Theorem}
\newtheorem{lemma}{Lemma}
\newtheorem{remark}{Remark}
\newtheorem{definition}{Definition}
\newtheorem{corollary}{Corollary}

\newcommand{\off}[1]{}

\usepackage{hyperref}
\usepackage[shortlabels]{enumitem}
\usepackage{cleveref}
\hypersetup{
    colorlinks=true,
    linkcolor=blue,
    filecolor=magenta,      
    urlcolor=cyan,
    pdftitle={Overleaf Example},
    pdfpagemode=FullScreen,
}

\allowdisplaybreaks
\usepackage{cuted}
\makeatletter
\newenvironment{breakablealgorithm}
  {
   \begin{center}
     \refstepcounter{algorithm}
     \hrule height.8pt depth0pt \kern2pt
     \renewcommand{\caption}[2][\relax]{
       {\raggedright\textbf{\ALG@name~\thealgorithm} ##2\par}%
       \ifx\relax##1\relax 
         \addcontentsline{loa}{algorithm}{\protect\numberline{\thealgorithm}##2}%
       \else 
         \addcontentsline{loa}{algorithm}{\protect\numberline{\thealgorithm}##1}%
       \fi
       \kern2pt\hrule\kern2pt
     }
  }{
     \kern2pt\hrule\relax
   \end{center}
  }
\makeatother

\begin{document}
\title{Coding-Based Hybrid Post-Quantum Cryptosystem for Non-Uniform Information\vspace{-0.2cm}\thanks{Parts of this work were accepted for publication at the IEEE International Symposium on Information Theory, ISIT 2024.}} 



\author{%
  \IEEEauthorblockN{Saar Tarnopolsky and
                    Alejandro Cohen}\\
  \IEEEauthorblockA{
                   Faculty of Electrical and Computer Engineering, Technion --- Institute of Technology, Haifa, Israel,\\Emails: saar.tar@campus.technion.ac.il and alecohen@technion.ac.il\vspace{-0.45cm}}
}

\maketitle

\begin{abstract}
    We introduce for non-uniform messages a novel hybrid universal network coding cryptosystem (NU-HUNCC) in the finite blocklength regime that provides Post-Quantum (PQ) security at high communication rates. Recently, hybrid cryptosystems offered PQ security by premixing the data using secure linear coding schemes and encrypting only a small portion of it. The data is assumed to be uniformly distributed, an assumption that is often challenging to enforce. Standard fixed-length lossless source coding and compression schemes guarantee a uniform output in \emph{normalized divergence}. Yet, this is not sufficient to guarantee security. We consider an efficient compression scheme uniform in \emph{non-normalized variational distance} for the proposed hybrid cryptosystem, that by utilizing a uniform sub-linear shared seed, guarantees PQ security.
    Specifically, for the proposed PQ cryptosystem, first, we provide an end-to-end practical coding scheme, NU-HUNCC, for non-uniform messages. Second, we show that NU-HUNCC is information-theoretic individually secured (IS) against an eavesdropper with access to any subset of the links and provide a converse proof against such an eavesdropper. Third, we introduce a modified security definition, individual semantic security under a chosen ciphertext attack (ISS-CCA1), and show that against an all-observing eavesdropper, NU-HUNCC satisfies its conditions. Finally, we provide an analysis of NU-HUNCC's high data rate, low computational complexity, and the negligibility of the shared seed size.
\end{abstract} 


\section{Introduction} \label{sec:into}
We consider the problem of high data rate Post-Quantum (PQ) secure communication over a noiseless multipath network for non-uniform messages. In this setting, the transmitter, Alice, wants to send confidential non-uniform messages to the legitimate receiver, Bob, over noiseless communication links. We consider the two
traditional eavesdroppers \cite{goldwasser2019probabilistic,forouzan2015cryptography,bloch2011physical}: 1) Information-Theory Eve (IT-Eve), which has unlimited computational power and can access any subset of the communication links  \cite{bloch2011physical}, and 2) Cryptographic Eve (Crypto-Eve), which has access to all links but is limited computationally \cite{goldwasser2019probabilistic}. We aim to ensure: 1) reliable communication between Alice and Bob, 2)  information-theoretic security against IT-Eve, and 3) PQ computational cryptographic security against Crypto-Eve \cite{bernstein2017post}.

Information-theoretic and computational cryptographic security are products of Shannon's seminal work on perfect secrecy \cite{Shannon1949} from 1949.
Shannon showed that perfect secrecy requires Alice and Bob to share a fresh secret key with entropy higher than or equal to the entropy of the confidential message. Using such keys is costly and often non-practical \cite{bloch2011physical,bernstein2017post}. Information-theoretic and computational cryptographic security offer more practical solutions for secure communication at the expense of some relaxation of the security conditions. Traditional information-theoretic security relies on the probabilistic nature of communication channels, whereas computational cryptographic security assumes that the eavesdropper has limited computational power. 

Wyner's work in 1975 on the wiretap channel \cite{WiretapWyner} introduced an information-theoretic secured coding scheme taking advantage of the eavesdropper's weaker observations of the transmissions over the channel. Wyner showed that by using a local source of randomness, IT-Eve remains ignorant of the confidential message. The information leakage to IT-Eve was defined by $\frac{1}{n}H(M|Z^n)$, where $M$ is the confidential message, $Z^n$ are IT-Eve's observations, and $H$ is the entropy function. Wyner showed that for his proposed code $\lim\limits_{n \rightarrow \infty} \frac{1}{n}H(M|Z^n) =  \lim\limits_{n \rightarrow \infty} \frac{1}{n}H(M)$, i.e. in the asymptotic regime, IT-Eve's observations, $Z^{n}$, provide insignificant information regarding the confidential message. This notion of information-theoretic security is called \emph{weak security}. The use of a local source of randomness in Wyner's proposed code results in a significant decrease of the communication rate \cite{bloch2011physical,liang2009physical,liang2009information,zhou2013physical,cohen2016wiretap}.
\textcolor{blue}{\off{In 1984, Ozarow and Wyner demonstrated that an eavesdropper who observes a subset $w$ out of $n$ transmitted symbols over the wiretap channel can be kept ignorant of the original message transmitted at a rate of $\frac{n-w}{n}$. This result was obtained by using group codes \cite{ozarow1984wire}.}}In 1977, Carliel et al. \cite{carleial1977note} managed to increase the communication rate compared to Wyner's \off{ and Ozarow's} coding scheme, by replacing the source of local randomness with another uniform confidential message. This approach was named \emph{individual secrecy (IS)}. In this setting, Alice sends $\ell$ confidential message to Bob, $\{M_i\}_{i=1}^{\ell}$. IS ensures IT-Eve is ignorant about any single message, i.e., in the finite regime $H(M_i|Z^n) = H(M_i)$, $\forall i \in \ell$. However, the eavesdropper may obtain some insignificant and controlled information about the mixture of all messages\cite{SMSM}. Recently, IS was considered in various communication models such as the broadcast channel, multicast multi-source networks, terahertz wireless, etc \cite{SMSM,JointIndividual2014,IndividualDegradedMU2015,SecrecyBroadcast,tan2019can,cohen2023absolute,yeh2023securing}, to efficiently increase data rates.

On the other hand, computational cryptographic security relies solely on the limited computational power of the eavesdropper. To securely transmit a message to Bob, Alice encrypts the message using a one-way function that is considered hard to invert. Crypto-Eve can't invert the function and decrypt the message in a reasonable time frame, but with the right shared secret key, inversion is possible for Bob \cite{forouzan2015cryptography}.

An example of a commonly used computationally secured cryptosystem is Rivest-Shamir-Adleman (RSA) \cite{kaltz2008introduction}, \cite{forouzan2015cryptography,RSA}. The security of RSA cryptosystem relies on the hardness\off{ of solving the RSA problem and} the integer factorization problem for a Crypto-Eve. However, recent developments in the capabilities of quantum computers significantly increase Crypto-Eve's computational capabilities. Integer factorization is a problem that can be solved in polynomial time by a quantum computer using Shor's algorithm \cite{QuantumRSA}, rendering RSA unsecured. Nevertheless, RSA is not the only cryptosystem that was affected by quantum computer-based algorithms. Throughout the years, an increasing number of cryptosystems have been demonstrated to be vulnerable when exposed to quantum computer base attacks \cite{QuantumRSA,sidelnikov1992insecurity,sendrier1998concatenated,minder2007cryptanalysis,monico2000using,landais2013efficient}. Those advancements triggered interest in PQ-secured cryptosystems i.e., cryptosystems that are secured against a quantum computer based attacks.

A cryptosystem that to this day remains computationally cryptographically secured against quantum computer-based algorithms is McEliecce \cite{mceliece1978public} and its probabilistic extensions, e.g.,  \cite{nojima2008semantic,dottling2012cca2,aguirre2019ind}. McEliecce is a coding-based cryptosystem first introduced in 1978. The security of McEliecce relies on the hardness of decoding a general linear code, a problem considered NP-hard \cite{berlekamp1973goppa,patterson1975algebraic}. However, McEliecce inherently suffers from low data rates of around $0.51$. Increasing the data rate of McElicce is possible, yet it is considered unsecured without a significant increase in its key size \cite{faugere2013distinguisher}.

\begin{figure*}
     \centering
     \begin{subfigure}[b]{0.49\textwidth}
         \centering
         \includegraphics[width=\textwidth]{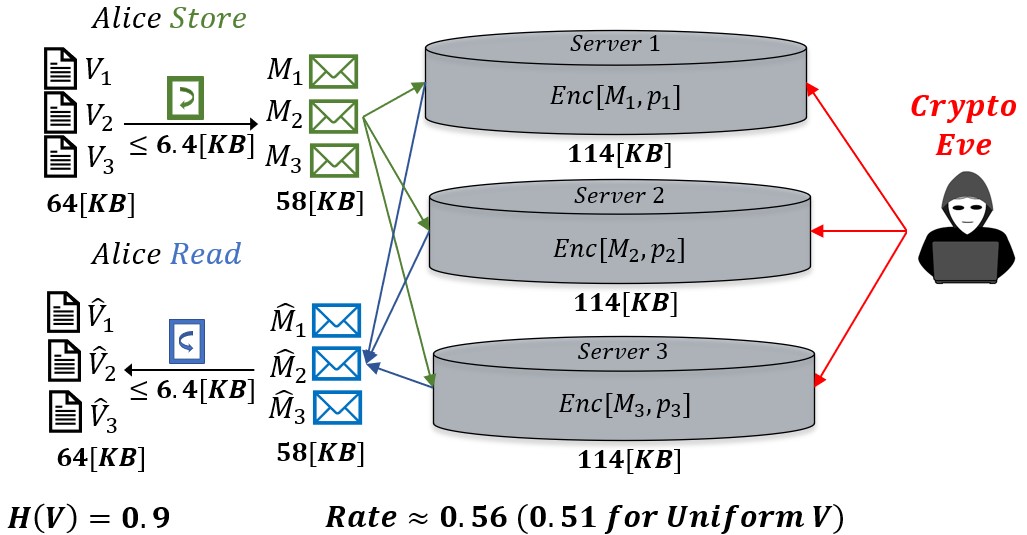}
         \caption{NUM}
         \label{fig:demo-NUM}
         \vspace{0.2cm}
     \end{subfigure}
        \hfill
     \begin{subfigure}[b]{0.49\textwidth}
         \centering
         \includegraphics[width=\textwidth]{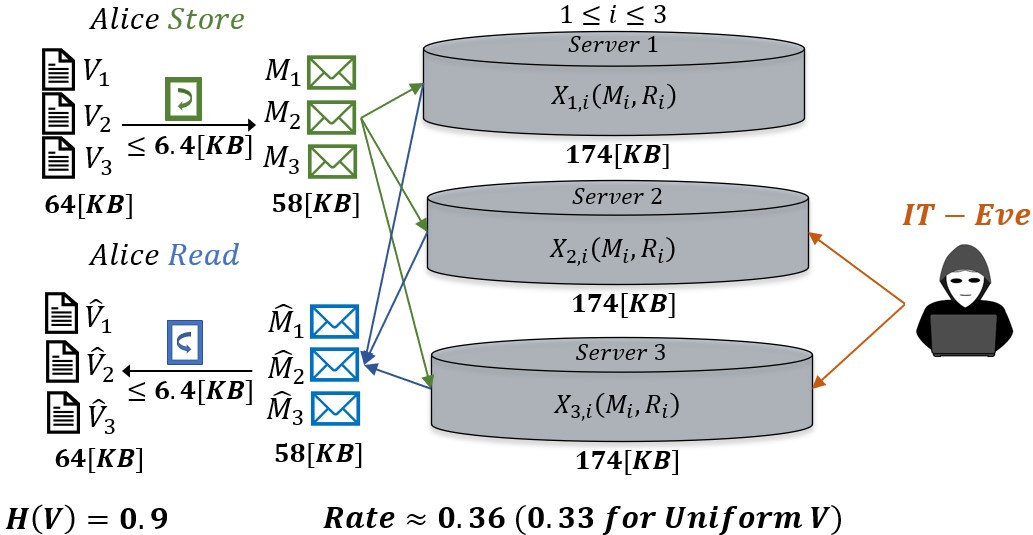}
         \caption{Compressed NC Wiretap Type II}
         \label{fig:demo-NC-WTC-2}
         \vspace{0.2cm}
     \end{subfigure}
        \hfill
    \begin{subfigure}[b]{0.49\textwidth}
         \centering
         \includegraphics[width=\textwidth]{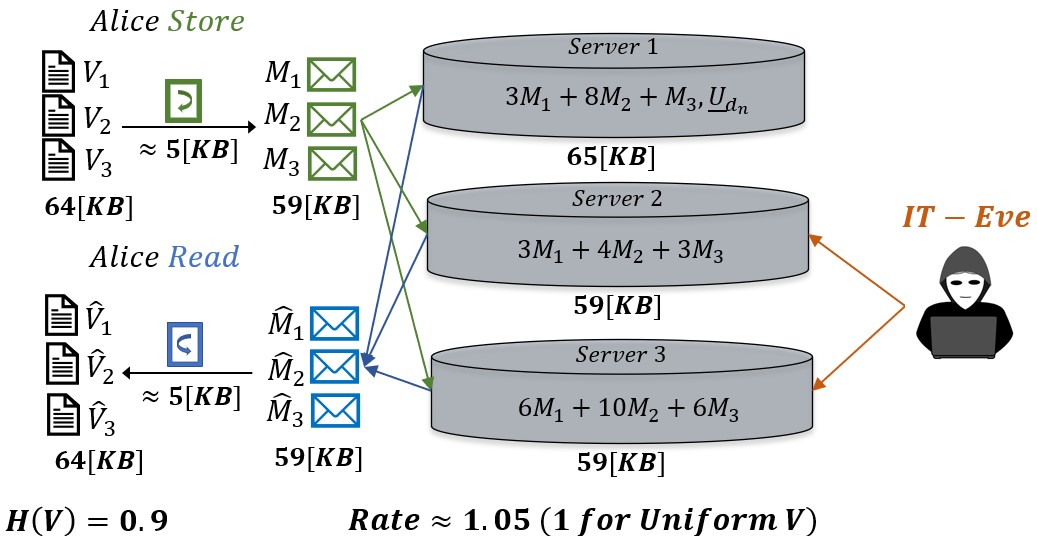}
         \caption{NU-Individual Security}
         \label{fig:demo-NU-Individual}
     \end{subfigure}
        \hfill
    \begin{subfigure}[b]{0.49\textwidth}
         \centering
         \includegraphics[width=\textwidth]{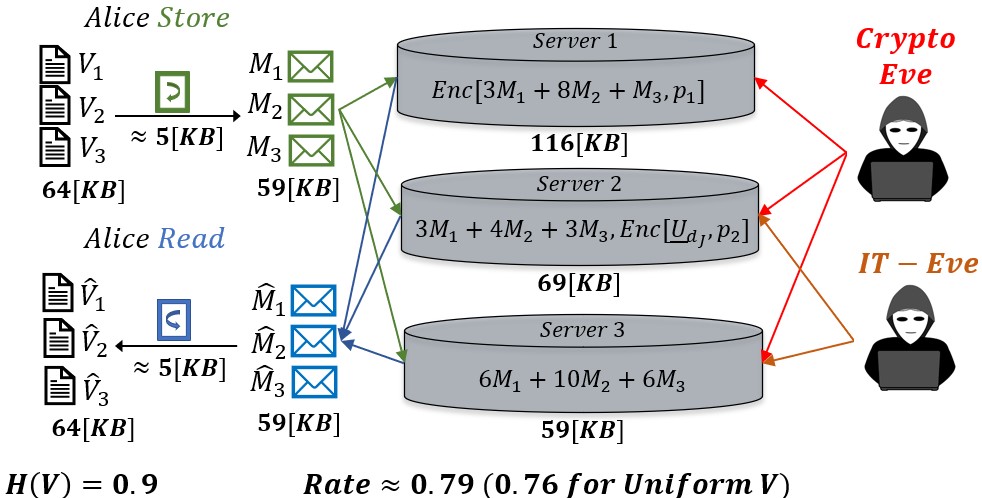}
         \caption{Hybrid NU-HUNCC}
         \label{fig:demo-NU-HUNCC}
     \end{subfigure}
        \hfill
        \caption{Secured storage solution of three files $\{V_i\}_{i=1}^{3}$ on three servers against all observing Crypto-Eve and IT-Eve which has access to two servers. (a) optimal compression of the source and encryption using the original McEliecce cryptosystem with a $[1024,524]$-Goppa codes \cite{mceliece1978public,berlekamp1973goppa} against Crypto-Eve, (b) optimal compression of the source and encoding using Network Coding Wiretap Type \uppercase\expandafter{\romannumeral2} \cite{el2007wiretap} against IT-Eve, (c) Proposed NU-IS using an almost uniform compression scheme with a uniform seed of negligible size along with an IS channel \cite{cohen2023absolute} code against IT-Eve and (d) Proposed NU-HUNCC using an almost uniform compression scheme with a uniform seed of negligible size along with an IS channel code \cite{cohen2023absolute} and encryption using the original McEliecce cryptosystem with a $[1024,524]$-Goppa codes \cite{mceliece1978public,berlekamp1973goppa} against both IT and Crypto-Eve's.}
        \label{fig:distributed_servers}
\end{figure*}

In a work by Cohen et al. \cite{HUNCC}, a novel hybrid approach combining information theory and computational cryptographic security is introduced and named HUNCC. HUNCC combines an IS channel coding scheme \cite{SMSM}, which premixes the messages, with a PQ cryptographic secure scheme that encrypts a small amount of the mixed data. The result is an efficient hybrid PQ secure communication scheme against both IT-Eve and Crypto-Eve attaining a data rate approaching one \cite{d2021post,cohen2022partial,woo2023cermet,kim2024crypto}.

To guarantee security, the information leaked by HUNCC is measured by the \emph{non-normalized mutual information/variational distance} between Eve's observations and the source messages.\off{ to demonstrate that } This leakage meets the cryptographic criterion for ineligibility as well. Although non-uniform messages can be pre-compressed, standard lossless fixed-length source coding guarantees uniformity in \emph{normalized KL-divergence}, but not in \emph{non-normalized distances} \cite{han2005folklore}.

Quantum computers significantly increase the computational capabilities of the eavesdropper, jeopardizing the security of confidential data. As a consequence, a need for PQ-secured cryptosystems arose. Fig.~\ref{fig:distributed_servers} demonstrates that existing cryptosystems that are considered PQ-secured either work at a low data rate, have high complexity, or rely on conditions that are hard-to-enforce, such as uniform input messages. High rate, low complexity practical PQ-secured cryptosystems remain an open problem.

\subsection{Main Contributions}

In this paper, for non-uniform messages, we introduce a novel hybrid universal network coding cryptosystem, coined NU-HUNCC. We propose hybrid source and channel coding schemes, with partial encryption on a small portion of the data, that are secured against both, IT and Crypto Eves. NU-HUNCC is universal in the sense that it can applied to any communication and information system, with any cryptosystem. In particular, the contributions of this work are as follows:

1) Against IT-Eve, PQ individual secure communication at a high data rate approaching capacity is obtained for non-uniform messages. In this setting, the non-uniform source messages are first compressed into an almost uniform outcome using a polar codes-based encoder with a negligible-sized shared seed \cite{chou2013data,chou2015polar,NegligbleCost}. Then, the compressed messages are mixed by a practical IS linear code \cite{silva2009universal,silva2011universal,SMSM}. The number of secured messages this scheme provides is upper bounded by the difference between the number of overall links and the number of links observed by IT-Eve. Conversely, we show that increasing the number of secured messages by one necessarily requires the number of links observed by IT-Eve to be decreased by one as well. Those results have an independent interest in information-theoretic security. 

\off{2) Converse against IT-Eve...}

2) Against Crypto-Eve, PQ cryptographically secured communication at high rates is obtained by encrypting a portion of the links using a semantically secure cryptosystem. We present a new strong definition for PQ computational cryptography security, individual semantic security against chosen ciphertext attack (ISS-CCA1). We show that the controlled information leaked from the proposed source encoder and the IS channel coding scheme for non-uniform messages \emph{meets the cryptographic criterion for ineligibility}.

\off{3) We show that NU-HUNCC remains IS and ISS-CCA1, against IT-Eve by the coding scheme and against Crypto-Eve including partial encryption, respectively. 
Using an IS code the information leakage to IT-Eve and Crypto-Eve is increased compared to a scheme operating on completely uniform messages. However, we show that the additional negligible controlled information leakage added due to the non-uniform distribution of the messages still meets the cryptographic criterion of ineligibility.}

3) We establish rate theorems for NU-HUNCC against both IT-Eve and Crypto-Eve. We show the increase in the communication rate that NU-HUNCC provides compared to other PQ cryptosystems. The rate analysis also includes the effect that the size of the shared seed has on the rate as a function of the message size. Although NU-HUNCC adds the use of a shared seed between the legitimate parties, the negligibility of its size results in it having an insignificant effect on the communication rate. \off{We evaluate NU-HUNCC performance, in terms of data rate and complexity, against both IT-Eve and Crypto-Eve.}\off{ and compare their performance to the following communication systems: a) Non-Unifrom McEliecce (NUM) - all the messages are compressed and then encrypted by the original McEliecce, b) Network coding Wiretap Type \uppercase\expandafter{\romannumeral2} (NC WTC \uppercase\expandafter{\romannumeral2}) - all messages are compressed and encoded by the network coding Wiretap Type \uppercase\expandafter{\romannumeral2} scheme \cite{el2007wiretap}.}We demonstrate the rate enhancement NU-HUNCC provides with an example of a distributed storage application as illustrated by Fig.~\ref{fig:distributed_servers}. Distributed storage servers provide secure and reliable storage services \cite{ghemawat2003google,decandia2007dynamo,chang2008bigtable,borthakur2011apache,calder2011windows}. In our example, Alice wants to store $3$ files over $\ell = 3$ servers and read them back after some time (in this scenario Bob is represented by Alice reading the files in the future). The files are from the source $\mathcal{V} \in \{0,1\}$, denoted by $\{V_i\}_{i=1}^{3}$ and each of them is of size $64$[KB] or $2^{19}$ bits, i.e. $|V_i| = |V| = 2^{19}$ bits ($64$[KB]). The entropy of the source is $H(V) = 0.9$. Alice tries to store the files by four methods:

3.a) \underline{NUM (see Fig.~\ref{fig:demo-NUM}):} First, Alice uses an optimal lossless compression scheme to compress the files\cite{shannon1948mathematical,ziv1978compression,cronie2010lossless}. The size of each compressed file is $58$[KB]. Each compressed file is separately encrypted by the original McEliecce based on $[1024,524]$-Goppa codes\footnote{Original version of McEliecce is not SS-CCA1, however, this is the most commonly used version of this cryptosystem. Thus, we selected to demonstrate NU-HUNCC performance enhancement, in terms of data rate, with this version of it. For SS-CCA1 guarantees, one can select one of his probabilistic extensions, e.g.,  \cite{nojima2008semantic,dottling2012cca2,aguirre2019ind}.}\label{footnote:original_mcceleicce} with a rate of $0.51$. This, method of storing the files is PQ secured against Crypto-Eve and obtains a total rate of about $0.56$. 

3.b) \underline{NC WTC \uppercase\expandafter{\romannumeral2} (see Fig.~\ref{fig:demo-NC-WTC-2})}: Alice uses an optimal lossless source encoder to compress each file to a size of $58$[KB]. The compressed files are encoded by NC WTC Type \uppercase\expandafter{\romannumeral2}. The compressed file $M_i$, $1 \leq i \leq 3$ is encoded into $X_i(M_i,R_i)$. For every three bits of $X_i(M_i,R_i)$, the first is stored on server 1 (denoted by $X_{1,i}(M_i,R_i)$), the second on server 2 (denoted by $X_{2,i}(M_i,R_i)$) and the third on server 3 (denoted by $X_{3,i}(M_i,R_i)$). This scheme is secured against IT-Eve with access to two out of the three servers, thus the data rate is $0.36$\footnote{To demonstrate NC WTC \uppercase\expandafter{\romannumeral2} we could have chosen to store one file over the three servers and remained with the same data rate.}.

3.c) \underline{NU-Individual Security (see Fig.~\ref{fig:demo-NU-Individual})}: First, Alice uses the specialized polar codes-based source encoder and a uniform seed \cite{chou2015polar,NegligbleCost} to compress the files into an almost uniform form. The compressed files, $\{M_i\}_{i=1}^{3}$, sizes are $59[KB]$ ($473147$ bits) each and the size of the seed used to compress each file is $1.56$[KB]. Alice injectivly maps the messages from $\mathbb{F}_{2}^{3 \times 473147}$ into $\mathbb{F}_{11}^{3 \times 43014}$.\off{ We note that in this demonstration we chose $q = 2$ for simplicity, yet it can be any prime number Alice chooses.} To encode the messages matrix it is multiplied by the generator matrix \cite{cohen2023absolute}
\begin{equation*}
    \underline{G}_{IS} = \begin{bmatrix} 3 & 3 & 6 \\ 8 & 4 & 10 \\ 1 & 3 & 6\end{bmatrix}.
\end{equation*}
Thus, the encoded message matrix, $\underline{X}$, is given by
\begin{align*}
    \underline{X} &= \underline{M}^{T} \underline{G}_{IS} = \\
    &[
        3M_1 + 8M_2 + M_3, 3M_1 + 4M_2 + 3M_3,\\
        & \quad\quad\quad\quad\quad\quad\quad\quad\quad\quad 6M_1 + 10M_2 + 6M_3
    ],
\end{align*}
where $\underline{M} = [M_1,M_2,M_3]$. The encoded file $X_i$ size is the same as of the message $M_i$ for $1 \leq i \leq 3$, since the IS code linearly mixes the messages. Each $X_i$ is stored over server $i$ for all three files and servers and the seed is stored over server 1. IT-Eve has access to two of the servers$^{\ref{fn:seed}}$\off{ (for simplicity of illustration, we assume IT-Eve does not have access to the server that stores the seedsee footnote~\ref{fn:seed} in Sec.~\ref{sec:encryption})}, thus each message is individually secured. The overall data rate in the proposed scheme against IT-Eve is $1.05$ since we use compression. The seed's size is negligible compared to the sizes of the compressed files.\off{and the IS channel code does not change the message size.}

3.d) \underline{NU-HUNCC (see Fig.~\ref{fig:demo-NU-HUNCC})}: Alice performs the compression and the mixing of the messages as in NU-Individual Security. After the mixing of the messages, Alice maps the codewords back to $\mathbb{F}_{2}^{3 \times 473147}$. Then, Alice encrypts one of the messages by the original McEliecce cryptosystem based on $[1024,524]$-Goppa codes s.t. after the encryption the size of the mixed file is $116$[KB] and it is stored over one of the servers. Without loss of generality let the encrypted file be the first file. Alice encrypts the uniform seed by the same encryption scheme as well. After the encryption, the size of the seed used to compress each of files is $3.27$[KB]. The encrypted seed is stored on the second server. The rest of the files are stored over the rest of the servers, s.t. each server stores one file. This hybrid scheme is secured against both IT and Crypto-Eve's. The overall data rate of the proposed scheme is $0.79$.

\off{\textcolor{blue}{Alice wants to store three files from the same source $\mathcal{V} \in \{0,1\}$ with a distribution $p_V$ and binary entropy $H(V) = 0.9$. We denote the $3$ files by $\{V_i\}_{i=1}^{3}$ s.t. $|V_i| = |V| = 2^{19}$ bits ($64$[KB]). First, Alice employs an almost uniform source encoder on all the messages. The size of the uniform seed Alice uses as part of the source coding is $d_J = 12876$ bits (see Sec.~\ref{sec:seed_length}). Thus, the size of each of the three almost uniform messages is $|M_i| = |M| = 473147$ (including the seed size). Now, Alice mixes the messages with a linear IS code \cite[Sec. \uppercase\expandafter{\romannumeral6}]{SMSM}. Alice injectivly maps the messages matrix from $\mathbb{F}_{2}^{3 \times 473147}$ into $\mathbb{F}_{2^524}^{3 \times 903}$. We note that in this demonstration we chose $q = 2$ for simplicity, yet it can be any prime number Alice chooses. To encode the messages matrix it is multiplied by the generator matrix $\underline{G}_{IS} = \begin{bmatrix} 3 & 3 & 6 \\ 8 & 4 & 10 \\ 1 & 3 & 6\end{bmatrix}$. Alice multiplies between the messages and the generator matrix and receives the codeword matrix}
\textcolor{blue}{
\begin{align*}
    \underline{X} &= \underline{M}^{T} \underline{G}_{IS} = \\
    &[
        3M_1 + 8M_2 + M_3, 3M_1 + 4M_2 + 3M_3,\\
        & \quad\quad\quad\quad\quad\quad\quad\quad\quad\quad 6M_1 + 10M_2 + 6M_3
    ].
\end{align*}
}
\textcolor{blue}{After obtaining the codeword matrix Alice Encrypts one of the links using $[1024,524]$-Goppa codes-based McEliecce cryptosystem~\ref{footnote:original_mcceleicce}\off{\footnote{Original version of McEliecce is not SS-CCA1, however, this is the most common. Thus, we selected to demonstrate NU-HUNCC performance enhancement, in terms of data rate, with it. For SS-CCA1 guarantees, one can select one of his probabilistic extensions, e.g.,  \cite{nojima2008semantic,dottling2012cca2,aguirre2019ind}.}}. Alice maps the codeword matrix back to $\mathbb{F}_{2}^{3 \times 473147}$. Each $524$ bits of the first line of $\underline{X}$ and the entire uniform sees matrix are encrypted into $1024$ bits. The encrypted first line of $\underline{X}$ is stored on server 1, the second line of $\underline{X}$ along with the encrypted seed is stored on server 2, and the third line of $\underline{X}$ is stored on server 3. Crypto-Eve which has access to all three of the servers can't obtain any significant information on each message individually. The total amount of information Alice stores is $192$[KB] whereas the total amount of information stored on the distributed servers is $241$[KB], resulting in a rate of approximately $0.8$. \off{In Fig.~\ref{fig:distributed_servers} }We compare the performance in terms of communication rate between NU-HUNCC and other secured communication schemes. A more comprehensive comparison of the communication rates and complexity of the communication systems are provided in Sec.~\ref{sec:rate_performance} and Sec.~\ref{sec:complexity} respectively.}}

4) Finally, we assess the computational complexity of the proposed communication scheme. Complexity is measured in this paper, as the number of binary operations required to perform encoding and decoding of all the messages. We show that NU-HUNCC, which encrypts only a subset of links using McEliece, exhibits a more efficient run-time complexity compared to NUM. NU-HUNCC's efficiency makes it a promising candidate for practical applications.

The remainder of this paper is structured in the following manner. Sec.~\ref{sec:sys} presents NU-HUNCC setting, while Sec.~\ref{sec:security_definitions} provides the comprehensive security definitions for IS and ISS-CCA1. In Sec.~\ref{sec:NU-HUNCC}, we introduce the encoding/decoding algorithm for NU-HUNCC. In Sec.~\ref{sec:main_results}, we provide the key findings and theorems of this paper. Sec.~\ref{sec:numerical_performance} offers numerical results demonstrating the performance of NU-HUNCC. The proofs of the theorems provided in the paper are given in Sec.~\ref{sec:linear-code-proof}, \ref{sec:converse}, \ref{sec:linear-partial-enc}, and Appendixes~\ref{sec:random-code-proof} and \ref{sec:random-partial-enc}. We conclude the paper is Sec.~\ref{sec:discussion}. 

\section{System Model} \label{sec:sys}
We consider a communication system where Alice wishes to transmit $\ell$ non-uniform confidential message\footnote{In this paper, we assume the messages are independent to focus on the key methods and results. However, our proposed solution can be easily shown to be valid for dependent sources by using joint source coding schemes \cite{slepian1973noiseless}.} over $\ell$ noiseless links, $\mathcal{L}=\{1,...,\ell\}$, to Bob, in the presence of an eavesdropper, Eve. The messages are taken from a DMS $\left(\mathcal{V},p_V\right)$ s.t. $\mathcal{V} \in \{0,1\}$. We denote the source message matrix by $\underline{V}_{\mathcal{L}} \in \mathbb{F}_{2}^{\ell \times n}$ when $n$ is the size of each source message.

Bob's observations are denoted by $\underline{Y}_{\mathcal{L}}$. Those observations, provide Bob reliable decode $\underline{V}_{\mathcal{L}}$ with high probability. That is, $\mathbb{P}(\underline{\hat{V}}_{\mathcal{L}}(\underline{Y}_{\mathcal{L}}) \neq \underline{V}_{\mathcal{L}}) \leq \epsilon_e$, where $\underline{\hat{V}}_{\mathcal{L}}(\underline{Y}_{\mathcal{L}})$ is the estimation of $\underline{V}_{\mathcal{L}}$. We consider two types of Eve: 1) IT-Eve, which observes any subset $\mathcal{W} \subset \mathcal{L}$ of the links s.t. $|\mathcal{W}| \triangleq w < \ell$, but is computationally unbounded, and 2) Crypto-Eve which observes all the links, but is bounded computationally. We denote IT-Eve's observations by $\underline{Z}_{\mathcal{W}}$ and Crypto-Eve's observations by $\underline{Z}_{\mathcal{L}}$.

\section{Security Definitions} \label{sec:security_definitions}
In this section, we provide the formal security definitions used throughout this paper.
\subsection{Security against IT-Eve}
Against IT-Eve, we consider information-theoretic security. For any subset of $k_s < \ell - w$ source messages, we use the notion of $k_s$ individual security ($k_s$-IS). We measure the leakage of information to the eavesdropper using \emph{non-normalized variational distance}, denoted by $\mathbb{V}(\cdot,\cdot)$. Additionally, we require the code to be reliable where the reliability is measured by the decoding error probability at Bob's. For a code to be $k_s$-IS we require the information leakage and error probability to be negligible. We now formally define $k_s$-IS:
\begin{definition} \label{def:IS}
    ($k_s$ Individual Security) Let $\underline{V}_{\mathcal{L}} \in \mathbb{F}^{\ell \times n}_{q}$ 
    be a set of $\ell$ confidential source messages Alice intends to send, $\underline{Y}_{\mathcal{L}}$ be Bob's observations of the encoded messages, and $\underline{Z}_{\mathcal{W}}$ be IT-Eve's observations of the encoded messages. We say that the coding scheme is $k_s$-IS if:
    \begin{enumerate}
        \item \underline{Security}: $\forall \epsilon_s > 0$, $\forall \mathcal{W} \subset{\mathcal{L}}$ s.t. $|\mathcal{W}| = w < \ell$, and $\forall \underline{V}_{\mathcal{K}_s} \subset \underline{V}_{\mathcal{L}}$ s.t. $|\mathcal{K}_s|=k_s < \ell - w$, it holds that $\mathbb{V}(p_{\underline{Z}_{\mathcal{W}}|{\underline{V}_{\mathcal{K}_s}=\underline{v}_{\mathcal{K}_s}}},p_{\underline{Z}_{\mathcal{W}}}) \leq \epsilon_s$.
        \item \underline{Reliability}: $\forall \epsilon_e > 0$ it holds that $\mathbb{P}(\hat{\underline{V}}_{\mathcal{L}}(\underline{Y}_{\mathcal{L}}) \neq \underline{V}_{\mathcal{L}}) \leq \epsilon_e$, where $\hat{\underline{V}}_{\mathcal{L}}(\underline{Y}_{\mathcal{L}})$ is the decoding estimation of the message matrix from Bob's observations. 
    \end{enumerate}
\end{definition}
Thus, IT-Eve that observes any subset of $w$ links in the network can't obtain any information about any set of $k_s < \ell - w$ individual messages, $\underline{V}_{\mathcal{K}_s}$. However, IT-Eve might be able to obtain some insignificant information about the mixture of all the messages. Yet, this negligible information is controlled \cite{SMSM,SCMUniform,bhattad2005weakly}. Bob can reliably decode the message matrix from his observations of the encoded messages.

\subsection{Security against Crypto-Eve}
Crypto-Eve may perform passive/active attacks against Alice and the ciphertexts she produces, to obtain information about confidential messages. Two of the most common attacks given in the literature are the chosen-plaintext attack (CPA) and the chosen-ciphertext attack (CCA) \cite{bellare1998relations}. In both attacks, Crypto-Eve is given a ciphertext from which she tries to obtain information about the plaintext. In CCA Crypto-Eve is active and prior to receiving the ciphertext, she can question a decryption oracle to provide her the plaintexts for a limited number of ciphertexts of her choice. However, she can't use this decryption oracle after receiving the test ciphertext \cite{bellare1998relations}. The information leakage is measured by Crypto-Eve's ability to distinguish between plaintexts given the test ciphertext provided by Alice. There are two ways in which this leakage is measured: 1) indistinguishability (IND) - by observing a ciphertext created from one of two possible plaintexts, Crypto-Eve can't distinguish between the two plaintexts better than a uniform coin-toss, 2) semantic security (SS) - by observing a ciphertext, Crypto-Eve can't obtain any information about the original plaintext or some function applied on it, that is more significant than the information she obtains from the plaintexts original probability distribution. Those two definitions are equivalent, i.e. a cryptosystem that is IND is also SS and vice versa \cite{goldwasser2019probabilistic}. The security level of a cryptosystem is defined by the combination of Crpyto-Eve's attack model and its information leakage.

In this paper, we introduce a new notion of security, individual semantic security against a chosen ciphertext attack (ISS-CCA1). This notion of security is based on SS-CCA1 cryptographic security \cite{goldwasser2019probabilistic,bellare1998relations}, and usually requires the encryption scheme to be probabilistic\footnote{\label{comm:pub_key} To focus on our main contributions, we choose to work with public key encryption schemes, but any other PQ probabilistic encryption schemes would have achieved similar results.}. To properly define ISS-CCA1 we start by providing the formal definitions of public key cryptosystems and SS-CCA1 \cite{bellare1998relations,dowsley2009cca2,aguirre2019ind}.

\begin{definition}\label{def:Public-key}
    A public key cryptosystem consists of three algorithms:
    \begin{enumerate}
        \item Key generation algorithm $Gen(c)$ with an input $c$ which generates a public key, $p_c$, and a secret key, $s_c$.
        \item An Encryption algorithm used by Alice which is taking a message $m$ and the public key, $p_c$ as an input, and outputs the ciphertext $\kappa$. We denote the encryption algorithm as $\kappa = Crypt(m,p_c)$.
        \item A polynomial time decryption algorithm taking the ciphertext, $\kappa$, and the secret key, $s_c$, and output the original message $m = Crypt^{-1}(Crypt(m,p_c),s_c)$.
    \end{enumerate}
\end{definition}

\off{The notion of semantic security (SS) in computational cryptographic security is the equivalent of Shannon's perfect secrecy against a computationally bounded eavesdropper\cite{goldwasser2019probabilistic}. \textcolor{blue}{Consequently, we provide the definition for SS-CCA1 \cite{bellare1998relations,dowsley2009cca2,aguirre2019ind}.}}

\begin{definition}\label{def:SS-CCA1}
    (SS-CCA1). Semantic security under chosen ciphertext attack (SS-CCA1) is a computational security notion defined by a game played between a challenger and an adversary. The game is played the following way:
    \begin{enumerate}
        \item The challenger generates a key pair using a security parameter $c$: $Gen(c) = (p_c,s_c)$ and shares $p_c$ with the adversary.
        \item The adversary outputs a pair $(\mathcal{M},S)$, where $\mathcal{M}$ is a messages space, and $S$ is a state.
        \item The adversary sends a polynomial number of ciphertexts to a decryption oracle that decrypts them and returns them to the adversary.
        \item Let $\mathcal{F} = \{f : \mathcal{M} \rightarrow \Sigma\}$ be the set of functions on the message space. For any value $\sigma \in \Sigma$ we denote by $f^{-1}(\sigma)$ the inverse image of $\sigma$ on the message space, $\{m \in \mathcal{M}|f(m)=\sigma \}$. We denote the probability for the most probable value for $f(m)$ as $p_{\sigma-max} = \max\limits_{\sigma \in \Sigma}\{ \sum_{m\in f^{-1}(\sigma)} p(m) \}$.
        \item The challenger samples a message $m \in \mathcal{M}$ according to its probability distribution.
        \item The challenger encrypts the message $\kappa = Enc(m,p_c)$, and sends it to the challenger.
        \item The adversary tries to find a pair $(f,\sigma)$ s.t. $\sigma=f(m)$.
    \end{enumerate}
    If the adversary finds a pair $(f,\sigma)$ s.t. $\sigma=f(m)$ he wins. The cryptosystem is considered semantically secured under chosen ciphertext attack if for any pair $(\mathcal{M},S)$ and for any function $f \in \mathcal{F}$, the advantage the adversary has over guessing according to the message space probability distribution is negligible. We denote a function $\epsilon(c)$ s.t. for every $d > 0 $, there exists an integer $c_d$ such that $\forall c > c_d$, the bound $\epsilon(c) < \frac{1}{c^d}$ holds. The function $\epsilon(c)$ is called a negligible function. The advantage of the adversary is measured by the adversary's probability to win: $p_{\sigma-max} + \epsilon(c)$. When $\epsilon(c)$ is a negligible function, the adversary's advantage is negligible as well.
\end{definition}

Now, we introduce the new notion of semantic security that is based on the definition of SS-CCA1, Definitions~\ref{def:IS}, and~\ref{def:Public-key}.

\begin{definition}\label{def:individuall-SS-CCA1}
    (ISS-CCA1). Individual semantic security under a chosen ciphertext attack is defined by the following game between an adversary and a challenger:
    \begin{enumerate}
        \item The challenger generates a key pair using a security parameter $c$: $Gen(c) = (p_c,s_c)$ and shares $p_c$ with the adversary.
        \item The adversary outputs a pair $(\mathcal{M},S)$, where $\mathcal{M} \in \mathbb{F}_{q}^{\ell}$ is the messages space, and $S$ is a state.
        \item  The adversary sends a polynomial number of ciphertexts to the challenger that decrypts them and returns them to the adversary.
        \item The adversary chooses an index $i^* \in \{1,...,\ell\}$ and tells it to the challenger.
        \item Let $\mathcal{F} = \{f : \mathcal{M} \rightarrow \Sigma\}$ be the set of functions on the message space. For any value $\sigma \in \Sigma$ we denote by $f^{-1}(\sigma)$ the inverse image of $\sigma$ on the message space, $\{m \in \mathcal{M}|f(m)=\sigma \}$. We denote the probability for the most probable value for $f(m)$ as $p_{\sigma-max} = \max\limits_{\sigma \in \Sigma}\{ \sum_{m\in f^{-1}(\sigma)} p(m) \}$.
        \item For every $i \in \{1,...,\ell\}$, we consider $m_i$ as an individual message. The challenger samples each $m_i$ from its probability distribution creating the message $m$. 
        \item The challenger encrypts all the messages $\kappa = Crypt(m,p_c)$ and sends it to the adversary.
        \item The adversary tries to find a pair $(f,\sigma)$ s.t. $\sigma = f(m_{i^*})$.
    \end{enumerate}
    If the adversary finds a pair $(f,\sigma)$ s.t. $\sigma=f(m_{i^{*}})$ he wins. The cryptosystem is considered ISS-CCA1 if for any pair $(\mathcal{M},S)$ and for any function $f \in \mathcal{F}$, the advantage the adversary has over guessing according to the message space probability distribution is negligible. We denote a function $\epsilon(c)$ s.t. for every $d > 0 $, there exists an integer $c_d$ s.t. $\forall c > c_d$, the bound $\epsilon(c) < \frac{1}{c^d}$ holds. The function $\epsilon(c)$ is called a negligible function. The advantage of the adversary is measured by the adversary's probability to win: $p_{\sigma-max} + \epsilon(c)$. When $\epsilon(c)$ is a negligible function, the adversary's advantage is negligible as well.
\end{definition}
This definition of ISS-CCA1 guarantees that Crypto-Eve's advantage over guessing an individual $m_i$ according to its distribution is negligible for all $1 \leq i \leq \ell$. 

\off{\section{Background}
\textcolor{blue}{In this section we provide a brief description of the coding and encryption schemes that are used as the building blocks for NU-HUNCC: 1) almost uniform source coding scheme, 2) Linear IS channel coding scheme, 3) IS random coding channel code, and 3) a SS-CCA1 cryptosystem.}

\subsection{Source Code}
For the almost uniform source coding a specialized version of a polar-codes-based source encoder is used \cite[proposition 4]{NegligbleCost}. Let $m \in \mathbb{N}$, and the source messages blocklength be $n = 2^m$. First, each row of the source message matrix, $\underline{V}_{\mathcal{L}}$, is separately encoded by the polarization transformation \cite{ArikanBase2009}. Let $\delta_n \triangleq 2^{-n^\beta}$ for $\beta \in [0,\frac{1}{2})$. We denote a polarized message by $A \in \mathbb{F}_2^n$, and by using $\delta_n$ the set of $n$ polarized bits are divided into three groups: 1) The bits with entropy almost 1 - $\mathcal{H}_V \triangleq \{ j \in [1,n]: H(A^{(j)}|A^{j-1}) > 1 - \delta_n \}$, 2) The bits with entropy almost 0 - $\mathcal{U}_V \triangleq \{ j \in [1,n]: H(A^{(j)}|A^{j-1}) < \delta_n \}$, and 3) The complementary group of bits with entropy in-between - $\mathcal{J}_V = (\mathcal{U}_V \cup \mathcal{H}_V)^{C}$. The compressed message is formed from $\mathcal{H}_V \cup \mathcal{J}_V$, the bits with non-negligible entropy, i.e., groups 1 and 3. For the compressed message to be almost uniform, a uniform seed of size $|\mathcal{J}_{V}| \triangleq d_{J}$ must be used on each message separately. The uniform seed one-time pads the bits from group $\mathcal{J}_V$. We denote the random seed matrix for all the messages by $\underline{U}_{d_{J}} \in \mathbb{F}_{2}^{\ell \times d_{J}}$. The size of the almost uniform message is $\tilde{n} = |\mathcal{H}_{V}| + d_{J}$ and the encoder for the $i$-th row the source message matrix denoted by $f_{s,n}$, operates as follows 
\begin{gather} \label{eq:src_code}
    f_{s,n} : (\underline{V}_{\mathcal{L},i},\underline{U}_{d_{J},i}) \in \mathbb{F}_{2}^{n} \times \mathbb{F}_{2}^{d_{J}} \rightarrow \underline{M}_{\mathcal{L},i} \in \mathbb{F}_{2}^{\tilde{n}}
\end{gather}
\begin{remark} \label{rm:polar}
    The size of the seed can be bounded by $n^{0.7214} \leq d_{J} \leq n^{0.7331}$. A more detailed analysis \off{of the seed's size}is given in Appendix~\ref{appendix:seed_length}. \off{In \cite{ArikanBase2009} it has been shown that $\lim \limits_{n \rightarrow \infty} \frac{|\mathcal{H}_V|}{n} = H(V)$. Thus, By choosing $n$ large enough, the compression is close to its optimal rate while the seed size is negligible compared to the compressed message size.}
\end{remark}
\textcolor{blue}{For a more detailed description of the source coding scheme we refer to \cite[Sec. ]{NegligbleCost} and to Appendix~\ref{}.}

\subsection{Linear IS channel code}
\textcolor{blue}{We use a linear $(w,\ell)$ over the field $\mathbb{F}_{2^\mu}$ where $\mu \geq \ell$ as given in \cite[Sec. ]{SMSM}. The code pre-mixed the messages using its generator matrix and the generator for the null space of the code. Let $w < \ell$ be the number of paths observed by IT-Eve and $k_s \triangleq \ell - w$ be the number of messages kept secured from IT-Eve. The generator matrix of the code is denoted by $\underline{G} \in \mathbb{F}_{2^\mu}^{w \times \ell}$ and the generator for the code's null space is denoted by $\underline{G}^{\star} \in \mathbb{F}_{2^\mu}^{k_s \times \ell}$. }
\subsection{Random IS channel code}
We use a random coding-based joint encoder as given in \cite[Sec. IV]{SMSM}. The channel code pre-mixes the messages by encoding the message matrix column by column.\off{The codebook is generated from an i.i.d. source $X ~ Bernouli(\frac{1}{2})$ s.t. the probability distribution of each codeword, $x^l$, is $P(x^{\ell}) = \prod_{i=1}^{\ell}P(x_i)$.} The random codebook is organized in a binning structure of $2^{k_s}$ bins s.t. each bin has $2^{k_w}$ codewords, where $k_w \triangleq \ell - k_s$. For each column in the message matrix, $k_s$ bits are used to choose a bin, and the other $k_w$ bits are used to choose a codeword inside the bin. Thus, we receive the codeword matrix $\underline{X}_{\mathcal{L}}$. The encoder of a column is denoted by $f_{c,\ell}$. The encoding of the $j$-th column operates as follows
\begin{gather} \label{eq:msg_code}
    f_{c,\ell} : \underline{M}_{\mathcal{L}}^{(j)} \in \mathbb{F}_{2}^{\ell} \rightarrow \underline{X}_{\mathcal{L}}^{(j)} \in \mathbb{F}_{2}^{\ell}
\end{gather}

Using the source and IS random coding channel scheme against IT-Eve, we provide the following achievability theorem.

\subsection{SS-CCA1}}

\off{
\section{Related Works}
\textcolor{blue}{
In 2021 Cohen et. al. \cite{HUNCC} offered the Hybrid Universal Network Coding Cryptosystem (HUNCC) as a PQ-secured scheme operating at a high communication rate. HUNCC considers the problem of sending $\ell$ messages over a multipath network with $\ell$ links in the presence of IT-Eve and Crypto-Eve. IT-Eve has access to any subset of size $w < \ell$ of the links. Crypto-Eve has access to all of the links, but her computational power is limited. By employing the linear IS code from \cite[Sec. \uppercase\expandafter{\romannumeral6}]{SMSM} and encrypting a subset of $1 \leq c < \ell$ links, HUNCC obtains PQ security against both IT-Eve and Crypto-Eve while maintaining a high communication rate. The PQ security of HUNCC heavily relies on the input messages being from a uniform distribution \cite{SMSM,HUNCC}. This assumption is not trivial and often hard to enforce.}

\textcolor{blue}{The main idea behind the IS channel codes used by HUNCC \cite{SMSM} is to use some subset of the messages as a source of randomness to protect the other messages as in the wiretap code \cite{WiretapWyner}. This idea was first introduced by Carliel and Hellman in \cite{carleial1977note} as an extension to Wyner's work on the wiretap channel and Ozarow's and Wyner's work on the wiretap channel Type \uppercase\expandafter{\romannumeral2} \cite{ozarow1984wire}. Instead of using a source of randomness, Carliel and Hellman suggested breaking the message into $m$ parts s.t. each piece will use the other pieces as a source of randomness. Considering a BSC($p$) channel to Eve and a noiseless channel to Bob, Carliel and Hellman showed that by ensuring $\ell h(p) \geq m$, where $\ell$ is the size of the message, each piece of the message can be transmitted securely at full capacity over the main channel. This result ensured a high communication rate in the expanse of reducing the security level of the communication.}

\textcolor{blue}{The idea of using messages as sources of randomness was further developed by Kobayashi and Yamamoto in \cite{SCMUniform}. In their work, Kobayashi and Yamamoto consider $m$ messages transmitted between Alice and Bob over the wiretap channel.
The messages are encoded by a stochastic encoder. Each group of $m-1$ messages serves as the source of local randomness that is used to protect the remaining message. Kobayashi and Yamamoto show the required conditions to keep Eve ignorant about any single message. However, the suggested coding scheme still relies on the messages being from a uniform distribution.}

\textcolor{blue}{IS was also addressed in the context of network coding \cite{bhattad2005weakly}. In a network coding setting, $n$ packets of size $m$ are sent over the network with an eavesdropper that has access to any $\mu$ of the packets. If the eavesdropper is ignorant about any individual packet, the code is considered IS. Linear network codes can attain IS when used applied over an extension field \cite{silva2011universal}. The linear code used by HUNCC \cite{SMSM,HUNCC} is a linear maximum rank distance (MRD) \cite{gabidulin1985theory,roth1991maximum} network code operating over an extension field that must be larger than the number of messages being sent to Bob. The code is structured similarly to a coset code s.t. one piece of the message chooses the coset and the other piece of the message chooses the codeword from the coset.}

\textcolor{blue}{A key requirement for a code to be IS is the input messages to be from a uniform distribution similar to the local source of randomness from the wiretap code \cite{WiretapWyner}. In a work by Bloch \cite{bloch2012secure}, wiretap codes using constrained randomization were suggested. In his paper, Bloch used a public message that is not necessarily uniform as the source of local randomness and showed the constraint required on the rate of the public and confidential messages to obtain strong secrecy. It was shown that by using a structured binning coding scheme and requiring that $\frac{1}{n}H_{2}(M_p) \geq I(X;Z|Q)$, where $H_{2}(M_p)$ is the Renayi entropy \cite{golshani2009some} and $Q$ is an auxiliary random variable, strong secrecy is attained.}

\textcolor{blue}{Uniform distribution of the input messages is hard to enforce whereas using non-uniform public messages reduces the communication rate since Renayi entropy is lower or equal to the standard entropy i.e., $H_2(M) \leq H(M)$. In a work by Chou \cite{chou2013data} the conditions for a completely uniform lossless source encoder were first introduced. In his work, Chou showed that for a lossless source encoder to output messages that are uniform in the variational distance, a sub-linear seed must be used. Furthermore, he offered a code construction that achieves uniformity in the variational distance by using a shared uniform seed. However, the proposed source encoder/decoder is based on typical sequence encoding/decoding, rendering them not practical.}

\textcolor{blue}{In \cite{NegligbleCost}, Chou et. al. showed a wiretap code construction attaining strong secrecy by using a public message that is losslessly compressed by an almost uniform source encoder with a seed shared between the encoder and the decoder. Furthermore, a practical lossless almost uniform source encoder based on polar codes was introduced. By using a sub-linear shared seed and a standard polar codes-based source code, a message uniform in variational distance is obtained. The achievable rate of this code differs from the achievable rate in \cite{SCMUniform} and \cite{SMSM} since there is no security requirement on the public message.}

\textcolor{blue}{The computational cryptographic security of HUNCC was addressed in \cite{HUNCC} and \cite{cohen2022partial}. In \cite{HUNCC} it was shown that HUNCC is $b$-bit secured, i.e. the amount of binary operation for Eve to decrypt the messages without the secret key is $2^b$. $b$ is a function of the cryptosystem HUNCC uses to encrypt $c$ of the links. In \cite{cohen2022partial} the linear network code for the channel is replaced by a random coding scheme from \cite{SMSM}. The random coding scheme is shown to be Individually IND-CCA1 (IIND-CCA1) i.e., each message is IND-CCA1 by itself.}

\textcolor{blue}{
The complexity of HUNCC, was addressed in \cite{d2021post}. As opposed to a system that encrypts all its links by McEliecce, it was shown that in HUNCC, decreasing the number of encrypted links down to one reduces the complexity of the communication significantly while maintaining PQ security against Crypto-
Eve.}
}

\section{Background} \label{sec:background}
In this section, we provide a description of the coding and encryption schemes used as the main building blocks for NU-HUNCC: 1) Almost uniform source coding scheme, 2) Linear IS channel coding scheme, 3) Non-linear IS channel coding scheme, and 4) PQ SS-CCA1 cryptosystem.

\subsection{Source Code} \label{sec:source_code}
A specialized source encoder with an outcome uniform in the non-normalized variational distance was introduced by Chou et. al. in \cite{chou2013data}. In their work, Chou et. al. showed that for a lossless source encoder to output messages that are uniform in the non-normalized variational distance, a sub-linear seed must be used. Furthermore, they offered a code construction that achieves uniformity in the variational distance by using a uniform seed shared between the encoder and the decoder. However, their proposed source coding scheme is based on typical sequences, rendering them not practical.

In \cite{chou2015polar,NegligbleCost}, Chou et. al. proposed a polar-codes based source encoder/decoder that utilizes a sub-linear shared uniform seed to obtain information uniform in \emph{non-normalized KL-divergence}. In \cite{NegligbleCost}, Chou et. al. showed strong secrecy is obtained over broadcast wiretap channel \cite{csiszar1978broadcast} by compressing a public message using the proposed almost uniform source encoder and using its outcome as a source of randomness for the wiretap code.\off{ The achievable rate obtained by this code differs from the rates obtained in \cite{SCMUniform} and \cite{SMSM} since there is no requirement for the security of the public message.}

We introduce here the specialized version of a polar-codes-based source encoder with an almost uniform outcome given in \cite[proposition 4]{NegligbleCost}. Let $m \in \mathbb{N}$ and the blocklength be $n = 2^m$. We denote by $\underline{G}_n$, the polarization transform as defined in \cite{ArikanBase2009} s.t. $\underline{G}_n = \underline{P}_n\begin{bmatrix} 1 & 0 \\ 1 & 1 \end{bmatrix}^{\otimes m} \in \mathbb{F}_{2}^{n \times n}$, where $\otimes$ is the Kronecker product, and $\underline{P}_n$ is the bit reversal matrix. 

First, the source message, $V_n \in \mathbb{F}_{2}^{n}$ is polarized by the polarization transformation: $A = V_n^T \cdot \underline{G}_n \in \mathbb{F}_{2}^{n}$.\off{ In our proposed scheme, each row of the source message matrix $\underline{V}_{\mathcal{L}}$ is separately encoded by the polarization transformation $\underline{G}_n$: $\underline{A}_n = 
     \underline{V}_\mathcal{L} \cdot \underline{G}_n$.} Let $\delta_n \triangleq 2^{-n^\beta}$, for $\beta \in [0,\frac{1}{2})$.\off{ For each row $A \subset \underline{A}_n$, the bits are divided into three groups:} The bits in $A$ are divided into three groups
\begin{equation} \label{eq:unreliable-group}
    \begin{aligned}
        1) \quad \mathcal{H}_V \triangleq \left\{ j \in [1,n]: H\left(A^{(j)}|A^{j-1}\right) > 1 - \delta_n \right\},
    \end{aligned}
\end{equation}
\begin{equation*}
    \begin{aligned}
     \hspace{-1.05cm}   2) \quad \mathcal{U}_V \triangleq \left\{ j \in [1,n]: H\left(A^{(j)}|A^{j-1}\right) < \delta_n \right\},
    \end{aligned}
\end{equation*}
\begin{equation} \label{eq:seed-group}
    \begin{aligned}
     \hspace{-4.00cm}  3) \quad  \mathcal{J}_V \triangleq (\mathcal{U}_V \cup \mathcal{H}_V)^{C}.
    \end{aligned}
\end{equation}
Traditionally, the compressed message in polar codes-based source coding is obtained from the concatenation of groups, $\mathcal{H}_V$, and $\mathcal{J}_V$, the bits with non-negligible entropy. However, the entropy of each bit from group $\mathcal{J}_V$ is not necessarily high, i.e. their distribution is not necessarily uniform. This implies that the distribution of the entire compressed message is almost uniform in \emph{normalized} KL-divergence.
To ensure the entire compressed message is uniform in the \emph{non-normalized} variational distance, the bits from group $\mathcal{J}_V$ undergo one-time padding \cite{Shannon1949,matt2013one} with a uniform seed of size $|\mathcal{J}_{V}| \triangleq d_{J}$. We denote, the uniform seed by $U_{d_{J}}$. \off{\textcolor{blue}{Each message is one-time padded by a different realization of the uniform seed.}}\off{A different seed is used for each row Thus, we denote the seed matrix by $\underline{U}_{d_{J}} \in \mathbb{F}_{2}^{\ell \times d_{J}}$.} Finally, \off{ the $i$-th row of }the almost uniform compressed message\off{ matrix} is given by
\begin{equation} \label{eq:output_src}
    \begin{aligned}
        \off{\underline{M}_{\mathcal{L},i}}M \triangleq \left[A[\mathcal{H}_V],A[\mathcal{J}_V] \oplus U_{d_{J}}\right].
    \end{aligned}
\end{equation}
Here $\oplus$ denotes the xor operation over $\mathbb{F}_2$. $A[\mathcal{H}_V]$ and $A[\mathcal{J}_V]$ are the bits in indexes $\mathcal{H}_V$ and $\mathcal{J}_V$ of $A$, respectively.
The size of each almost uniform message obtained at the outcome of the source coding scheme is $\tilde{n} = |\mathcal{H}_V| + d_{J}$. We denote the encoder of each message by $f_{s,n}$ and it operates as follows
\begin{equation} \label{eq:src_code}
\begin{aligned}
    f_{s,n} : \off{(\underline{V}_{\mathcal{L},i},\underline{U}_{d_{J},i}) \in \mathbb{F}_{2}^{n} \times \mathbb{F}_{2}^{d_{J}}} (V_n, U_{d_J}) \in \mathbb{F}_{2}^{n} \times \mathbb{F}_{2}^{d_J} \rightarrow \off{\underline{M}_{\mathcal{L},i}} M \in \mathbb{F}_{2}^{\tilde{n}}.
    \end{aligned}
\end{equation}
\begin{remark} \label{rm:polar}
    The size of the seed can be bounded by $n^{0.7214} \leq d_{J} \leq n^{0.7331}$. A more detailed analysis is given in Sec.~\ref{sec:seed_length}.
\end{remark}

The decoding process is divided into two parts. First, Bob one-time pads the bits from the group $\mathcal{J}_V$ using the seed shared with him by Alice. Second, Bob uses a successive cancellation decoder to reliably decode the original message \cite{ArikanBase2009,cronie2010lossless}. We denote the decoding function by $g_{s,n}$ and it operates as follows
\begin{equation} \label{eq:src_decode}
\begin{aligned}
    g_{s,n} : (M,U_{d_J}) \in \mathbb{F}_{2}^{\tilde{n}} \times \mathbb{F}_{2}^{d_J} \rightarrow V_n \in \mathbb{F}_{2}^{n}.
\end{aligned}    
\end{equation}
\off{In the suggested scheme, this decoding process is performed separately on each row of the message matrix. \textcolor{blue}{We provide the error probability analysis is Sec.~\ref{sec:linear-code-reliability} and Sec.~\ref{sec:random-code-reliability}.}}
\subsection{Channel Code}
For the channel coding, we provide two coding schemes. The first is a practical linear IS channel code from \cite[Sec. \uppercase\expandafter{\romannumeral6}]{SMSM}. The second is a non-linear IS channel code from \cite[Sec. \uppercase\expandafter{\romannumeral4}]{SMSM} which we provide for completeness. In Section~\ref{subsec:diss}, we compare and discuss the advantages/disadvantages of each of these secure channel coding schemes given in this section.  

\subsubsection{Linear IS channel code} \label{sec:linear-IS-code}

\off{We use the joint linear encoder as given in  We describe the joint linear encoder given in\cite[Sec. \uppercase\expandafter{\romannumeral6}]{SMSM}.} Let $w < \ell$ be the number of links observed by IT-Eve, and $k_s \leq \ell - w$ be the number of messages kept secure from IT-Eve. The code operates over the extension field $\mathbb{F}_{2^\mu}$, s.t. $\mu \geq \ell$ is a fixed integer.\off{ each $\mu$ bits of the messages are considered to be a symbol over $\mathbb{F}_{2^\mu}$ }\footnote{The coding should be over the extension field $\mathbb{F}_{q^\mu}$. We chose $q = 2$ throughout this paper to focus our paper on our main results. Any other choice of prime $q$ would provide the same results.}.\off{ We denote by $\tilde{n}^{*} = \frac{\tilde{n}}{\mu}$ the number of symbols in each row of the message matrix s.t. $\underline{M}_{\mathcal{L}} \in \mathbb{F}_{2^\mu}^{\ell \times \tilde{n}^{*}}$.} We provide here the detailed codebook generation for this code:

\underline{\textit{Codebook Generation:}} First, we denote the linear code over $\mathbb{F}_{2^\mu}$ with dimensions $\ell - k_s$ and length $\ell$ by $\mathcal{C}$. Let $\underline{H} \in \mathbb{F}_{2^\mu}^{k_s \times \ell}$ and $\underline{G} \in \mathbb{F}_{2^\mu}^{(\ell - k_s) \times \ell}$ be the parity check matrix and generator matrix of the code, respectively. We consider $2^{\mu k_s}$ cosets of size $2^{\mu (\ell - k_s)}$. In addition, let $\underline{G}^{\star} \in \mathbb{F}_{2^\mu}^{k_s \times \ell}$ span the null space of the code $\mathcal{C}$ i.e. $\underline{G}^{\star}$ is composed from $k_s$ linearly independent rows from $\mathbb{F}_{2^\mu}^\ell \setminus \mathcal{C}$. The linear code $\mathcal{C}$ is considered a Maximum Rank Distance (MRD) code \cite{gabidulin1985theory,roth1991maximum}\footnote{MRD codes were used in previous works as network codes and have been shown to achieve IS \cite{silva2009universal,silva2011universal}.}.\off{ In our setting, we consider multiple messages transmitted over a multipath (or a single path) between legitimate receivers. This code have been shown to be IS over this setting as well \cite{SMSM}. However, the messages in \cite{SMSM} are from a uniform distribution, an assumption that we omit in our setting.}

\underline{\textit{Encoding:}} Each message, $M \in \mathbb{F}_{2^\mu}^{\ell}$ is divided into two parts\off{ Each column of the message matrix $\underline{M}_{\mathcal{L}}$ 
 is divided into two parts}: $M_{k_s} \in \mathbb{F}_{2^\mu}^{k_s}$ and $M_{w} \in \mathbb{F}_{2^\mu}^{\ell-k_s}$. $M_{k_s}$ are the symbols considered IS.\off{ whereas $M_{w}$ are used instead of a local source of randomness. For each column of the message matrix, } $M_{k_s}$ is used to choose a coset out of $2^{\mu k_s}$ possible cosets, and $M_w$ is used to choose the codeword within the coset. The encoding process can be simplified using the generator matrix $\underline{G}$ and the matrix $\underline{G}^{\star}$. We denote the encoding matrix by $\underline{G}_{IS} = \begin{bmatrix} \underline{G}^{\star} \\ \underline{G} \end{bmatrix} \in \mathbb{F}_{2^\mu}^{\ell \times \ell}$. The codeword $X \in \mathbb{F}_{2^\mu}^{\ell}$  is obtained from the multiplication of the message \off{The codeword matrix $\underline{X}_{\mathcal{L}} \in \mathbb{F}_{2^\mu}^{\ell \times \tilde{n}^{*}}$ is obtained from the multiplication of the message matrix }with the IS matrix $\underline{G}_{IS}$. The length of the codeword remains the same as the length of the message since the links are noiseless and error correction is not required  \cite{SMSM}. We denote the encoding function by $f_{c,\ell}$ s.t.
 \begin{equation} \label{eq:msg_code}
 \begin{aligned}
    \off{f_{c,\ell}(\underline{M}_{\mathcal{L}}) = \underline{M}_{\mathcal{L}}^{T} \cdot \underline{G}_{IS}.}f_{c,\ell}(M) = M^T \cdot \underline{G}_{IS}.
\end{aligned}
\end{equation}
\underline{\textit{Decoding:}}\off{ The codeword matrix $\underline{X}_{\mathcal{L}}$ is directly obtained by the legitimate receiver.} To compute \off{each}$M_{k_s}$ from the codeword\off{ matrix}, the parity check matrix, $\underline{H}$, is used. It was shown in \cite{SMSM} that $M_{k_s} = \underline{H} \cdot X$. To obtain $M_w$ we first introduce the matrix $\underline{\tilde{G}} \in \mathbb{F}_{2^\mu}^{w \times \ell}$ which upholds $\underline{\tilde{G}} \cdot \underline{G}^{T} = 0$ and $\tilde{\underline{G}}\cdot \underline{G}^{\star T} = \underline{I}$. As shown in \cite{SMSM}, we obtain $M_w$ from $M_w = \underline{\tilde{G}}\cdot X$.\off{\footnote{After obtaining $M_{k_s}$ which represents the coset of the codeword, $M_w$ can be obtained from the index of the codeword inside the coset. However, matrix multiplication using $\underline{H}$ and $\tilde{\underline{G}}$ is much more efficient.}}

\off{\textcolor{blue}{Considering IT-Eve observes $w$ links, this code guarantees the IS of at most $k_s = \ell - w$ confidential messages. From, now on we assume $k_s = \ell - w$ for the linear IS code $\mathcal{C}$.}}

\subsubsection{Non-Linear IS channel Code} \label{sec:random-IS-code}
\off{The IS random channel coding scheme is applied separately on each column of the message matrix $\underline{M}_{\mathcal{L}}$.} Let $w < \ell$ be the number of links observed by IT-Eve, and $k_s \leq \ell - w - \ell\epsilon$ be the number of messages kept secured from IT-Eve s.t. $\ell\epsilon$ is an integer. We denote $k_w \triangleq w + \ell\epsilon$. Each \off{column } message, $M \in \mathbb{F}_{2}^{\ell}$, \off{of the message matrix }is divided into two parts of sizes $k_s$ and $k_w$. The first part is denoted by $M_{k_s} \in \mathbb{F}_2^{k_s}$, and the second is denoted by $M_{k_w} \in \mathbb{F}_2^{k_w}$. We now give the detailed code construction for the IS non-linear channel code\off{ applied by Alice on each $j$-th column}.

\textit{\underline{Codebook Generation}}: Let $P(x) \sim Bernouli(\frac{1}{2})$. There are $2^{k_s}$ possible messages for $M_{k_s}$ and $2^{k_w}$ possible messages for $M_{k_w}$. For each possible message $M_{k_s}$, generate $2^{k_w}$ independent codewords $x^{\ell}(e)$, $1 \leq e \leq 2^{k_w}$, using the distribution $P(X^{\ell}) = \prod_{j=1}^{\ell}P(X_j)$. Thus, we have $2^{k_s}$ bins, each having $2^{k_w}$ possible codewords. The length of the codeword remains the same as the length of the message since the links are noiseless \cite{SMSM} and error correction is not required.\off{, e.g., as given in \cite{cohen2022partial}.}

\textit{\underline{Encoding}}: For each message, $M \in \mathbb{F}_{2}^{\ell}$, \off{ column in the message matrix,} $k_s$ bits are used to choose the bin, while the remaining $k_w$ bits are used to select the codeword inside the bin.

\textit{\underline{Decoding}}: \off{Bob decodes each column separately. }Bob looks for $X$ in the codebook, if the codeword appears only once in the codebook, then $M_{k_s}$ is the index of the bin in which the codeword was found, and $M_{k_w}$ is the index of the codeword inside the bin. However, if the codeword appears more than once, a decoding error occurs. The probability of a codeword appearing more than once in a codebook decreases exponentially as a function of $\ell$ \cite[Sec. 3.4]{bloch2011physical}. \off{The reliability analysis of the code is given in \cite[Section \uppercase\expandafter{\romannumeral4}]{SMSM}.}

\subsection{Computationally secure cryptosystem} \label{sec:encryption}
We focus on public key cryptosystems and specifically, on the original version of McEliecce based on Goppa codes. A randomized version of this code is considered SS-CCA1 and PQ secured \cite{nojima2008semantic,dottling2012cca2,aguirre2019ind}. We note that Goppa codes can be replaced by other coding schemes. However, those versions of McEliecce are not considered PQ-secured \cite{niederreiter1986knapsack,sidelnikov1992insecurity,sendrier1998concatenated,sidelnikov1994public,minder2007cryptanalysis,monico2000using,londahl2012new,landais2013efficient}. We note that for the proposed scheme, McEliecce can be replaced by any other PQ secured cryptosystem, while the scheme will maintain its PQ security. Nevertheless, we chose to demonstrate our proposed scheme using McEliecce as it is a well-known PQ secure cryptosystem with efficient encryption and decryption mechanism \cite{mceliece1978public}.

\off{\textcolor{blue}{\underline{\textit{Goppa codes}}:}
\textcolor{blue}{Goppa codes are linear error correction codes \cite{berlekamp1973goppa}. The code is defined by a randomly chosen irreducible polynomial $g(z)$ of degree $t$ over $GF(2^d)$, where $d$ is defined later. The set of elements $L$ is defined as the elements of $GF(2^d)$ that are not roots of $g(z)$. The number of elements in $L$ is denoted by $n_g \triangleq |L| = 2^d$. The set of codewords is defined as the solutions to the equation $\sum_{i=1}^{n_g}\frac{c_{g,i}}{z - L_i} = 0 \mod g(z)$. The number of the solutions to those equations defines the dimensions of the code and it is denoted by $c_g$. The length of the code is $n_g$ and it upholds the inequality $c_g \geq n_b - mt$. This set of equations can be viewed as parity check equations and define the parity check matrix $\underline{H}_g \in \mathbb{F}_{2}^{n_g - c_g \times n_g}$. The generator matrix of Goppa codes is denoted by $\underline{G}_g \in \mathbb{F}_{2}^{c_g \times n_g}$. This Goppa codes can correct up to $t$ errors and can be efficiently decoded by a decoding algorithm denoted by $D_g$.}}

The public key for Goppa codes based McEliecce is the multiplication product of the following matrices: \off{\footnote{In our setting we consider a SS-CCA1 version of McEliecce which has to be probabilistic.}, the generator matrix of Goppa codes along with two additional matrices are used to create the public key:} 1) The generator matrix of the Goppa codes denoted by $\underline{G}_g \in \mathbb{F}_{2}^{c_g \times n_g}$ where $c_g$ and $n_g$ are respectively the dimensions and the length of the Goppa codes, 2) A random dense nonsingular matrix $\underline{S} \in \mathbb{F}_{2}^{c_g \times c_g}$, and 3) A random permutation matrix $\underline{P} \in \mathbb{F}_{2}^{n_g \times n_g}$. The encryption scheme operates as follows:

\underline{\textit{McEliecce Cryptosystem}}:
\begin{itemize}
    \item \underline{Public key generation}: The public key is generated by Bob. Bob first generates the generator matrix of the Goppa codes according to $c_g$ and $n_g$. The public key is then $\underline{G}^{pub} =\underline{S}\underline{G}_g\underline{P}$, s.t. $\underline{G}^{pub} \in \mathbb{F}_{2}^{c_g \times n_g}$. Both Alice and Eve have access to $\underline{G}^{pub}$.
    \item \underline{Secret key generation}: The private key consists of $(S,D_g,P)$ where $D_g$ is an efficient decoding algorithm for Goppa codes.
    \item \underline{Encryption}: Alice randomly chooses a vector $a \in \mathbb{F}_{2}^{n_g}$ of weight $t$. The encryption of the message $m \in \mathbb{F}_{2}^{c_g}$ is given by: $\kappa = Crypt(m,\underline{G}^{pub}) = m\underline{G}^{pub} \oplus a$.
    \item \underline{Decryption}: To decrypt the message, Bob first calculates $\kappa P^{-1} = m\underline{S}\underline{G}_g \oplus aP^{-1}$. Now, Bob applies the efficient decoding algorithm $D_g$. Since $aP^{-1}$ has a weight of $t$ it can be considered as an error vector, thus $D_g$ can reliably retrieve $m\underline{S}$ s.t. $m\underline{S} = D_g(\kappa P^-1) = D_g(m\underline{S}\underline{G}_g)$. $\underline{S}$ is inevitable, thus Bob obtains $m = Crypt^{-1}(Crypt(m,\underline{G}^{pub}),(\underline{S},D_g,\underline{P}))$.
\end{itemize}

\off{\textcolor{blue}{Goppa codes can be replaced by other codes. However, those versions of McEliecce are not considered PQ-secured \cite{niederreiter1986knapsack,sidelnikov1992insecurity,sendrier1998concatenated,sidelnikov1994public,minder2007cryptanalysis,monico2000using,londahl2012new,landais2013efficient}. McEliecce cryptosystem can also be replaced by other PQ-secured cryptosystems. Nevertheless, we chose McEliecce because of its encryption and decryption efficiency \cite{mceliece1978public}.}}

\section{Non-Uniform Hybrid Universal Network Coding Cryptosystem (NU-HUNCC)}\label{sec:NU-HUNCC}
\begin{figure*}[htbp]
  \centering
  \includegraphics[width=1\textwidth]{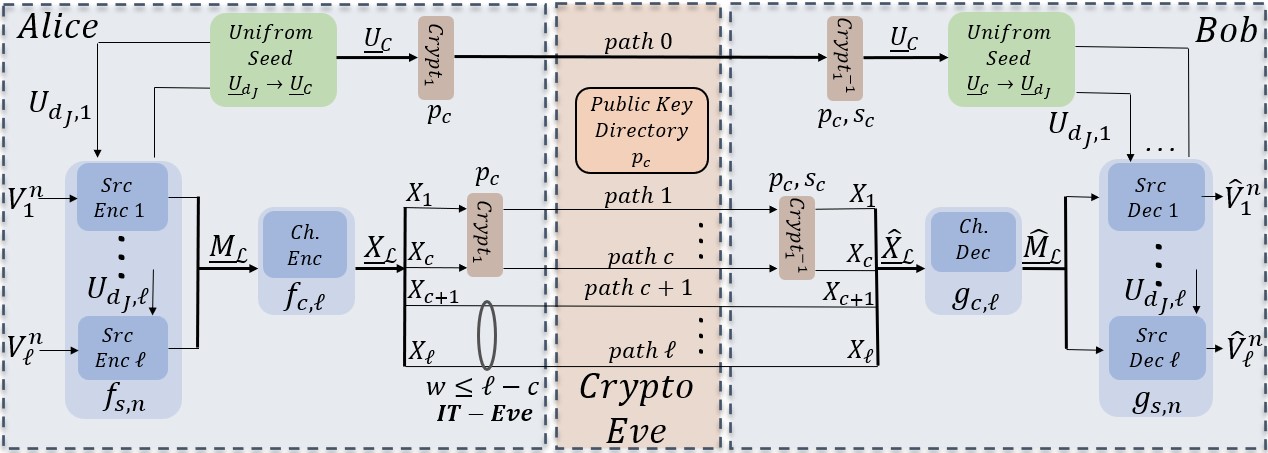}
  \caption{NU-HUNCC cryptosystem with $\ell$ noiseless communication links and two types of Eve's: IT-Eve with access to $w < \ell$ links, and Crypto-Eve with access to all the links. The lossless almost uniform compression is done by the polar codes-based encoder from \cite{NegligbleCost}. $c$ of the links are encrypted by a PQ public-key SS-CCA1 cryptosystem. The mixing of the messages is done by either the linear or non-linear IS network code scheme from \cite{SMSM,cohen2022partial}. The uniform seed is encrypted as well and shared by a separate link. In practice, the encrypted seed is concatenated to the $c$ encrypted messages.}
  \label{fig:NU-HUNCC}
\end{figure*}

In this section, we present NU-HUNCC scheme, a PQ-secured coding and encryption scheme for non-uniform messages. An illustration of the scheme is given in Fig.~\ref{fig:NU-HUNCC}. The full description of the encryption/decryption algorithm is given in Algorithm~\ref{algo:NU-HUNCC}.\off{  
In this section, we present our proposed NU-HUNCC scheme against IT-Eve and Crypto-Eve as illustrated in Fig.~\ref{fig:NU-HUNCC}. We present the main results of our work and the PQ secure code construction for non-uniform messages.} NU-HUNCC provides a way to securely and efficiently transmit non-uniform messages at high data rates between Alice and Bob\off{, while maintaining PQ security}. We present three main novelties in this paper: 1) An efficient end-to-end communication scheme for non-uniform messages, 2) The proposed scheme is $k_s$-IS against IT-Eve, and 3) ISS-CCA1 PQ secure against Crypto-Eve.

NU-HUNCC cryptosystem consists of three main parts: 1) A lossless almost uniform source coding scheme using a uniform seed shared efficiently with the source decoder, 2) A joint linear IS message channel coding scheme to pre-mix the almost uniform messages, and 3) An SS-CCA1 encryption scheme used to encrypt $1 \leq c < \ell$ links chosen by Alice.\footnote{The uniform seed can be shared between Alice and Bob in the expanse of some rate. Against Crypto-Eve, the seed as we demonstrate in this section is shared over one of the encrypted links. Against IT-Eve, the seed can be shared using wiretap coding techniques \cite[Chapter 4]{bloch2011physical}.\label{fn:seed}} \off{\textcolor{blue}{For completeness, in addition to the linear IS code, we provide a random coding based joint IS coding scheme and show that by encrypting a subset $c$ of the links, it also attains ISS-CCA1 against Crypto-Eve.}}

\off{Now we present our proposed NU-HUNCC scheme.}

\subsection{NU-HUNCC - scheme description} \label{sec:linear-NU-HUNCC-scheme-desc}
\off{We provide a detailed description of the encoding/decoding process as presented in Algorithm~\ref{algo:NU-HUNCC}.  We consider a non-uniform DMS $(\mathcal{V},p_V)$ s.t. $\mathcal{V} \in \{0,1\}$. Alice obtains $\ell$ non-uniform source messages of size $n$ from the DMS $(\mathcal{V},p_V)$ denoted by $\{V_i\}_{i=1}^{\ell} \in \mathbb{F}_{2}^{n}$. Alice organizes the messages into a source message matrix, denoted by $\underline{V}_{\mathcal{L}} \in \mathbb{F}_{2}^{\ell \times n}$.} First, Alice employs the nearly uniform source encoder from Sec.~\ref{sec:source_code} (lines 1-7 of Algorithm~\ref{algo:NU-HUNCC}). Each message from the source message matrix is individually encoded using $f_{s,n}$ as given in~\eqref{eq:src_code}. The encoding process for each message involves a distinct realization of a uniform seed of size $d_J$. The uniform seed matrix, denoted by $\underline{U}_{d_J} \in \mathbb{F}_{2}^{\ell \times d_J}$, contains the seeds utilized to encode $\underline{V}_{\mathcal{L}}$. The source encoder yields an almost uniform message matrix, $\underline{M}_{\mathcal{L}} \in \mathbb{F}_{2}^{\ell \times \tilde{n}}$, where $\tilde{n} = |\mathcal{H}_V| + d_J$. Each row in the message matrix $\underline{M}_{\mathcal{L}}$ represents an almost uniform message.

After obtaining the almost uniform messages, Alice mixes the messages using the linear IS channel code from Sec.~\ref{sec:linear-IS-code}. The linear IS channel codes operate over the extension field $\mathbb{F}_{2^{\mu}}$. Since $\mathbb{F}_2^{\mu}$ is a vector space over $\mathbb{F}_2$ via the isomorphism $\mathbb{F}_{2}^{\mu} \rightarrow \mathbb{F}_{2^\mu}$, each $\mu$ bit of the $i$ -th row of $\underline{M}_{\mathcal{L}}$ can be injectively mapped onto a symbol over $\mathbb{F}_{2^\mu}$. Thus, for the channel coding process, we consider the message matrix $\underline{M}_{\mathcal{L}} \in \mathbb{F}_{2^\mu}^{\ell \times \lceil \tilde{n} / \mu \rceil}$ (line 9 of Algorithm~\ref{algo:NU-HUNCC}). \off{where each $\mu$ bits were mapped into a symbol over $\mathbb{F}_{2^\mu}$.} We note that the mapping is performed on each row separately, implying that the rows of the message matrix maintain their independence.

Now, let $1 \leq c < \ell$ be the number of links Alice intends to encrypt and let $w \triangleq \ell - c$ and $k_s \triangleq c$. Let $\mathcal{C}$ be the linear $(w,\ell)$ IS code from Sec.~\ref{sec:linear-IS-code}. We consider  $\underline{G} \in \mathbb{F}_{2^\mu}^{w \times \ell}$, the generator matrix for the code and $\underline{G}^{\star} \in \mathbb{F}_{2^\mu}^{c \times \ell}$, the generator matrix for the null space of the code. The IS channel encoding matrix is given by $\underline{G}_{IS} = \begin{bmatrix} \underline{G}^{\star} \\ \underline{G} \end{bmatrix} \in \mathbb{F}_{2^\mu}^{\ell \times \ell}$. The message matrix is then encoded into the codeword matrix using $f_{c,\ell}$ as in~\eqref{eq:msg_code} s.t. $\underline{X}_{\mathcal{L}}^{T} = \underline{M}_{\mathcal{L}}^{T}\cdot \underline{G}_{IS} \in \mathbb{F}_{2^\mu}^{\ell \times \lceil \tilde{n} / \mu \rceil}$ (lines 8-10 in Algorithm~\ref{algo:NU-HUNCC}).

To ensure the scheme's security against Crypto-Eve, Alice encrypts $c$ links. Let $Crypt^{\dag}_1$ be a SS-CCA1 public key cryptosystem as per Definitions~\ref{def:Public-key} and~\ref{def:SS-CCA1}, respectively. $Crypt^{\dag}_1$ maps each $\mu c$ bits into $\mu (c + r)$ bits where the extra bits represent the added randomness of size $\mu r$ required for $Crypt^{\dag}_1$ to be SS-CCA1 \cite{goldwasser2019probabilistic}\footnote{Encrypting the messages in the presented manner simplifies the security proofs as given in Sec.\ref{sec:linear-code-proof}. Practically, it's also feasible to execute the encryption on each row separately and maintain ISS-CCA1 of NU-HUNCC. The tradeoffs between the encryption alternatives are discussed in Sec.~\ref{sec:linear-code-proof}.}.

Each symbol in the codeword matrix is mapped back into a $\mu$ bits vector s.t. after the mapping the cardinality of the codeword matrix is given by $\underline{X}_{\mathcal{L}} \in \mathbb{F}_{2}^{\ell \times \tilde{n}}$. Without loss of generality, we assume that the encrypted links chosen by Alice are the first $c$ links. Using the encryption scheme and the public key of $Crypt_{1}^{\dag}$ Alice encrypts blocks of $\mu c$ bits from the first $c$ rows of the codeword matrix into a ciphertext of size $\mu (c + r)$. To ensure the security of the scheme, Alice also uses $Crypt_{1}^{\dag}$ to encrypt the uniform seed matrix $\underline{U}_{d_J}$. As part of the encryption process, Alice rearranges the uniform seed matrix into $\underline{U}_{C} \in \mathbb{F}_{2}^{c \times \gamma}$ s.t. $\gamma \triangleq \frac{\ell d_J}{c}$ (line 17 of Algorithm~\ref{algo:NU-HUNCC}). We denote by $\underline{X}^{(j)} = [X^{(j \mu + 1)},...,X^{(j \mu + \mu + 1)}] \in \mathbb{F}_{2}^{c \times \mu}$ the group of $c$ rows and $\mu$ columns from $\underline{X}_{\mathcal{L}}$ where $j \in \{1,...,\lceil \tilde{n} / \mu \rceil\}$. Additionally, we denote by $\underline{U}_{C}^{(k)}=[U_{C}^{(k \mu + 1)},...,U_{C}^{(k \mu + \mu + 1)}] \in \mathbb{F}_{2}^{c \times \mu}$ a group of $c$ rows and $\mu$ columns from $\underline{U}_{C}$ where $k \in \{1,...,\lceil \gamma / \mu \rceil \}$. $Crypt_1^{\dag}$ is employed on: 1) each group $j$ of $\mu$ columns from $\underline{X}_{\mathcal{L}}$, and 2) each group $k$ of $\mu$ columns from $\underline{U}_{C}$ (lines 13-16 and lines 18-21 of Algorithm~\ref{algo:NU-HUNCC})
\begin{multline} \label{eq:linear-encryption}
    Crypt_{1}^{\dag} : \underline{X}^{(j)} \in \mathbb{F}^{c \times \mu}_{2} \rightarrow \underline{Y}^{(j)} \in \mathbb{F}^{(c+r) \times \mu}_{2},\\ 
    \text{and } \underline{U}^{(k)}_{C} \in \mathbb{F}^{c \times \mu}_{2} \rightarrow \underline{Y}^{(\lceil \tilde{n} / \mu \rceil\ +k)} \in \mathbb{F}^{(c+r) \times \mu}_{2},
\end{multline}
where the additional $\mu r$ bits of $\underline{Y}^{(j)}$ and the entire $\underline{Y}^{(\lceil \tilde{n} / \mu \rceil\ +k)}$ are concatenated to the $c$ encrypted links (lines 23-26 of Algorithm~\ref{algo:NU-HUNCC}). We denote by $Crypt_{2}^{\dag}$ the encryption of the $j$-th column of $\underline{M}_{\mathcal{L}} \in \mathbb{F}_{2^\mu}^{\ell \times \lceil \tilde{n} / \mu \rceil}$ s.t.:
\begin{gather} \label{eq:linear_encryption_crypt2}
    Crypt_{2}^{\dag} : \underline{M}_{\mathcal{L}}^{(j)} \in \mathbb{F}^{\ell}_{2^{\mu}} \rightarrow \underline{Y}_{\mathcal{L}}^{(j)} \in \mathbb{F}^{\ell+r}_{2^{\mu}},
\end{gather}
where we injectively mapped Bob's observations over $\mathbb{F}_{2}^{\ell \times \tilde{n}}$ into $\underline{Y}_{\mathcal{L}} \in \mathbb{F}_{2^\mu}^{\ell \times \lceil \tilde{n} / \mu \rceil}$. $Crypt_{2}^{\dag}$ combines the IS linear channel code and the SS-CCA1 PQ-secured encryption scheme $Crypt_{1}^{\dag}$. We denote the encrypted links indexes by $\mathcal{D}$ and non-encrypted links indexes by $\mathcal{D}^{C} = \mathcal{L} \setminus \mathcal{D}$.
Thus, for the $j$-th column Bob's observations are $ \underline{Y}_{\mathcal{L}}^{(j)} = [Crypt_{1}^{\dag}(f_{c,\ell}(\underline{M}_{\mathcal{L}}^{(j)})[\mathcal{D}],p_{\mu c}),f_{c,l}(\underline{M}_{\mathcal{L}}^{(j)})[\mathcal{D}^{C}]]$, where $p_{\mu c}$ is the public key for $Crypt_{1}^{\dag}$ as given in Definition~\ref{def:Public-key}.

We now detail the decoding process at Bob's. First, Bob decrypts the encrypted links and the seed using his secret key $s_{\mu c}$ and the decryption algorithm $Crypt_1^{\dag -1}$\footnote{Against IT-Eve the decryption stage is not required.} (lines 32-40 in Algorithm~\ref{algo:NU-HUNCC}). For each $j \in \{1,...,\lceil \tilde{n} / \mu \rceil\}$ and $k \in \{1,...,\lceil \gamma / \mu \rceil \}$:
\begin{multline} \label{eq:decryption}
    Crypt_{1}^{\dag -1} : \underline{Y}^{(j)} \in \mathbb{F}^{(c+r) \times \mu}_{2} \rightarrow \underline{X}^{(j)} \in \mathbb{F}^{c \times \mu}_{2},\\ 
    \text{and } \underline{Y}^{(\lceil \tilde{n} / \mu \rceil\ +k)} \in \mathbb{F}^{(c+r) \times \mu}_{2} \rightarrow \underline{U}^{(k)}_{C} \in \mathbb{F}^{c \times \mu}_{2}.
\end{multline}
After obtaining the decrypted shared seed $\underline{U}_{C}$, Bob rearranges the matrix back to its original form $\underline{U}_{d_J}$ (line 41 of Algorithm~\ref{algo:NU-HUNCC}). Bob maps the decrypted codeword matrix back to $\underline{X}_{\mathcal{L}} \in \mathbb{F}_{2^\mu}^{\ell \times \lceil \tilde{n} / \mu \rceil}$ (line 43 from Algorithm~\ref{algo:NU-HUNCC}). To derive the message matrix, Bob applies the decoding algorithm of the linear IS channel code, $\mathcal{C}$. 

\begin{breakablealgorithm} \label{algo:NU-HUNCC}
\caption{Non Uniform Hybrid Network Coding Cryptosystem (NU-HUNCC)}\label{alg:cap}
\begin{algorithmic}[1]
\State \textbf{Input:} $\ell$ messages of size $n$ from the source $\mathcal{V} \in \{0,1\}$ arranged in the source message matrix $\underline{V}_{\mathcal{L}} \in \mathbb{F}_{2}^{\ell \times n}$
\State \underline{\textbf{Encoding at Alice:}}
\State \textbf{Source Coding:}
\off{\State Polarize the messages: $\underline{A}_n = \underline{V}_{\mathcal{L}} \cdot \underline{G}_n$}
\State Draw the uniform matrix $\underline{U}_{d_J} \in \mathbb{F}_{2}^{\ell \times d_J}$
\For{\textbf{each} row $i \in \{1,...,\ell\}$ in $\underline{V}_{\mathcal{L}}$}
    \State $ \underline{M}_{\mathcal{L},i} = f_{s,n}(\underline{V}_{\mathcal{L},i},\underline{U}_{d_{J},i})\off{\left[A_i[\mathcal{H}_V],A_i[\mathcal{J}_V] \oplus \underline{U}_{d_{J},i}\right]}$
\EndFor
\State \textbf{Linear IS Channel Coding:}
\State For $\mu \geq \ell$ map $\underline{M}_{\mathcal{L}}$ from $\mathbb{F}_{2}^{\ell \times \tilde{n}}$ to $\mathbb{F}_{2^\mu}^{\ell \times \lceil \tilde{n} / \mu \rceil}$
\State $\underline{X}_{\mathcal{L}}^{T} = \underline{M}_{\mathcal{L}}^{T} \cdot \underline{G}_{IS}$
\State \textbf{Public-key Encryption:}
\State Map $\underline{X}_{\mathcal{L}}$ from $\mathbb{F}_{2^\mu}^{\ell \times \lceil \tilde{n} / \mu \rceil}$ to $\mathbb{F}_{2}^{\ell \times \tilde{n}}$
\For{\textbf{each} $j \in \{1,...,\lceil \tilde{n} / \mu \rceil\}$ in $\underline{X}_{\mathcal{L}}$}
    \State $\underline{X}^{(j)} \in \mathbb{F}^{c \times \mu}_{2} \gets [X^{(j \mu + 1)},...,X^{(j \mu + \mu + 1)}] \in \mathbb{F}_{2}^{c \times \mu}$
    \State $\underline{Y}^{(j)} = Crypt_{1}^{\dag}(\underline{X}^{(j)},p_{\mu c})$
\EndFor
\State $\underline{U}_{C} \in \mathbb{F}_{2}^{c \times \gamma} \gets \underline{U}_{d_J} \in \mathbb{F}_{2}^{\ell \times d_J}$
\For{\textbf{each} $k \in \{1,...,\lceil d_J / \mu \rceil\}$ in $\underline{U}_{C}$}
    \State $\underline{U}_{C}^{(k)} \in \mathbb{F}^{c \times \mu}_{2} \gets [U_{C}^{(k \mu + 1)},...,U_{C}^{(k \mu + \mu + 1)}] \in \mathbb{F}_{2}^{c \times \mu}$
    \State $\underline{Y}^{(\lceil \tilde{n} / \mu \rceil\ +k)} = Crypt_{1}^{\dag}(\underline{U}_{C}^{(k)},p_{\mu c})$
\EndFor
\State \underline{\textbf{Transmission of the encoded messages:}}
\State \textbf{$c$ encrypted links:}
\For{\textbf{each} $j \in \{1,...,\lceil \tilde{n} / \mu \rceil\}$ in $\underline{X}_{\mathcal{L}}$}
    \State Transmit $\underline{Y}^{(j)}$
    \State Concatenate the $r \mu$ extra bits to the end of the transmissions
\EndFor
\For{\textbf{each} $k \in \{1,...,\lceil d_J / \mu \rceil\}$ in $\underline{U}_{C}$}
    \State Concatenate $\underline{Y}^{(\lceil \tilde{n} / \mu \rceil\ +k)}$ to end of the transmission
\EndFor
\State \underline{\textbf{Decoding at Bob:}}
\State \textbf{Public-key Decryption:}
\For{\textbf{each} $j \in \{1,...,\lceil \tilde{n} / \mu \rceil\}$ in $\underline{X}_{\mathcal{L}}$}
    \State $\underline{X}^{(j)} = Crypt_{1}^{\dag -1}(\underline{Y}^{(j)},s_{\mu c})$
        \State $[X^{(j \mu + 1)},...,X^{(j \mu + \mu + 1)}] \in \mathbb{F}_{2}^{c \times \mu} \gets \underline{X}^{(j)} \in \mathbb{F}^{c \times \mu}_{2}$
\EndFor
\For{\textbf{each} $k \in \{1,...,\lceil d_J / \mu \rceil\}$ in $\underline{U}_{C}$}
    \State $\underline{U}_{C}^{(k)} = Crypt_{1}^{\dag -1}(\underline{Y}^{(\lceil \tilde{n} / \mu \rceil\ +k)},s_{\mu c})$
    \State $[U_{C}^{(k \mu + 1)},...,U_{C}^{(k \mu + \mu + 1)}] \in \mathbb{F}_{2}^{c \times \mu} \gets \underline{U}_{C}^{(k)} \in \mathbb{F}^{c \times \mu}_{2}$
\EndFor
\State $\underline{U}_{d_J} \in \mathbb{F}_{2}^{\ell \times d_J} \gets \underline{U}_{C} \in \mathbb{F}_{2}^{c \times \gamma}$
\State \textbf{Linear IS Channel Decoding:}
\State Map $\underline{X}_{\mathcal{L}}$ from $\mathbb{F}_{2}^{\ell \times \tilde{n}}$ to $\mathbb{F}_{2^\mu}^{\ell \times \lceil \tilde{n} / \mu \rceil}$
\State $\underline{M}_c = \underline{H} \cdot \underline{X}_{\mathcal{L}}$
\State $\underline{M}_w = \underline{\tilde{G}} \cdot \underline{X}_{\mathcal{L}}$
\State $\underline{M}_{\mathcal{L}} \in \mathbb{F}_{2^\mu}^{\ell \times \lceil \tilde{n} / \mu \rceil} \gets \begin{bmatrix} \underline{M}_c \\ \underline{M}_w \end{bmatrix} \in \mathbb{F}_{2^\mu}^{\ell \times \lceil \tilde{n} / \mu \rceil}$
\State \textbf{Source Decoding:}
\State Map $\underline{M}_{\mathcal{L}}$ from $\mathbb{F}_{2^\mu}^{\ell \times \lceil \tilde{n} / \mu \rceil}$ to $\mathbb{F}_{2}^{\ell \times \tilde{n}}$
\For{\textbf{each} row $i \in \{1,...,\ell\}$ in $\underline{M}_{\mathcal{L}}$}
    \State $ \underline{\hat{V}}_{\mathcal{L},i} = g_{s,n}(\underline{M}_{\mathcal{L},i}, \underline{U}_{d_{J},i})\off{(\left[\underline{M}_{\mathcal{L},i}[\mathcal{H}_V],\underline{M}_{\mathcal{L},i}[\mathcal{J}_V] \oplus \underline{U}_{d_{J},i}\right])}$
\EndFor
\end{algorithmic}
\end{breakablealgorithm}
We denote the first $c$ rows of the message matrix by $\underline{M}_{c} \in \mathbb{F}_{2^\mu}^{c \times \lceil \tilde{n} / \mu \rceil}$. Respectively, we denote the complementary part of the message matrix by $\underline{M}_{w} \in \mathbb{F}_{2^\mu}^{w \times \lceil \tilde{n} / \mu \rceil}$. First, Bob uses the parity-check matrix $\underline{H} \in \mathbb{F}_{2}^{c \times \ell}$ to obtain the first $c$ symbols in each column of the message matrix, i.e. $\underline{M}_{c} = \underline{H}\cdot \underline{X}_{\mathcal{L}}$ (line 44 of Algorithm~\ref{algo:NU-HUNCC}). Then, by using $\underline{\tilde{G}} \in \mathbb{F}_{2}^{w \times \ell}$, Bob obtains the remaining $w$ symbols of each message $\underline{M}_w = \underline{\tilde{G}} \cdot \underline{X}_{\mathcal{L}}$ (line 45 of Algorithm~\ref{algo:NU-HUNCC}). Since the transmission is noiseless, the linear decoding process is error-free. To obtain the source messages, Bob maps the symbols of $\underline{M}_{\mathcal{L}}$ back into bit vectors of size $\mu$ and obtains the message matrix over the binary field $\underline{M}_{\mathcal{L}} \in \mathbb{F}_{2}^{\ell \times \tilde{n}}$ (line 48 of Algorithm~\ref{algo:NU-HUNCC}).

Finally, Bob employs the decoding function $g_{s,n}$ as given in~\eqref{eq:src_decode} on each row. First, Bob one-time pads the bits from group $\mathcal{J}_V$ of each row of the message using the rows of the seed matrix $\underline{U}_{d_J}$. Then, Bob uses successive cancellation decoding on each message and finally obtains the decoded source message matrix $\underline{\hat{V}}_{\mathcal{L}} \in \mathbb{F}_{2}^{\ell \times \tilde{n}}$ (lines 49-51 of Algorithm~\ref{algo:NU-HUNCC}).

\subsection{Non-Linear IS Channel Code}
In this section we replace the linear IS code with the non-linear IS code given in Sec.~\ref{sec:random-IS-code}. We provide a description of NU-HUNCC using the non-linear IS code, focusing on the channel coding and decoding as it holds the main difference between the two schemes.

First, Alice uses the almost uniform source compression from Sec.~\ref{sec:source_code} and obtains the almost uniform message matrix $\underline{M}_\mathcal{L} \in \mathbb{F}_{2}^{\ell \times \tilde{n}}$. Let $1 \leq c < \ell$ be the number of encrypted links and $w \triangleq \ell - c$. Alice chooses $k_s \leq \ell - w -\ell \epsilon$ s.t. $k_s$ will be the number of IS messages and $k_w \triangleq \ell - k_s$. For each column $j \in \{1,...,\tilde{n}\}$ the non-linear IS channel code premixes the almost uniform messages using a codebook generated as described in Sec.~\ref{sec:random-IS-code} s.t. $2^{k_s}$ is the number of bins and $2^{k_w}$ is the number of codewords per bin. The codeword matrix obtained from encoding each of the columns is denoted by $\underline{X}_{\mathcal{L}} \in \mathbb{F}_{2}^{\ell \times \tilde{n}}$.

To encrypt the codeword matrix, the PQ SS-CCA1 public key ceryptosystem, $Crypt_1$, is used. $Crypt_1$ is employed on $c$ bits at a time and maps them into $c+r$ bits, where the $r$ additional bits are the added randomness required for $Crypt_1$ to be SS-CCA1 \cite{goldwasser2019probabilistic}. \off{We note that since the encoding is done column by column, then the encryption must also be done column by column. With the linear IS channel code, we have more flexibility in the order of the encryption. We discuss this issue in a more detailed manner in Sec.~\ref{}.} $Crypt_1$ is employed on: 1) $c$ bits of each $j$-th column, $j\in\{1,\ldots,\tilde{n}\}$, of the encoded matrix $\underline{X}_{\mathcal{L}}$ which are denoted by $X^{(j)}$, and 2) each column $k\in \{1,\ldots,\gamma\}$ of the seed matrix, $\underline{U}_{C}^{(k)}$. Thus, for the $j$-th column of $\underline{X}_{\mathcal{L}}$ and $k$-th column of $\underline{U}_{C}$, we have:
\begin{multline} \label{eq:encryption}
    Crypt_1 : X^{(j)} \in \mathbb{F}^c_{2} \rightarrow Y^{(j)} \in \mathbb{F}^{c+r}_{2},\\ 
    \text{and } \underline{U}^{(k)}_{C} \in \mathbb{F}^c_{2} \rightarrow Y^{(\tilde{n}+k)} \in \mathbb{F}^{c+r}_{2},
\end{multline}
where the additional $r$ bits of $Y^{(j)}$ and the entire $Y^{(\tilde{n}+k)}$ are concatenated to the $c$ encrypted links between Alice and Bob. We denote the encryption of the $j$-th column of the message matrix $\underline{M}_{\mathcal{L}}$ as $Crypt_2$. s.t.
\begin{gather} \label{eq:encryption_crypt2}
    Crypt_2 : \underline{M}_{\mathcal{L}}^{(j)} \in \mathbb{F}^{\ell}_{2} \rightarrow \underline{Y}_{\mathcal{L}}^{(j)} \in \mathbb{F}^{\ell+r}_{2}.
\end{gather}
$Crypt_2$ combines the source and IS channel coding scheme with the PQ SS-CCA1 encryption scheme. We denote the encrypted links indexes as $\mathcal{D}$ and non-encrypted links indexes as $\mathcal{D}^{C} = \mathcal{L} \setminus \mathcal{D}$. Thus, for the $j$-th column Bob's observations are $ \underline{Y}_{\mathcal{L}}^{(j)} = [Crypt_1(f_{c,\ell}(\underline{M}_{\mathcal{L}}^{(j)})[\mathcal{D}],p_c),f_{c,l}(\underline{M}_{\mathcal{L}}^{(j)})[\mathcal{D}^{C}]]$, where $p_c$ is the public key for $Crypt_1$ as given in Definition~\ref{def:Public-key}.

For the decoding process, Bob first decrypts each of the $c$ encrypted bits separately (including the shared seed) using the secret key $s_c$\off{\footnote{Against IT-Eve the decryption stage is not required.}} obtaining the decrypted codeword matrix and shared uniform seed matrix. Bob decodes each column of the encoded message matrix by looking for: 1) The bin in which the codeword resides, and 2) The offset of the codeword inside its corresponding bin as described in Sec.~\ref{sec:random-IS-code}. Then, for each row separately, Bob employs the decoder $g_{s,n}$ given in \eqref{eq:src_decode}. Bob one-time pads the bits from group $\mathcal{J}_V$ using the decrypted seed, and employs successive cancellation decoder as given in \cite{ArikanBase2009,cronie2010lossless} to reliably decode each source message and obtain the decoded source messages matrix $\underline{\hat{V}}_{\mathcal{L}}$.

\section{Main Analytical Results}\label{sec:main_results}
In this section, we show that NU-HUNCC is IS against IT-Eve and PQ ISS-CCA1 against Crypto-Eve. We give achievability theorems for two coding schemes (linear and non-linear) and a converse theorem against IT-Eve as well. Additionally, we provide the communication rate achieved by the schemes against both IT-Eve and Crypto-Eve.\off{The proof for the achievability of NU-HUNCC with the IS linear channel code is given in Sec.~\ref{sec:linear-code-proof} whereas its ISS-CCA1 proof is given in Sec.~\ref{sec:linear-partial-enc}. The achievability of NU-HUNCC with the random IS channel code is given in Sec.~\ref{sec:random-code-proof} whereas its ISS-CCA1 proof is given in Sec.~\ref{sec:random-partial-enc}. The converse proof is given Sec~\ref{sec:converse}.}

We start with the achievability theorems for NU-HUNCC against IT-Eve. I.e., for efficient NU-IS scheme proposed regardless of the computational encryption scheme against Crypto-Eve. First, an achievability theorem is given for the scheme employing the IS linear channel code from Sec.~\ref{sec:linear-IS-code}.
\begin{theorem} \label{DirectLinear}
   Assume a noiseless multipath communication $(\ell,w)$. NU-HUNCC with a linear IS code reliably delivers with high probability $\ell$ non-uniform messages from a DMS $(\mathcal{V},p_V)$ to the legitimate receiver, s.t. $\mathbb{P}(\underline{\hat{V}}_{\mathcal{L}}(\underline{Y}_{\mathcal{L}}) \neq \underline{V}_{\mathcal{L}}) \leq \epsilon_e$, while keeping IT-Eve ignorant of any set of $k_s \leq \ell - w$ messages individually, s.t. $\mathbb{V}(p_{\underline{Z}_{\mathcal{W}}|{\underline{V}_{\mathcal{K}_s}=\underline{v}_{\mathcal{K}_s}}},p_{\underline{Z}_{\mathcal{W}}}) \leq \epsilon_s$, whenever $d_{J} \geq |\mathcal{J}_V|$,\off{, $\ell\epsilon = o(\ell)$,} $\ell$ is \off{lower bounded by $\omega(\tilde{n}^{\frac{2}{t}})$ for some $t \geq 1$ and}upper bounded by $o(2^{n^\beta}/\tilde{n})$, and $\mu \geq \ell$ where $\mathbb{F}_{2^\mu}$ is the extension field over which the code operates. \off{ The IS communication rate obtained is $R = \frac{1}{\frac{|\mathcal{H}_V|}{n} + \frac{2d_{J,n}}{n}}$.}
\end{theorem}

{\em Proof:} The proof for Theorem~\ref{DirectLinear} is provided in Sec.~\ref{sec:linear-code-proof}. The information leakage obtained from Theorem~\ref{DirectLinear} is upper bounded by $\epsilon_s = 2\sqrt{2\tilde{n}\ell 2^{-n^{\beta}}}$ whereas the error probability is upper bounded by $\epsilon_e = \ell 2^{-n^\beta}$. Thus, for both $\epsilon_s$ and $\epsilon_e$ to be negligible we require $\ell$ to be upper bounded by $o\left(2^{n^{\beta}}/\tilde{n}\right)$.\off{The information leakage and the error probability of the code are provided there as well. We note that for NU-HUNCC using the linear IS channel code, we consider the number of protected messages to be $k_s = \ell - w = c$.\off{, whereas in the random code, $k_s$ must be smaller than $c$ for the code to be IS.} } We now provide the achievability theorem for the scheme employing the non-linear IS channel code from Sec.~\ref{sec:random-IS-code}:

\begin{theorem} \label{Direct}
   Assume a noiseless multipath communication $(\ell,w)$. NU-HUNCC with non-linear IS code reliably delivers with high probability $\ell$ non-uniform messages from a DMS $(\mathcal{V},p_V)$ to the legitimate receiver, such that $\mathbb{P}(\underline{\hat{V}}_{\mathcal{L}}(\underline{Y}_{\mathcal{L}}) \neq \underline{V}_{\mathcal{L}}) \leq \epsilon_e$, while keeping IT-Eve ignorant of any set of $k_s \leq \ell - w - \ell\epsilon$ messages individually, s.t. $\mathbb{V}(p_{\underline{Z}_{\mathcal{W}}|{\underline{V}_{\mathcal{K}_s}=\underline{v}_{\mathcal{K}_s}}},p_{\underline{Z}_{\mathcal{W}}}) \leq \epsilon_s$, whenever $d_{J} \geq |\mathcal{J}_V|$, $\ell\epsilon = o(\ell)$, and $\ell$ is lower bounded by $\omega(\tilde{n}^{\frac{2}{t}})$ for some $t \geq 1$ and upper bounded by $o(2^{n^\beta}/\tilde{n})$.
   \off{ The IS communication rate obtained is $R = \frac{1}{\frac{|\mathcal{H}_V|}{n} + \frac{2d_{J,n}}{n}}$.}
\end{theorem}
\off{\textit{Proof(Rate)}: Alice wants to send $\ell n$ symbols to Bob. The message to Bob contains the following information: 1) $\ell$ premixed messages of total size $\tilde{n} \cdot \ell$, 2) $\ell$ uniform seed realizations of size $d_{J,n}$ each. Dividing the numerator and denominator by $n$, gives the required result. \qed}

{\em Proof:} The proof for Theorem~\ref{Direct} is provided in Appendix.~\ref{sec:random-code-proof}. The information leakage obtained from Theorem~\ref{DirectLinear} is upper bounded by $\epsilon_s = \tilde{n}\ell^{-\frac{t}{2}} + 2\sqrt{2\tilde{n}\ell 2^{-n^{\beta}}}$ whereas the error probability is upper bounded by $\epsilon_e = \tilde{n}O(2^{-\ell}) + \sqrt{2\tilde{n}\ell 2^{-n^{\beta}}} + \ell 2^{-n^\beta}$. Thus, for both $\epsilon_s$ and $\epsilon_e$ to be negligible, we require $\ell$ to be upper bounded by $o\left(2^{n^{\beta}}/\tilde{n}\right)$ and lower bounded by $\omega\left(\tilde{n}^{\frac{2}{t}}\right)$.
\begin{remark}
    The achievability result in Theorem~\ref{Direct} using the non-linear code from Sec.~\ref{sec:random-IS-code} requires the number of messages to be upper and lower bounded. The bounds are determined both by the secrecy and reliability constraints. The lower bound is determined by the expression $\tilde{n} \ell^{-\frac{t}{2}}$ in $\epsilon_s$ and the upper bound is determined by the expression $\sqrt{2\tilde{n}\ell 2^{-n^{\beta}}}$ which resides in both $\epsilon_s$ and $\epsilon_e$. By choosing $\ell \epsilon = \lceil t \log{\ell} \rceil$ to be an integer for any $t \geq 1$ and taking $k_s \leq \ell - w - \ell \epsilon$ Alice can guarantee reliable and secure communication for non-uniform messages.
\end{remark}
\off{The information leakage and the error probability of the code are provided there as well. For NU-HUNCC using the IS non-linear channel coding scheme, the number of IS messages must be $k_s < \ell - w$ instead of the linear IS code where we can protect $k_s = \ell - w$ messages.} \off{We also note, that as demonstrated in the proofs, the number of messages $\ell$ in Theorem~\ref{Direct} is lower and upper bounded due to the reliability and security constraints respectively.}

The converse result for NU-HUNCC against IT-Eve, under IS constraint (Definition~\ref{def:IS}), is given by the following theorem.

\begin{theorem} \label{Converse}
     Assume a noiseless multipath communication $(\ell,w)$. If a coding scheme exists s.t. Alice delivers to Bob any $|\mathcal{K}_s| \triangleq k_s$ source messages from the DMS $(\mathcal{V},p_V)$ in the presence of IT-Eve in an IS manner, i.e.
     \begin{enumerate}
         \item $I(\underline{V}_{\mathcal{K}_s};\underline{Z}_{\mathcal{W}}) \leq \epsilon_s$ (Secure),
         \item $\mathbb{P}(\underline{\hat{V}}_{\mathcal{L}}(\underline{Y}_{\mathcal{L}}) \neq \underline{V}_{\mathcal{L}}) \leq \epsilon_e$ (Reliable),
         \item $\frac{d_n}{n} \leq \epsilon_d$ (Seed).
     \end{enumerate}
     Then, one must have:
     \begin{enumerate}
         \item $H(\underline{V}_{\mathcal{L}}) \leq I(\underline{X}_{\mathcal{L}};\underline{Y}_{\mathcal{L}})$,
         \item $H(\underline{V}_{\mathcal{K}_s}) \leq I(\underline{X}_{\mathcal{L}};\underline{Y}_{\mathcal{L}}) - I(\underline{X}_{\mathcal{L}};\underline{Z}_{\mathcal{W}})$,
         \item $H(\underline{V}_{\mathcal{K}_w}) \geq I(\underline{X}_{\mathcal{L}};\underline{Z}_{\mathcal{W}})$,
     \end{enumerate}
     where $|\mathcal{K}_w| \triangleq k_w = \ell - k_s$.
     \off{Consider $\ell$ DMS's $(V_i,p_{V_i})$, $1 \leq i \leq \ell$ that are being transmitted over $\ell$ \off{noisy}links, each of them described by the channel $(\mathcal{X},p_{YZ|X},\mathcal{Y} \times \mathcal{Z})$ as described in \ref{sec:sys:general}. Also, consider two disjoint subsets of the links: $\mathcal{K}_s$ and $\mathcal{K}_w$ such that $|\mathcal{K}_s| + |\mathcal{K}_w| = \ell$. If there exists a sequence of codes $\{C_n\}_{n \geq 1}$ such that \ref{IS-Conditions} holds, then:
     \[
    \left\{
\begin{aligned}
    &H(\underline{V}_{\mathcal{L}}) \leq I(\underline{X}_{\mathcal{L}};\underline{Y}_{\mathcal{L}}) \\
    &H(\underline{V}_{\mathcal{K}_s}) \leq I(\underline{X}_{\mathcal{L}};\underline{Y}_{\mathcal{L}}) - I(\underline{X}_{\mathcal{L}};\underline{Z}_{\mathcal{L}})\\
    &H(\underline{V}_{\mathcal{K}_w}) \geq I(\underline{X}_{\mathcal{L}};\underline{Z}_{\mathcal{L}})
\end{aligned}
\right.
\]}
\end{theorem}

{\em Proof:} The proof for Theorem~\ref{Converse} is given in Sec.~\ref{sec:converse}.

The information rate obtained by NU-HUNCC against IT-Eve is given by the following Theorem
\begin{theorem} \label{IT-Eve-Rate}
    Consider the setting of NU-HUNCC with an encoding scheme as in Theorem~\ref{DirectLinear} or Theorem~\ref{Direct}, the information rate of NU-HUNCC is
    \begin{gather*} \label{eq:IT-Eve-Rate}
        R = \frac{1}{\frac{|\mathcal{H}_V|}{n} + \frac{2d_{J}}{n}}.
    \end{gather*}
\end{theorem}

\textit{Proof}: Alice wants to send $\ell n$ bits to Bob. The messages to Bob contain the following information: 1) $\ell$ encoded confidential messages of size $\tilde{n} = |\mathcal{H}_V| + d_J$ each, and 2) seed of size $\ell d_{J}$. Dividing the numerator and denominator by $\ell n$, we obtain the result in Theorem~\ref{IT-Eve-Rate} on NU-HUNCC data rate against IT-Eve. \qed

So far, we have considered IT-Eve as the eavesdropper and claimed the source and channel coding schemes are IS. Now, we consider Crypto-Eve as a potential all-observing eavesdropper. By the following theorems, we claim our proposed hybrid coding and encryption scheme is ISS-CCA1. First, we claim that the encryption scheme considering the linear IS channel code from Sec.~\ref{sec:linear-IS-code} is ISS-CCA1 according to Definition~\ref{def:individuall-SS-CCA1}.
\begin{theorem} \label{IndividualLinear-SS-CCA1}
    Assume the setting in Theorem~\ref{DirectLinear}. Let $Crypt_1^{\dag}$ be an SS-CCA1 secured cryptosystem as described in \eqref{eq:linear-encryption}. Then, NU-HUNCC is ISS-CCA1 secure.
\end{theorem}
{\em Proof:} The proof of Theorem~\ref{IndividualLinear-SS-CCA1} is given in Sec.~\ref{sec:linear-partial-enc}.

Now, we claim that the encryption scheme considering the non-linear coding IS channel code from Sec.\ref{sec:random-IS-code} is ISS-CCA1 as well.
\begin{theorem} \label{Individual-SS-CCA1}
    Assume the setting in Theorem~\ref{Direct}. Let $Crypt_{1}$ be an SS-CCA1 secured cryptosystem as described in \eqref{eq:encryption}. Then, NU-HUNCC is ISS-CCA1 secure.
\end{theorem}
{\em Proof:} The proof of Theorem~\ref{Individual-SS-CCA1} is given in Appendix~\ref{sec:random-partial-enc}.

The information rate for both coding schemes (linear and non-linear) against Crypto-Eve remains the same and is given by the following theorem.
\begin{theorem} \label{Crypto-Eve-Rate}
    Consider the setting of the NU-HUNCC with an encoding scheme as in Theorem~\ref{Direct}, the information rate of NU-HUNCC is
    \begin{gather*} \label{eq:Crypto-Eve-Rate}
        R = \frac{1}{\frac{|\mathcal{H}_V|}{n}(1 +\frac{r}{\ell})+ \frac{d_{J}}{n}(2+\frac{r}{\ell} + \frac{r}{c})},
    \end{gather*}
with convergence rate of $\mathcal{O}(1/\ell)$.
\end{theorem}
\textit{Proof}: Alice wants to send $\ell n$ bits to Bob. The messages to Bob contain the following information: 1) $\ell-c$ not encrypted messages of total size $\tilde{n}(\ell-c)$, 2) $c$ encrypted messages of total size $\tilde{n}(c+r)$, and 3) encrypted seed of size $\ell d_{J} \cdot \frac{c+r}{c}$. Dividing the numerator and denominator by $\ell n$, we obtain the result in Theorem~\ref{Crypto-Eve-Rate} on NU-HUNCC data rate. \qed

\begin{remark}
    Considering the size of the messages, $n$, as constant the optimal rate of NU-HUNCC is given by Theorem~\ref{IT-Eve-Rate} and can be achieved by increasing the number of links, $\ell$.
\end{remark}
In Sec.~\ref{sec:rate_performance} we provide numerical results demonstrating the rate enhancement of NU-HUNCC compared to other PQ-secured systems (see Table~\ref{table:1}). We show that by choosing large $n$ and increasing $\ell$ the rate of NU-HUNCC converges to its upper bound. Specifically, we show that by choosing $n=128$[KB], then for ten communication links ($\ell=10)$, the rate of NU-HUNCC is within $7\%$ margin of its upper and within $10\%$ margin of the optimal possible rate, $\frac{1}{H(V)}$.

\begin{remark}
    The security theorems provided in this section assume a multipath communication network with $\ell$ links. However, NU-HUNCC coding scheme can be applied on any communication or information system since the security of both the linear and non-linear codes does not rely on the network structure \cite{SMSM}. In this sense, NU-HUNCC is a universal coding scheme. We show our results on a multipath network merely for simplicity of representation. 
\end{remark}

\subsection{Discussion}\label{subsec:diss}
The IS non-linear channel code given in Sec.\ref{sec:random-IS-code} holds similarities with other binning-based codes. Wyner's wiretap code from \cite{WiretapWyner} is also based on a binning structure. However, in Wyner's wiretap code, a source of local randomness is used to choose the codeword inside the bin, and the confidential message is used to choose the bin itself. The use of a source of local randomness resulted in a significant degradation in the communication rate between the legitimate parties since the randomness has to be shared between them along with the confidential message. In the non-linear code, the source of local randomness is replaced by other messages. Thus the transmission can attain the full capacity of the network.

The wiretap code proposed by Chou et. al. in \cite{NegligbleCost} is also based on a binning structure. Chou et. al. considered the transmission of a confidential message over the wiretap channel by utilizing a public message as a source of local randomness. The proposed codebook is organized in bins s.t. the bin is chosen by the confidential message and the codeword inside the bin is chosen by the public message. Before utilizing the public message as a source of randomness, the message is compressed by an almost uniform source coding scheme that uses a shared seed of negligible size\off{ whereas the confidential message is compressed by a standard lossless compression scheme}. The proposed scheme achieves \emph{strong secrecy}. On the other hand, our setting, achieves \emph{IS} by encoding several confidential messages. i.e. the public message from Chou et. al. is replaced by confidential messages. However, to obtain IS in the presence of non-uniform confidential messages we must compress all the confidential messages by an almost uniform source encoder and a seed with a negligible size as done on the public message in Choe et. al. proposed scheme \cite{NegligbleCost}.

\off{\textcolor{blue}{The Secure Multiplex Coding scheme from \cite{SCMUniform} also considers the transmission of multiple confidential messages over a wiretap channel. The security of the messages can be obtained by using deterministic or stochastic random coding-based encoders that mix the messages s.t. they protect each other. In our setting, the messages are not from a uniform distribution. The encoder we propose mixes between parts of the messages at a time by operating column-wise on the message matrix $\underline{M}_{\mathcal{L}}$.}}

Both the linear and non-linear codes have their merits and demerits. The non-linear IS channel code operates over the binary field.\off{ The structure of the code depends solely on the structure of the multipath and the number of links observed by IT-Eve.} IS is obtained by bounding the number of links in the multipath between $\omega\left(\tilde{n}^{\frac{2}{t}}\right) \leq \ell \leq o\left(\frac{2^{n^\beta}}{\tilde{n}}\right)$ for any $t \geq 1$. However, using the non-linear code requires both Alice and Bob to store the entire codebook of size $2^\ell$ codewords. On the contrary, the linear IS code is structured against the deterministic scenario where IT-Eve observes no more than $w = \ell - k_s$. The encoding and decoding of the code are performed using linear transformations requiring Alice and Bob to store a finite set of matrices each of size smaller than $\ell \times \ell$. Nonetheless, to obtain the secrecy constraint, the number of links is bounded by $\ell \leq \mu$. This condition requires the code to operate over a large field, resulting in higher computation complexity. From a cryptographic perspective, the information leakage from both codes attains the cryptographic criterion for ineligibility as shown in Sec.~\ref{sec:linear-partial-enc} and Appendix~\ref{sec:random-partial-enc}.

\off{\textcolor{blue}{In \cite{cohen2022partial}, Cohen et al proposed a version of the IS random code that provides error correction in the presence of noisy channels. To make the code IS in the presence of noise over the channels, the codeword length was increased from $\ell$ by the entropy of the channel's noise to provide error correction capabilities.}}

\off{\textcolor{blue}{The information rate against IT-Eve and Crypto-Eve given in Theorems~\ref{IT-Eve-Rate} and~\ref{Crypto-Eve-Rate} is for the final length regime. However, by taking $n$ to infinity we can show the IT-Eve's and Crypto-Eve's rate converges to $\frac{1}{H(V)}$ where $H(V)$ is the entropy of the source. Since the seed's size is negligible compared to the message size and since in polar codes $\lim \limits_{n \rightarrow \infty} \frac{|\mathcal{H}_V|}{n} = H(V)$.}}

\section{Performance Analysis} \label{sec:numerical_performance}
In this section, we provide the performance analysis of NU-HUNCC. Our focus is on three performance criteria: secure communication rate, encoding-decoding complexity, and uniform seed length. We show the effect of the number of communication links, the number of encrypted links, the message size, and the probability distribution of the source on the data rate of NU-HUNCC.

Additionally, we provide a comparison between NU-HUNCC and other secure communication systems in terms of communication rate and complexity. We choose to demonstrate our results by using the original McEliecce cryptosystem with $[1024,524]$-Goppa codes (Sec.~\ref{sec:encryption}) as $Cryprt_1^{\dag}$ from \eqref{eq:linear-encryption}.\off{ We use this version of McEliecce although it is not SS-CCA1 since it is the most commonly used version of this cryptosystem.} We start our analysis by presenting the seed length compared to the message size.

\subsection{Seed Length} \label{sec:seed_length}

The optimal size of the seed was defined in \cite{NegligbleCost,chou2013data}. For a message of size $n$ bits, the optimal size of the seed is 1) $d_{opt} = O(k_n\sqrt{n})$, $\forall k_n$ s.t. $\lim\limits_{n \rightarrow \infty} k_n \rightarrow \infty$, and 2) $d_{opt} = \Omega(\sqrt{n})$. However, it was shown by Chou et al. in \cite{NegligbleCost} that the seed size for the polar source coding scheme from Sec.~\ref{sec:source_code} is sub-linear, i.e. $d_{opt} = o(n)$.\off{ and is considered sub-optimal.}
\off{However, the size of the seed for the polar codes source encoder/decoder from Sec.~\ref{sec:source-code} is sub-optimal.}

\begin{table*}
\centering
\label{table:table1}
\resizebox{\linewidth}{!}{%
\begin{tabular}{>{\centering\hspace{0pt}\vspace{0.2cm}}S{m{0.160\linewidth}}>{\centering\hspace{0pt}}S{m{0.280\linewidth}}>{\centering\hspace{0pt}}S{m{0.100\linewidth}}>{\centering\hspace{0pt}}S{m{0.140\linewidth}}>{\centering\arraybackslash\hspace{0pt}}S{m{0.120\linewidth}}} 
\toprule
                   & \SetCell{font=\normalsize}NU-HUNCC                                                                                                                        & \SetCell{font=\normalsize}NUM             & \SetCell{font=\normalsize}NU-IS                                                       & \SetCell{font=\normalsize}NC-WTC Type \uppercase\expandafter{\romannumeral2}                     \\ 
\hline
\SetCell{{font=\normalsize}}Eavesdropper       & \SetCell{font=\normalsize}Crypto-Eve $\And$ IT-Eve                                                                                                                      & \SetCell{font=\normalsize}Crypto-Eve     & \SetCell{font=\normalsize}IT-Eve                                                   & \SetCell{font=\normalsize}IT-Eve                            \\ 
\SetCell{font=\normalsize}Encrypted Channels & \SetCell{font=\normalsize}$c$                                                                                                                             & \SetCell{font=\normalsize}$\ell$          & \SetCell{font=\normalsize}$0$                                                      & \SetCell{font=\normalsize}$0$                               \\ 
\SetCell{font=\normalsize}Use of seed        & \SetCell{font=\normalsize}Yes                                                                                                                             & \SetCell{font=\normalsize}No              & \SetCell{font=\normalsize}Yes                                                      & \SetCell{font=\normalsize}No                                \\ 
\SetCell{font=\normalsize}Use of keys        & \SetCell{font=\normalsize}Yes                                                                                                                             & \SetCell{font=\normalsize}Yes             & \SetCell{font=\normalsize}No                                                      & \SetCell{font=\normalsize}No                                \\ 

\SetCell{font=\normalsize}Rate               & $1\Big/\left(\frac{|\mathcal{H}_V|}{n}\left(1 +\frac{r}{\ell}\right)+ \frac{d_{J}}{n}\left(2+\frac{r}{\ell} + \frac{r}{c}\right)\right)$ & $c_g\Big/n_g$ & $1\Big/\Big(\frac{|\mathcal{H}_V|}{n} + \frac{2d_{J}}{n}\Big)$ & $n(\ell-w)\Big/|\mathcal{H}_V|\ell$ \\
\bottomrule
\end{tabular}
}
\caption{The table illustrates the key attributes of the four coding and encryption schemes discussed in this paper. The overall rate of the encryption schemes depends on the probability distribution of the source, the size of the seed, the number of links, the number of encrypted links, and the rate of the underlying encryption scheme. We note that for NUM, the rate depends only on $c_g$, the size of the plaintext, and $n_g$, the size of the ciphertext.}
\label{table:1}
\end{table*}

The length of the seed is directly affected by the polarization rate which determines the portion of the bits that remain unpolarized after the polarization transform, i.e. the size of the group $\mathcal{J}_v$ \eqref{eq:seed-group}. This size is often referred to as the gap to capacity \cite{hassani2010scaling,wang2023sub} since it determines how close the channel code gets to the optimal capacity of the communication channel. Polar codes for source and channel coding are directly linked, \cite{cronie2010lossless}, thus the gap to capacity is equal to the gap of the polar codes for source coding to the optimal compression rate \cite[Corollary 3.16]{wang2021complexity}. In \cite{hassani2010scaling} it was shown that the number of unpolarized bits can be bounded by
\vspace{-0.15cm}
\begin{equation} \label{eq:SeedBound}
    \begin{aligned}
        n^{0.7214} \leq d_{J} \leq n^{0.7331}.
    \end{aligned}
\end{equation}
This bound is sub-optimal since for $k_n = n^{0.2}$, we have that $d_J = \Omega(n^{0.7})$. Yet, this overhead for the proposed NU-HUNCC scheme with non-uniform messages is still negligible for sufficiently large $n$ as illustrated by Fig. \ref{fig:SeedSize}.
\vspace{-0.3cm}
\begin{figure}[htbp]
    \centering
    \includegraphics[width=1\linewidth]{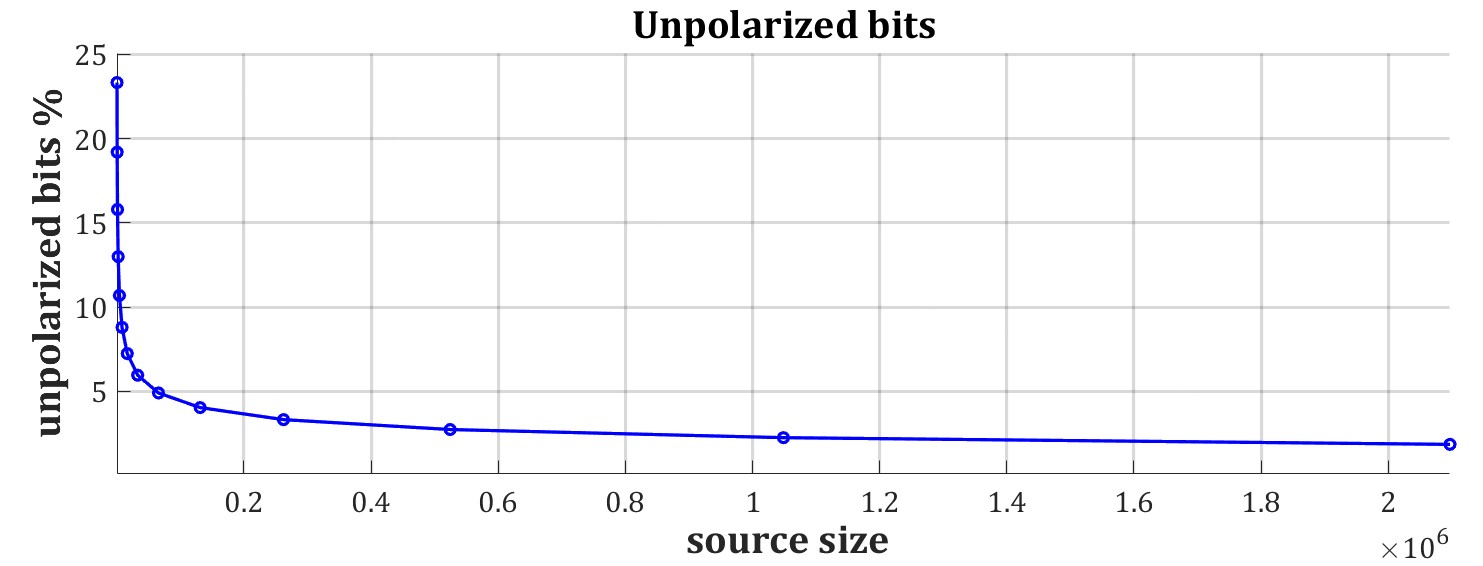}
    \caption{Numerical simulation of the seed size for a source $(\mathcal{V},p_V)$ with entropy $H(V) = 0.9$. For messages with a size greater than $2^{18}$ bits, the seed size already decreases to about $2.2\%$ of the compressed message size.}
    \label{fig:SeedSize}
\end{figure}
\off{
This bound is sub-optimal since for $k_n = n^{0.2}$, we have that $d_J = \Omega(n^{0.7})$. However, numeric simulations show that for $n=2^{19}$ bits, the length of the seed is already $2\%$ of the message size and $2.2\%$ of the compressed message size considering the entropy of the source is $H(V) = 0.9$.}

\subsection{Secure Data Rate} \label{sec:rate_performance}
To evaluate the communication rate of NU-HUNCC, we draw comparisons with other secured communication systems: 1) Non-Uniform McEliecce (NUM) - The source messages are compressed using an optimal lossless source encoder \cite{shannon1948mathematical,ziv1978compression,cronie2010lossless}. All the messages are then encrypted by the original McEliecce cryptosystem with $[1024,524]$-Goppa codes. The rate of Non-Uniform McEliecce depends on the rate of the Goppa codes and the entropy of the source. 2) Network Coding Wiretap Type \uppercase\expandafter{\romannumeral2} (NC WT Type \uppercase\expandafter{\romannumeral2}) - The source messages are compressed using an optimal lossless source encoder \cite{shannon1948mathematical,ziv1978compression,cronie2010lossless}. Then the messages are encoded using the NC WT Type \uppercase\expandafter{\romannumeral2} coding scheme \cite{el2007wiretap}. The rate depends on the entropy of the source, the total number of links, and the number of links observed by IT-Eve. 3) NU-IS - The source messages are compressed using the almost uniform source encoder from Sec.~\ref{sec:source_code}. Then the messages are mixed using the linear IS channel code from Sec.~\ref{sec:linear-IS-code}. The rate is given by Theorem~\ref{IT-Eve-Rate} and it is a function of the entropy of the source and the message size. \off{NU-HUNCC and NUM are both secured against Crypto-Eve whereas NU-HUNCC, IS, and NC WT Type \uppercase\expandafter{\romannumeral2} schemes are secured against IT-Eve. }Table~\ref{table:1} describes the general differences between the secure communication systems considered in this section.
\off{
\begin{table*}[htbp]
\centering
\resizebox{\textwidth}{!}{%
\begin{tabular}{|c|c|c|c|c|}
 \hline
   & NU-HUNCC & NUM & IS & NC-WTC type 2 \\
  \hline
  Eavesdropper & Crypto-Eve & Crypto-Eve & IT-Eve & IT-Eve \\
  \hline
  Encrypted Channels & $c$ & $\ell$ & $\ell$ & $\ell$  \\
  \hline
  Use of seed & Yes & No & Yes & No \\
  \hline
  Use of keys & Yes & Yes & Yes & No \\
  \hline
  Rate & $\frac{1}{\frac{|\mathcal{H}_V|}{n}(1 +\frac{r}{\ell})+ \frac{d_{J}}{n}(2+\frac{r}{\ell} + \frac{r}{c})}$ & $\frac{c}{c+r}$ & $\frac{1}{\frac{|\mathcal{H}_V|}{n} + \frac{2d_{J}}{n}}$ & $\frac{\frac{|\mathcal{H}_V|}{n}c}{\ell}$ \\
  \hline
\end{tabular}}
\caption{The table illustrates the key attributes of the four systems analyzed in this paper, employing security notions such as Individually Secure (IS), Computationally Secure (CS), and Individually Computationally Secure (IC). NU-HUNCC outperforms the McEliece Cryptosystems, achieving better rates, and ensuring IS communication against Weak Eve and IC communication against Strong Eve.}
\label{table:1}
\end{table*}}

\off{
\begin{table*}[htbp]\label{table:table1}
\centering
\resizebox{\textwidth}{!}{%
\begin{tabular}{|c|c|c|c|c|}
\hline
& NU-HUNCC & NUM & IS & NC-WTC type 2 \\
\hline
Eavesdropper & Crypto-Eve & Crypto-Eve & IT-Eve & IT-Eve \\
\hline
Encrypted Channels & $c$ & $\ell$ & $0$ & $0$  \\
\hline
Use of seed & Yes & No & Yes & No \\
\hline
Use of keys & Yes & Yes & Yes & No \\
\hline
Rate & \makecell{$\vspace{1pt}\dfrac{1}{\dfrac{|\mathcal{H}_V|}{n}\left(1 +\dfrac{r}{\ell}\right)+ \dfrac{d_{J}}{n}\left(2+\dfrac{r}{\ell} + \dfrac{r}{c}\right)}$} & \makecell{$\vspace{2pt}\dfrac{c}{c+r}$} & \makecell{$\vspace{1pt}\dfrac{1}{\dfrac{|\mathcal{H}_V|}{n} + \dfrac{2d_{J}}{n}}$} & \makecell{$\vspace{1pt}\dfrac{|\mathcal{H}_V|c}{n\ell}$} \\
\hline
\end{tabular}}%

\caption{The table illustrates the key attributes of the four systems analyzed in this paper, employing security notions such as Individually Secure (IS), Computationally Secure (CS), and Individually Computationally Secure (IC). NU-HUNCC outperforms the McEliece Cryptosystems, achieving better rates, and ensuring IS communication against Weak Eve and IC communication against Strong Eve.}
\label{table:1}
\end{table*}}

The communication rate of the described systems is affected by: 1) The messages size (in bits), 2) The number of links, and 3) The probability distribution of the source. In our analysis, we consider messages of sizes $256 - 2097152$ [bits] ($32$[B]-$256$[KB]), $4-10$ communication links, and binary entropies $0.7-0.95$ of the source.

\begin{figure}[htbp]
    \centering
    \includegraphics[width=1\linewidth]{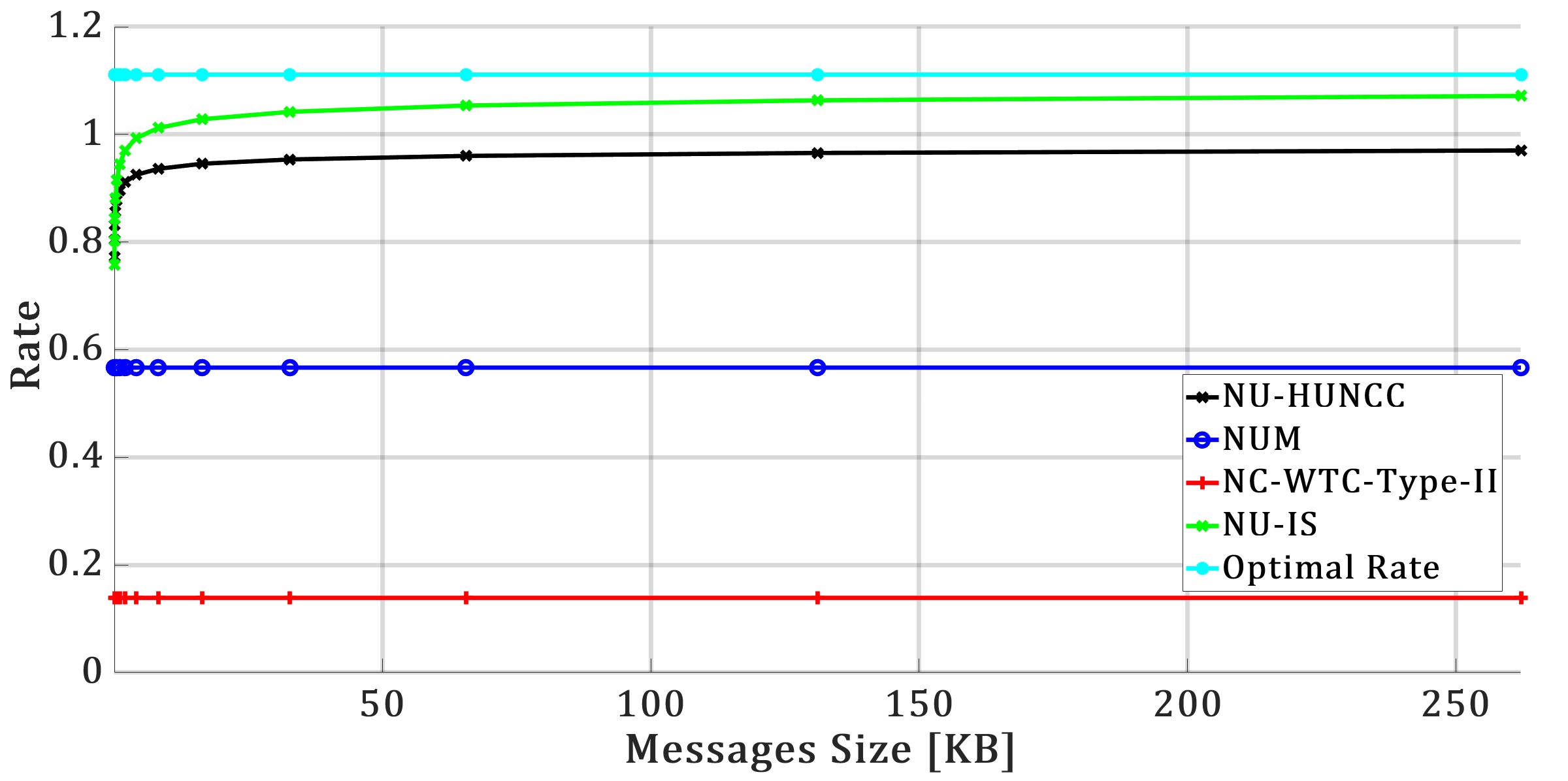}
    \caption{Numerical simulation of the communication rate as a function of the message size for eight communication links ($\ell=8$), one encrypted link ($c=1$) and source entropy of $0.9$ ($H(V) = 0.9$). NU-IS and NU-HUNCC both achieve better rates than NC WTC Type \uppercase\expandafter{\romannumeral2} and NUM, respectively.}
    \label{fig:RateAsF(Messages)}
\end{figure}

In Fig.~\ref{fig:RateAsF(Messages)} we illustrate the increase in the data rate as a function of the message size for eight communication links ($\ell=8$), one encrypted link ($c=1$), and source entropy of $0.9$ ($H(V) = 0.9$). The optimal communication rate in this case is given by $1 / H(V) \approx 1.1$. The data rate of NUM depends on the optimal compression rate of the source and the rate of McEliecce. Thus, NUM obtains a constant rate of $0.56$ regardless of the message size. However, for NU-HUNCC, the communication rate increases with the message size. NU-HUNCC does not reach the optimal communication rate due to the encryption performed over one of the links and the use of the uniform seed. From Theorem~\ref{Crypto-Eve-Rate} the optimal rate NU-HUNCC can obtain considering source entropy of $H(V) = 0.9$ and encryption by the original McEliecce with a $[1024,524]$-Goppa codes is approximately $0.99$ (where from $c/(c+r)$, we have $r = (1024-524)/524 \approx 0.95$). From Theorem~\ref{IT-Eve-Rate} the optimal rate NU-IS can obtain considering source entropy of $H(V) = 0.9$ is approximately $1.1$. As depicted in Fig.~\ref{fig:RateAsF(Messages)}, by choosing a message of size $64[KB]$, NU-HUNCC obtains a data rate of approximately $0.96$, and NU-IS obtains a data rate of $1.01$, both within a $5\%$ margin of their theoretical upper bound.

\begin{figure}[htbp]
    \centering
    \includegraphics[width=1\linewidth]{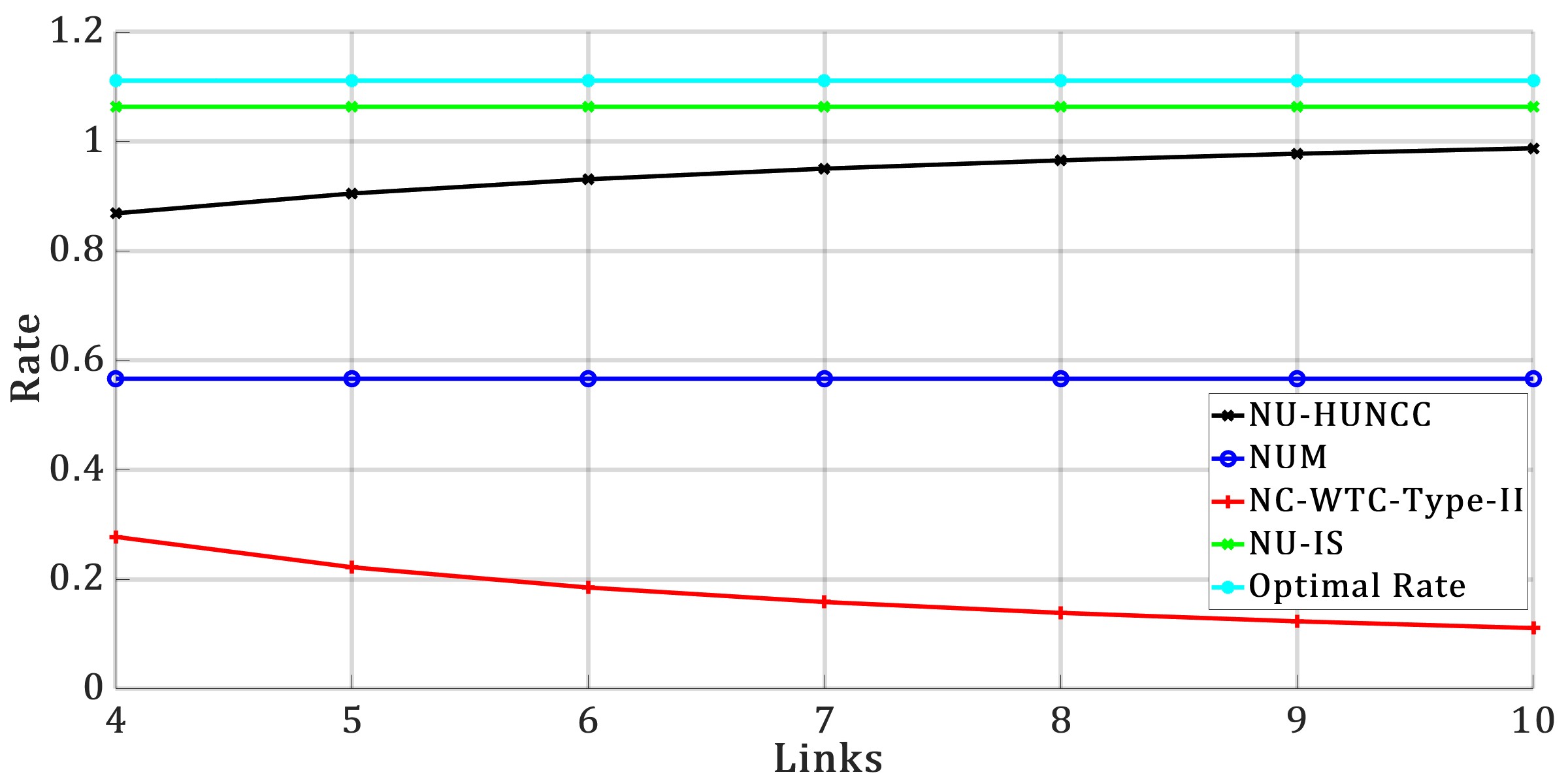}
    \caption{Numerical simulation of the communication rate as a function of the number of links for messages of size $128$[KB], one encrypted link ($c=1$), and a source entropy of $0.9$ ($H(V) = 0.9$). NU-IS and NU-HUNCC both achieve better rates than NC WTC Type \uppercase\expandafter{\romannumeral2} and NUM respectively.}
    \label{fig:RateAsF(Links)}
\end{figure}

In Fig.~\ref{fig:RateAsF(Links)} we show the communication rate as a function of the number of communication links for messages of size $128$[KB], one encrypted link ($c=1$), and source entropy of $0.9$ ($H(V) = 0.9$). Both NU-IS and NU-HUNCC operate with a communication rate higher than of NC WT Type \uppercase\expandafter{\romannumeral2} and NUM, respectively. NU-HUNCC's communication rate increases as the number of communication links grows since the number of encrypted links remains unchanged. The presence of the seed prevents both NU-HUNCC's and NU-IS's data rates from reaching their respective optimal values. NU-IS maintains a constant data rate since its rate does not depend on the number of communication links (see Theorem~\ref{IT-Eve-Rate} and Table~\ref{table:1}). NU-IS's constant data rate as depicted in Fig.~\ref{fig:RateAsF(Links)} is $1.06$ where the gap to the optimal rate of $1.1$ is due to the use of the uniform seed. On the other hand, NU-HUNCC's data rate increases as a function of the number of communication links, since the number of encrypted links remains the same. As depicted in Fig.~\ref{fig:RateAsF(Links)}, for ten communications links ($\ell=10$), NU-HUNCC's data rate is $0.987$, within a $7\%$ margin of its upper bound, and $10\%$ margin the optimal possible rate of $1.1$.

\begin{figure}[htbp]
    \centering
    \includegraphics[width=1\linewidth]{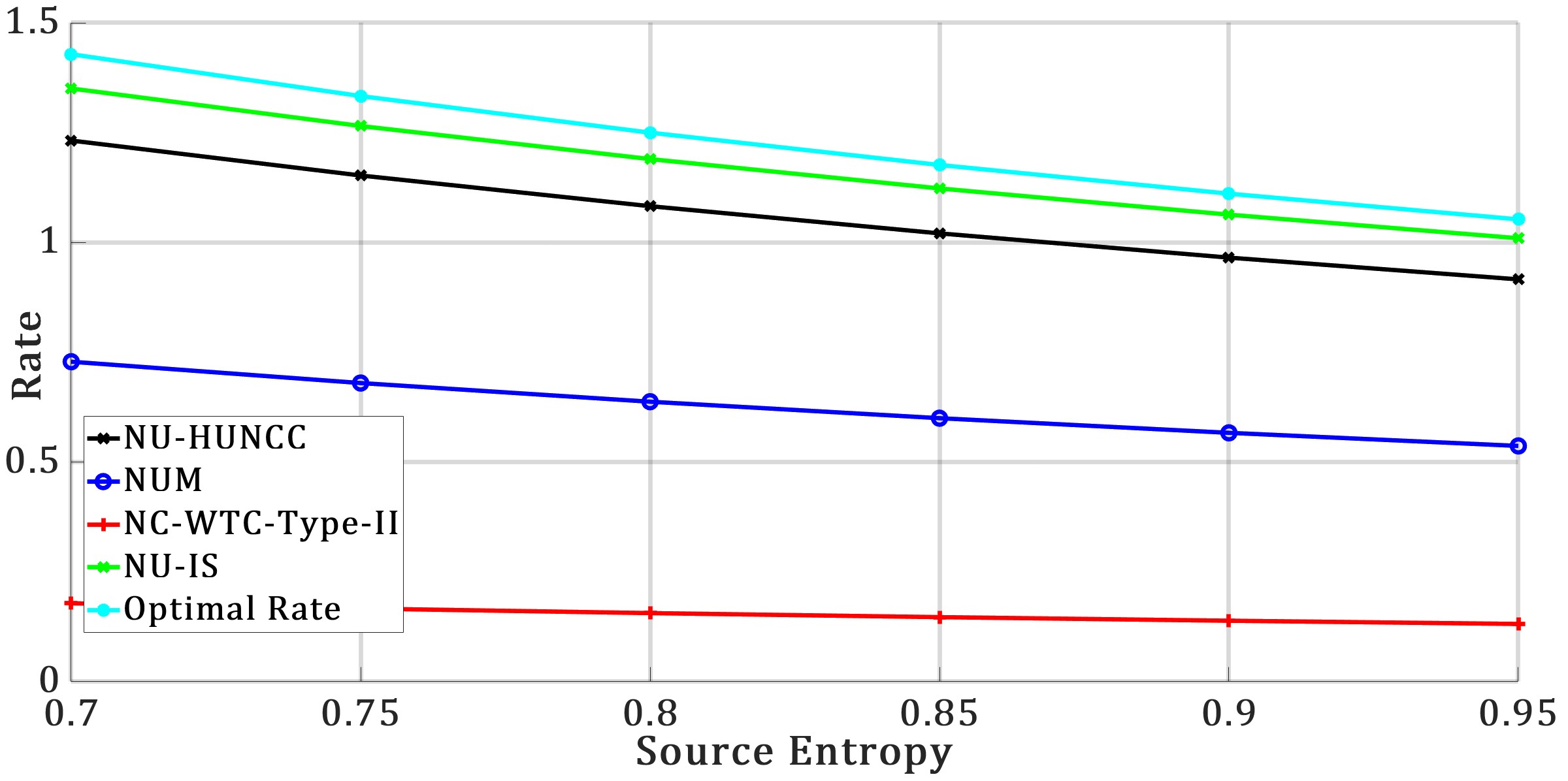}
    \caption{Numerical simulation of the communication rate as a function of the source entropy for eight communication links ($\ell = 8$), one encrypted link ($c = 1$), and message size of $128$[KB]. NU-IS and NU-HUNCC both achieve better rates than NC WTC Type \uppercase\expandafter{\romannumeral2} and NUM, respectively.}
    \label{fig:RateAsF(Entropy)}
\end{figure}

In Fig.~\ref{fig:RateAsF(Entropy)} we show the effect of the source message distribution on the communication rate. The communication rate decreases as with the increase of the source entropy as expected. When approaching $H(V) = 0.95$, NU-IS's data rate is $1.009$ which is within a $4\%$ margin from its upper bound. NU-HUNCC's data rate is $0.92$ which is within a $9\%$ margin of its upper bound and a $12-14\%$ of the optimal rate. In comparison, NUM's gap to the optimal communication rate is $51\%$.

\subsection{Computational Complexity} \label{sec:complexity}
In this section, we evaluate and compare the complexities of NUM and NU-HUNCC. We measure the complexity of each system by the number of binary operations required to complete the encryption and decryption of $\ell$ messages. First, we analyze the complexity of each step of the encoding/decoding process separately: 

1) \underline{Source coding/decoding:} For NU-HUNCC, we consider the almost uniform polar codes-based source encoder from Sec.~\ref{sec:source_code}. For NUM we consider a standard polar codes-based source encoder \cite{korada2010polar} \footnote{We use a regular lossless source encoder since original NUM has no requirements for the distribution of the input messages to be uniform.}. The specialized encoder, used by NU-HUNCC, requires extra steps of one-time padding of a subset of the bits from the compressed message to make it almost uniform. We will show that those additional binary operations do not have a major effect on the overall complexity of the encoding and decoding process.\off{  1) Source Encoding/Decoding:} The complexity of the polar encoding/decoding process is given in \cite[Theorem 5]{ArikanBase2009}. Given a source message of size $n$ bits, the encoding process requires $3n \log_2(n) / 2$ binary operations while the decoding process takes $n \log_2(n) / 2$ binary operations. The number of binary operations required for the one-time padding using the seed is $2n^{0.7331}$ binary operations for encoding and decoding, given by \eqref{eq:SeedBound}.

2) \underline{Linear IS channel code:} The encoding and decoding of the linear code is done by matrix multiplication (Sec.~\ref{sec:NU-HUNCC}, as given in lines $9-10$ and lines $43-45$, of Algorithm~\ref{algo:NU-HUNCC}). Overall the encoding and decoding process together sums up to $\tilde{n}\ell^2(\ell-1)$ \cite{d2021post}.

3) \underline{Original McEliecce with $[1024,524]$-Goppa codes:} We denote by $t$ the number of errors inserted by the code. In the original $[1024,524]$-Goppa codes for the McEliecce cryptosystem, we have $t=50$. We use the notations given in Sec.~\ref{sec:encryption}, $n_g$ and $c_g$, for the length of the code and the length of the message respectively. Thus, $n_g = 1024$ and $c_g = 524$. We denote the rate of the code by $\eta = c_g / n_g \approx 0.51$. As analyzed in \cite{d2021post}, the encryption process costs $\eta t n_g \log_{2}(n_g) / 2 \approx 130560$ binary operations per $c = 524$ bits. The decryption process costs $(3 - 2 \eta) t n_g \log_{2}(n_g) \approx 1013760$ binary operations per $c = 524$ bits.

\begin{figure}[htbp]
    \centering
    \includegraphics[width=1\linewidth]{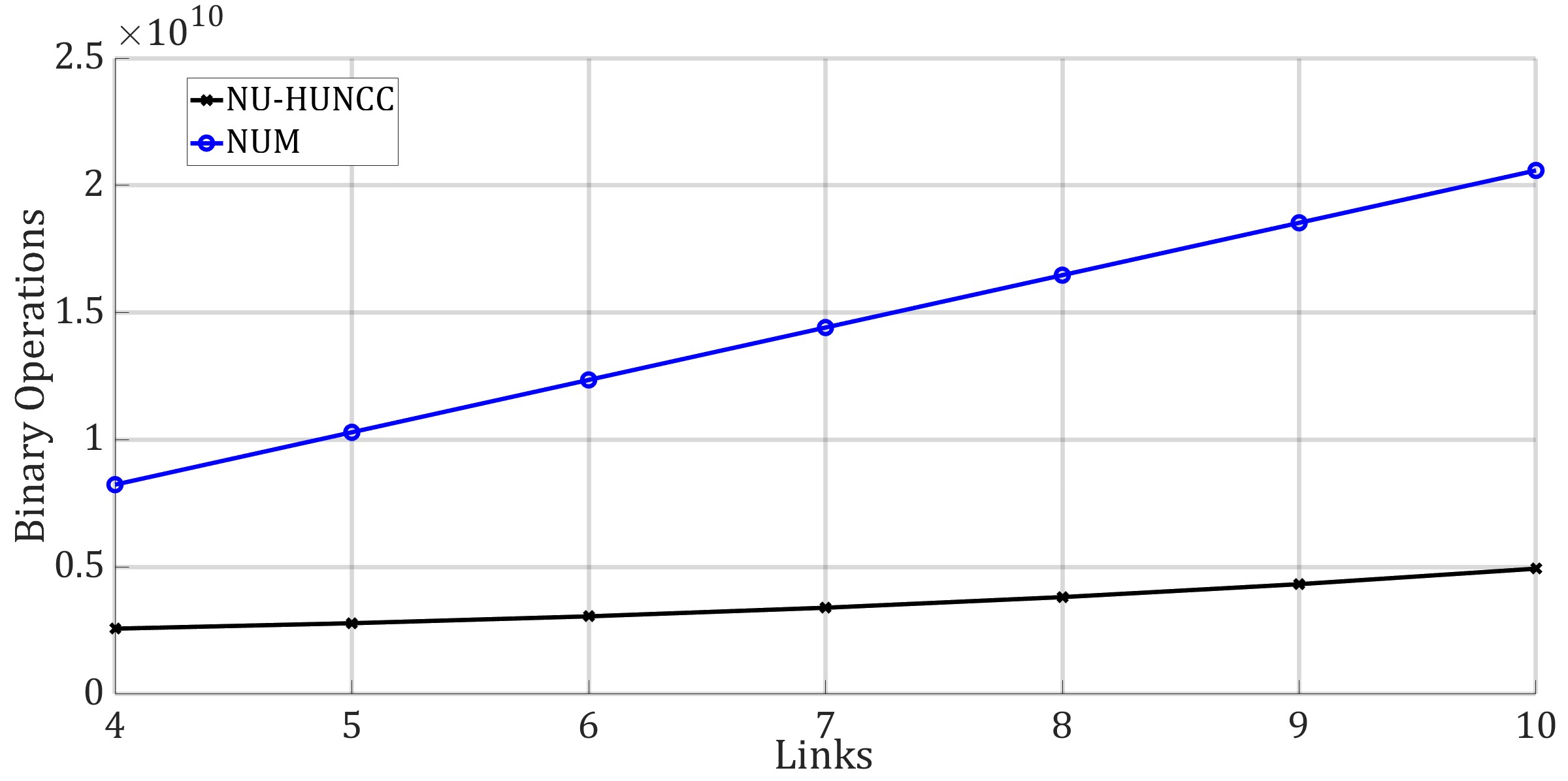}
    \caption{Numerical simulation of the complexity as a function of the number of communication links for a message of size $128$[KB] and source entropy of $0.9$ ($H(V) = 0.9$).}
    \label{fig:OperationsAsF(Links)}
\end{figure}

In Fig.~\ref{fig:OperationsAsF(Links)} we show a complexity comparison between the two systems considered in this section with $4-10$ links, where the number of encrypted links for NU-HUNCC is $c = 1$. The uniform seed used by NU-HUNCC is transmitted over the encrypted link as well (lines $17-20$ in Algorithm~\ref{algo:NU-HUNCC}). It can be seen that NUM which encryptes all of the links by McEliecce requires significantly more binary operations as the number of links grows, making it less efficient, compared to NU-HUNCC.

\section{Secure Individual Linear code\\ against IT-Eve \\ (Proof of Theorem~\ref{DirectLinear})} \label{sec:linear-code-proof}

We give here the full reliability and secrecy analysis of the $k_s$-IS linear coding scheme (proof of Theorem~\ref{DirectLinear}). We assume a noiseless multipath communication system with $\ell$ links and an eavesdropper, IT-Eve, with access to any subset $\mathcal{W} \subset \mathcal{L}$ of the links s.t. $|\mathcal{W}| \triangleq w < \ell$. Alice wants to send $\ell$ confidential messages to Bob while keeping IT-Eve ignorant about any set of $k_s \leq \ell - w$ messages individually.\off{We assume Alice uses the source coding scheme from Sec.~\ref{appendix:src_code} and the channel coding scheme from Appendix~\ref{appendix:msg_encoder}. We assume Alice uses the coding scheme from Sec.~\ref{sec:linear-NU-HUNCC-scheme-desc}.}

As described in Section~\ref{sec:NU-HUNCC}, Alice encodes $\ell$ source messages from a DMS $(\mathcal{V},p_V)$ (see lines $2 - 10$ in Algorithm~\ref{algo:NU-HUNCC}). The source message matrix $\underline{V}_{\mathcal{L}}$ is compressed using a polar codes-based source encoder from Sec.~\ref{sec:source_code}. The almost uniform messages at the output of the source encoder are of size $\tilde{n} = |\mathcal{H}_V| + d_{J}$, where $d_{J} = |\mathcal{J}_V|$. The message matrix at the output of the source encoder is denoted by $\underline{M}_\mathcal{L} \in \mathbb{F}_2^{\ell \times \tilde{n}}$. Finally, the message matrix is injectively mapped onto an extension field over $\mathbb{F}_{2^\mu}$ where $\mu \geq \ell$. We start by showing the code is reliable.

\subsection{Reliability with a Linear IS Channel Code} \label{sec:linear-code-reliability}

The reliability of the proposed end-to-end source-channel coding schemes is a consequence of the reliability of the source and channel decoders over the $\ell$ rows of the source matrix and $\lceil \tilde{n} / \mu \rceil$ columns of the message matrix.

The decoding of the messages is done by Bob as described by lines $42 - 51$ in Algorithm~\ref{algo:NU-HUNCC}. First, each column of the encoded message matrix is separately decoded. The decoded messages matrix is fully obtained from multiplying the codeword matrix $\underline{X}_{\mathcal{L}}$ by the parity check matrix $\underline{H}$ and by $\underline{\tilde{G}}$. Since Bob's observations of the codeword matrix are noiseless, then the decoding process is error-free. Thus, the proposed source-channel decoding error probability is made up of the source decoding error probability alone.

After obtaining the decoded message matrix, Bob can decode each row of the matrix separately to recover the source messages. Since Bob has the shared seed, he first one-time pads the bits from group $\mathcal{J}_V$~\eqref{eq:seed-group} as described in Sec.~\ref{sec:source_code}. Bob continues by employing successive cancellation decoding \cite{cronie2010lossless}\cite{arikan2009rate}, s.t., the error probability for the $i$-th row is given by
\begin{equation} \label{eq:single_row_pe}
\mathbb{P}\left[\underline{\hat{V}}_{\mathcal{L},i} \neq g_{s,n}\left(f_{s,n}(\underline{V}_{\mathcal{L},i},\underline{U}_{d_J,i}),\underline{U}_{d_J,i}\right)\right] \leq 2^{-n^{\beta}},
\end{equation}
for $\beta \in [0,\frac{1}{2})$ (Sec.~\ref{sec:source_code}). 

Now, using the union bound on all the rows in the recovered message matrix, we obtain the source-channel coding scheme error probability
\begin{align}
    &\mathbb{P}\left[V_{\mathcal{L}} \neq \underline{\hat{V}}_{\mathcal{L}}\right] = \nonumber \\
    &\mathbb{P}\left(\bigcup_{i=1}^{\ell} \left\{\underline{\hat{V}}_{\mathcal{L},i} \neq g_{s,n}\left(f_{s,n}(\underline{V}_{\mathcal{L},i},\underline{U}_{d_J,i}),\underline{U}_{d_J,i}\right)\right\}\right) \overset{(a)}{\leq} \ell2^{-n^{\beta}}, \label{eq:UnionError}
\end{align}
where (a) is from \eqref{eq:single_row_pe} and the union bound.

Finally, by requiring $\ell$ to be upper bounded by $o(2^{n^\beta})$ the error probability becomes negligible.

\subsection{Information Leakage against IT-Eve with Linear Code}\label{sec:linear-leakage-proof}
Recall the set $\mathcal{K}_s \subset \mathcal{L}$ where $|\mathcal{K}_s| \triangleq k_s = \ell - w$, and the set  $\mathcal{K}_w \triangleq \mathcal{L} \setminus \mathcal{K}_s$. We consider $\underline{M}_{\mathcal{K}_s} \subset \underline{M}_{\mathcal{L}}$ as the subset of the secured messages and  $\underline{M}_{\mathcal{K}_w} \subset \underline{M}_{\mathcal{L}} \setminus \underline{M}_{\mathcal{K}_s}$ as the rest of the messages\footnote{Considering the notations from Sec.~\ref{sec:linear-IS-code}, we have that $\underline{M}_{\mathcal{K}_w} \triangleq \underline{M}_w$. The change in notations is for compatibility with the proofs given in Sec.~\ref{sec:linear-code-proof} and Appendix~\ref{sec:random-code-proof}.}. The distribution of $\underline{Z}_{\mathcal{W}}$ induced by the uniform message matrix is denoted by $\tilde{p}_{\underline{Z}_{\mathcal{W}}}$ or $\tilde{p}_{\underline{Z}_{\mathcal{W}}|\underline{M}_{\mathcal{K}_s}=\underline{m}_{k_s}}$.
For any $\underline{m}_{k_s} \in \underline{\mathcal{M}}_{k_s} \in \mathbb{F}_{2^\mu}^{k_s}$
\begin{alignat}{1}
    &\mathbb{V}\left(p_{\underline{Z}_{\mathcal{W}}|\underline{M}_{\mathcal{K}_s}=\underline{m}_{k_s}},p_{\underline{Z}_{\mathcal{W}}}\right) \label{FullSecrecy}\\
    & \quad \overset{(a)}{\leq} \mathbb{V}\left(p_{\underline{Z}_{\mathcal{W}}|\underline{M}_{\mathcal{K}_s}=\underline{m}_{k_s}},\tilde{p}_{\underline{Z}_{\mathcal{W}}|\underline{M}_{\mathcal{K}_s}=\underline{m}_{k_s}}\right) \label{eq:CondSec} \\
    & \quad\quad + \mathbb{V}\left(\tilde{p}_{\underline{Z}_{\mathcal{W}}|\underline{M}_{\mathcal{K}_s}=\underline{m}_{k_s}},\tilde{p}_{\underline{Z}_{\mathcal{W}}}\right) \label{eq:UnifSec} \\
    & \quad\quad + \mathbb{V}\left(\tilde{p}_{\underline{Z}_{\mathcal{W}}},p_{\underline{Z}_{\mathcal{W}}}\right), \label{eq:FullZ}
\end{alignat}
where (a) is from the triangle inequality. Since this is true for all  $\underline{m}_{k_s} \in \underline{\mathcal{M}}_{k_s}$, from now on we omit the equality $\underline{M}_{\mathcal{K}_s} = \underline{m}_{k_s}$ for ease of notation.

Now, we bound each of the expressions \eqref{eq:CondSec}-\eqref{eq:FullZ}, starting with \eqref{eq:CondSec}
\begin{align*}
    & \mathbb{V}\left(p_{\underline{Z}_{\mathcal{W}}|\underline{M}_{\mathcal{K}_s}},\tilde{p}_{\underline{Z}_{\mathcal{W}}|\underline{M}_{\mathcal{K}_s}}\right) \\
    & \quad = \sum_{\underline{z}_{w}} \left|p_{\underline{Z}_{\mathcal{W}}|\underline{M}_{\mathcal{K}_s}}(\underline{z}_{w}|\underline{m}_{k_s})-\tilde{p}_{\underline{Z}_{\mathcal{W}}|\underline{M}_{\mathcal{K}_s}}(\underline{z}_{w}|\underline{m}_{k_s})\right| \\
    & \quad = \sum_{\underline{z}_{w}}\left|\sum_{\underline{m}_{k_w}}\left(p_{\underline{Z}_{\mathcal{W}}\underline{M}_{\mathcal{K}_w}|\underline{M}_{\mathcal{K}_s}}(\underline{z}_{w},\underline{m}_{k_w}|\underline{m}_{k_s}) \right. \right. \\
    & \quad\quad\quad\quad\quad\quad\quad \left. \left. - \tilde{p}_{\underline{Z}_{\mathcal{W}}\underline{M}_{\mathcal{K}_w}|\underline{M}_{\mathcal{K}_s}}(\underline{z}_{w},\underline{m}_{k_w}|\underline{m}_{k_s})\right)\right| \\
    & \quad = \sum_{\underline{z}_{w}}\left|\sum_{\underline{m}_{k_w}}\left(p_{\underline{Z}_{\mathcal{W}}|\underline{M}_{\mathcal{K}_w}\underline{M}_{\mathcal{K}_s}}(\underline{z}_{w}|\underline{m}_{k_w},\underline{m}_{k_s}) \right. \right. \\
    & \quad\quad\quad\quad\quad\quad\quad\quad\quad\quad\quad \left. \left.\cdot p_{\underline{M}_{\mathcal{K}_w}|\underline{M}_{\mathcal{K}_s}}(\underline{m}_{k_w}|\underline{m}_{k_s}) \right. \right. \\
    & \quad\quad\quad\quad\quad \left. \left. -p_{\underline{Z}_{\mathcal{W}}|\underline{M}_{\mathcal{K}_w}\underline{M}_{\mathcal{K}_s}}(\underline{z}_{w}|\underline{m}_{k_w},\underline{m}_{k_s}) \right. \right. \\
    & \quad\quad\quad\quad\quad\quad\quad\quad\quad\quad\quad \left. \left. \cdot \tilde{p}_{\underline{M}_{\mathcal{K}_w}|\underline{M}_{\mathcal{K}_s}}(\underline{m}_{k_w}|\underline{m}_{k_s})\right)\right| \\
    & \quad = \sum_{\underline{z}_\mathcal{W}} p_{\underline{Z}_{\mathcal{W}}|\underline{M}_{\mathcal{K}_w}\underline{M}_{\mathcal{K}_s}}(\underline{z}_{w}|\underline{m}_{k_w},\underline{m}_{k_s}) \cdot \\ 
    & \quad\quad\quad\quad \left|\sum_{\underline{m}_{k_w}} \left(p_{\underline{M}_{\mathcal{K}_w}|\underline{M}_{\mathcal{K}_s}}(\underline{m}_{k_w}|\underline{m}_{k_s}) \right. \right. \\
    &  \quad\quad\quad\quad\quad\quad\quad\quad\quad \left. \left. -\tilde{p}_{\underline{M}_{\mathcal{K}_w}|\underline{M}_{\mathcal{K}_s}}(\underline{m}_{k_w}|\underline{m}_{k_s})\right)\right| \\
    & \quad \overset{(a)}{\leq} \sum_{\underline{z}_\mathcal{W}} \sum_{\underline{m}_{k_w}} p_{\underline{Z}_{\mathcal{W}}|\underline{M}_{\mathcal{K}_w}\underline{M}_{\mathcal{K}_s}}(\underline{z}_{w}|\underline{m}_{k_w},\underline{m}_{k_s}) \cdot \\
    & \quad\quad\quad \left|p_{\underline{M}_{\mathcal{K}_w}|\underline{M}_{\mathcal{K}_s}}(\underline{m}_{k_w}|\underline{m}_{k_s})-\tilde{p}_{\underline{M}_{\mathcal{K}_w}|\underline{M}_{\mathcal{K}_s}}(\underline{m}_{k_w}|\underline{m}_{k_s})\right| \\
    & \quad \overset{(b)}{=} \sum_{\underline{m}_{k_w}} \left|p_{\underline{M}_{\mathcal{K}_w}}(\underline{m}_{k_w}) - \tilde{p}_{\underline{M}_{\mathcal{K}_w}}(\underline{m}_{k_w})\right| \\
    & \quad = \mathbb{V}\left(p_{\underline{M}_{\mathcal{K}_w}},p_{U_{\underline{M}_{\mathcal{K}_w}}}\right) \leq \mathbb{V}\left(p_{\underline{M}_{\mathcal{L}}},p_{U_{\underline{M}_{\mathcal{L}}}}\right),
\end{align*}
where (a) holds from the triangle inequality, (b) holds from the independence between messages and $p_{U_{\underline{M}_{\mathcal{K}_w}}}$ is the uniform distribution of the uniform matrix $\underline{\tilde{M}}_{\mathcal{K}_w}$. We continue with bounding \eqref{eq:FullZ}
\begin{multline*}
    \begin{aligned}
    &\mathbb{V}\left(\tilde{p}_{\underline{Z}_{\mathcal{W}}},p_{\underline{Z}_{\mathcal{W}}}\right) \\
       & \quad= \sum_{\underline{z}_{w}}\left|\tilde{p}_{\underline{Z}_{\mathcal{W}}}(\underline{z}_{w}) - p_{\underline{Z}_{\mathcal{W}}}(\underline{z}_{w})\right| \\
       & \quad = \sum_{\underline{z}_{w}}\left|\sum_{\underline{m}_{k_w},\underline{m}_{k_s}}\left(\tilde{p}_{\underline{Z}_{\mathcal{W}} \underline{M}_{\mathcal{K}_w} \underline{M}_{\mathcal{K}_s}}(\underline{z}_{w},\underline{m}_{k_w},\underline{m}_{k_s})  \right. \right. \\ 
       & \quad\quad\quad\quad\quad\quad\quad\quad\quad \left. \left. -p_{\underline{Z}_{\mathcal{W}} \underline{M}_{\mathcal{K}_w} \underline{M}_{\mathcal{K}_s}}(\underline{z}_{w},\underline{m}_{k_w},\underline{m}_{k_s})\right)\right| \\
       & \quad = \sum_{\underline{z}_{w}}\left|\sum_{\underline{m}_{\ell}}\left(p_{\underline{Z}_{\mathcal{W}}|\underline{M}_{\mathcal{L}}}(\underline{z}_{w}|\underline{m}_{\ell})  \cdot p_{U_{\underline{M}_{\mathcal{L}}}}(\underline{m}_{\ell}) \right. \right. \\
       & \quad\quad\quad\quad\quad\quad \left. \left.- p_{\underline{Z}_{\mathcal{W}} | \underline{M}_{\mathcal{L}}}(\underline{z}_{w}|\underline{m}_{\ell}) \cdot p_{\underline{M}_{\mathcal{L}}}(\underline{m}_{\ell})\right)\right| \\
       & \quad \overset{(a)}{\leq} \sum_{\underline{m}_{\ell}}\sum_{\underline{z}_{w}} p_{\underline{Z}_{\mathcal{W}}|\underline{M}_{\mathcal{L}}}(\underline{z}_{w}|\underline{m}_{\ell}) \cdot \left| p_{U_{\underline{M}_{\mathcal{L}}}}(\underline{m}_{\ell}) - p_{M_{\mathcal{L}}}(\underline{m}_{\ell})\right| \\
       & \quad = \sum_{\underline{m}_{\ell}} \left|p_{U_{\underline{M}_{\mathcal{L}}}}(\underline{m}_{\ell}) - p_{\underline{M}_{\mathcal{L}}}(\underline{m}_{\ell})\right|  = \mathbb{V}\left(p_{\underline{M}_{\mathcal{L}}},p_{U_{\underline{M}_{\mathcal{L}}}}\right),
    \end{aligned}
\end{multline*}
where inequality (a) follows from the triangle inequality.

We now bound $\mathbb{D}(p_{\underline{M}_{\mathcal{L}}}||p_{U_{\underline{M}_{\mathcal{L}}}})$ and obtain 
\begin{equation*}
\begin{aligned}
    &\mathbb{D}\left(p_{\underline{M}_{\mathcal{L}}}||p_{U_{\underline{M}_{\mathcal{L}}}}\right) \\
    & \quad = \sum_{\underline{m}_{\ell}} p_{\underline{M}_{\mathcal{L}}}(\underline{m}_{\ell}) \cdot \log_{2} \left(\frac{p_{\underline{M}_{\mathcal{L}}}(\underline{m}_{\ell})}{p_{U_{\underline{M}_{\mathcal{L}}}}(\underline{m}_{\ell})}\right) \\
    & \quad \overset{(a)}{=} \log_{2} (2^{\ell \cdot \tilde{n}}) - H\left(\underline{M}_{\mathcal{L}}\right) \\
    & \quad = \ell \tilde{n} - H\left(\underline{M}_{\mathcal{L}}\right) \\
    & \quad \overset{(b)}{=} \ell \tilde{n} - \ell \cdot \sum_{j=1}^{\tilde{n}}  H\left(M^{(j)}|M^{j-1}\right) \\
    & \quad \overset{(c)}{\leq} \ell \tilde{n} - \ell  \tilde{n} (1-\delta_n) = \ell  \tilde{n}  \delta_n \leq \ell  \tilde{n} 2^{-n^\beta},
\end{aligned}
\end{equation*}
where (a) holds from the one-to-one mapping between the message matrix over $\mathbb{F}_{2^\mu}^{\ell \times \lceil \tilde{n} / \mu \rceil}$ and $\mathbb{F}_{2^\mu}^{\ell \times \tilde{n}}$. (b) holds from the independence between messages, and (c) is from using the source coding scheme from Sec.~\ref{sec:source_code}.

Now, we invoke the Pinsker inequality \cite{1053968} to bound the variational distance between the distribution $p_{\underline{M}_{\mathcal{L}}}$ and the uniform distribution $p_{U_{\underline{M}_{\mathcal{L}}}}$. Thus, we have
\begin{equation*}
    \begin{aligned}
        \frac{1}{2}\mathbb{V}^2\left(p_{\underline{M}_{\mathcal{L}}},p_{U_{\underline{M}_{\mathcal{L}}}}\right) \leq  \mathbb{D}\left(p_{\underline{M}_{\mathcal{L}}}||p_{U_{\underline{M}_{\mathcal{L}}}}\right) \leq \ell \tilde{n} 2^{-n^\beta},
    \end{aligned}
\end{equation*}
s.t. we bound the expressions in \eqref{eq:CondSec} and \eqref{eq:FullZ} by
\begin{equation} \label{eq:uniform-bound}
    \begin{aligned}
        \mathbb{V}\left(p_{\underline{M}_{\mathcal{L}}},p_{U_{\underline{M}_{\mathcal{L}}}}\right) \leq \sqrt{2 \ell \tilde{n} 2^{-n^\beta}}.
    \end{aligned}
\end{equation}

We now bound \eqref{eq:UnifSec}. In \cite[Sec. \uppercase\expandafter{\romannumeral6}]{SMSM} it was shown that for a uniform message matrix $\underline{\tilde{M}}_{\mathcal{L}}\in \mathbb{F}_{2^\mu}^{\ell \times \lceil \tilde{n} / \mu \rceil}$ with $\mu \geq \ell$ each column $j$ is $k_s$-IS s.t. $I(\underline{\tilde{M}}_{\mathcal{K}_s}^{(j)};\underline{\tilde{Z}}_{\mathcal{W}}^{(j)}) = 0$. Since the message matrix is from a uniform distribution, its columns are independent. Thus, we also have that $I(\underline{\tilde{M}}_{\mathcal{K}_s};\underline{\tilde{Z}}_{\mathcal{W}}) = 0$. By employing Pinsker's inequality \cite{1053968}, we have that
\begin{equation*}
\begin{aligned}
    \frac{1}{2}\mathbb{V}^2\left(\tilde{p}_{\underline{Z}_{\mathcal{W}},\underline{M}_{\mathcal{K}_s}},\tilde{p}_{\underline{Z}_{\mathcal{W}}}\tilde{p}_{\underline{M}_{\mathcal{K}_s}}\right) \leq I(\underline{\tilde{M}}_{\mathcal{K}_s};\underline{\tilde{Z}}_{\mathcal{W}}) = 0,
\end{aligned} 
\end{equation*}
from which, we conclude

\begin{equation} \label{eq:HelpTotalVariation}
\begin{aligned}
    \mathbb{V}\left(\tilde{p}_{\underline{Z}_{\mathcal{W}},\underline{M}_{\mathcal{K}_s}},\tilde{p}_{\underline{Z}_{\mathcal{W}}}\tilde{p}_{\underline{M}_{\mathcal{K}_s}}\right) = 0.
\end{aligned} 
\end{equation}
On the other hand, we have that
\begin{equation*}
\begin{aligned}
    &\mathbb{V}\left(\tilde{p}_{\underline{Z}_{\mathcal{W}},\underline{M}_{\mathcal{K}_s}},\tilde{p}_{\underline{Z}_{\mathcal{W}}}\tilde{p}_{\underline{M}_{\mathcal{K}_s}}\right) \\
    & \quad = \sum_{\underline{z}_w, \underline{m}_{k_s}} \left|\tilde{p}_{\underline{Z}_{\mathcal{W}},\underline{M}_{\mathcal{K}_s}}(\underline{z}_{w},\underline{m}_{k_s})-\tilde{p}_{\underline{Z}_{\mathcal{W}}}(\underline{z}_{w})\tilde{p}_{\underline{M}_{\mathcal{K}_s}}(\underline{m}_{k_s})\right| \\
    & \quad = \sum_{\underline{z}_w, \underline{m}_{k_s}} \tilde{p}_{\underline{M}_{\mathcal{K}_s}}(\underline{m}_{k_s})\left|\tilde{p}_{\underline{Z}_{\mathcal{W}}|\underline{M}_{\mathcal{K}_s}}(\underline{z}_{w}|\underline{m}_{k_s})-\tilde{p}_{\underline{Z}_{\mathcal{W}}}(\underline{z}_{w})\right| \\
    & \quad = \sum_{\underline{m}_{k_s}} \tilde{p}_{\underline{M}_{\mathcal{K}_s}}(\underline{m}_{k_s}) \mathbb{V}\left(\tilde{p}_{\underline{Z}_{\mathcal{W}}|\underline{M}_{\mathcal{K}_s}=\underline{m}_{k_s}},\tilde{p}_{\underline{Z}_{\mathcal{W}}}\right).
\end{aligned}
\end{equation*}
Thus, from \eqref{eq:HelpTotalVariation}, for all $\underline{m}_{k_s} \in \underline{\mathcal{M}}_{\mathcal{K}_s}$
\begin{equation} \label{eq:HelpLemmaBoundLinear}
\begin{aligned}
    \mathbb{V}\left(\tilde{p}_{\underline{Z}_{\mathcal{W}}|\underline{M}_{\mathcal{K}_s}=\underline{m}_{k_s}},\tilde{p}_{\underline{Z}_{\mathcal{W}}}\right) = 0.
\end{aligned} 
\end{equation}

Now, we return to \eqref{FullSecrecy}. By substituting \eqref{eq:CondSec}-\eqref{eq:FullZ} with \eqref{eq:uniform-bound} and \eqref{eq:HelpLemmaBoundLinear}, we show that
\off{\[
\mathbb{V}\left(p_{\underline{Z}_{\mathcal{W}} | \underline{M}_{\mathcal{K}_s}=\underline{m}_{k_s}},p_{\underline{Z}_{\mathcal{W}}}\right) \leq  \tilde{n} \ell^{-\frac{t}{2}} + 2 \sqrt{2 \ell \tilde{n} 2^{-n^\beta}}.
\]}
\begin{equation} \label{eq:linear-msg-bound}
    \begin{aligned}
        \mathbb{V}\left(p_{\underline{Z}_{\mathcal{W}} | \underline{M}_{\mathcal{K}_s}=\underline{m}_{k_s}},p_{\underline{Z}_{\mathcal{W}}}\right) \leq 2 \sqrt{2 \ell \tilde{n} 2^{-n^\beta}}.
    \end{aligned}
\end{equation}

We have shown that the linear channel coding scheme employed on $\underline{M}_{\mathcal{L}}$ is $k_s$-IS. To conclude the leakage proof against IT-Eve, we show that $\underline{V}_{\mathcal{L}}$ is $k_s$-IS by bounding $\mathbb{V}\left(p_{\underline{Z}_{\mathcal{W}} | \underline{V}_{\mathcal{K}_s}=\underline{v}_{k_s}},p_{\underline{Z}_{\mathcal{W}}}\right)$, and showing that for any $\underline{v}_{k_s} \in \underline{\mathcal{V}}_{\mathcal{K}_s}$ the
information leakage becomes negligible. Thus, we have
\begin{align} \label{eq:final_leakage}
    &\mathbb{V}\left(p_{\underline{Z}_{\mathcal{W}} | \underline{V}_{\mathcal{K}_s}=\underline{v}_{k_s}},p_{\underline{Z}_{\mathcal{W}}}\right) \nonumber \\
    & \quad = \sum_{\underline{z}_w} \left|p_{\underline{Z}_{\mathcal{W}}|\underline{V}_{\mathcal{K}_s}}(\underline{z}_{w}|\underline{v}_{k_s})-p_{\underline{Z}_{\mathcal{W}}}(\underline{z}_{w})\right| \nonumber \\
    & \quad = \sum_{\underline{z}_{w}}\left|\sum_{\underline{m}_{k_s}} p_{\underline{Z}_{\mathcal{W}}|\underline{M}_{\mathcal{K}_s}\underline{V}_{\mathcal{K}_s}}(\underline{z}_{w}|\underline{m}_{k_s}\underline{v}_{k_s}) \cdot \right. \nonumber \\
    & \quad\quad\quad\quad\quad\quad\quad\quad\quad\quad \left. p_{\underline{M}_{\mathcal{K}_s}|\underline{V}_{\mathcal{K}_s}}(\underline{m}_{k_s}|\underline{v}_{k_s}) \right.  \nonumber \\ 
    & \quad\quad\quad\quad\quad \left. - p_{\underline{Z}_{\mathcal{W}}}(\underline{z}_{w})\cdot \sum_{\underline{m}_{k_s}} p_{\underline{M}_{\mathcal{K}_s}|\underline{V}_{\mathcal{K}_s}}(\underline{m}_{k_s}|\underline{v}_{k_s})\right| \nonumber \\
    & \quad \overset{(a)}{\leq} \sum_{\underline{m}_{k_s}} \sum_{\underline{z}_{w}} p_{\underline{M}_{\mathcal{K}_s}|\underline{V}_{\mathcal{K}_s}}(\underline{m}_{k_s}|\underline{v}_{k_s}) \cdot \nonumber \\
    & \quad\quad\quad\quad\quad\quad \left|p_{\underline{Z}_{\mathcal{W}}|\underline{M}_{\mathcal{K}_s}\underline{V}_{\mathcal{K}_s}}(\underline{z}_{w}|\underline{m}_{k_s}\underline{v}_{k_s}) - p_{\underline{Z}_{\mathcal{W}}}(\underline{z}_{w})\right| \nonumber \\
    & \quad = \sum_{\underline{m}_{k_s}} p_{\underline{M}_{\mathcal{K}_s}|\underline{V}_{\mathcal{K}_s}}(\underline{m}_{k_s}|\underline{v}_{k_s}) \cdot \nonumber \\
    & \quad\quad\quad\quad\quad\quad \mathbb{V}\left(p_{\underline{Z}_{\mathcal{W}} | \underline{V}_{\mathcal{K}_s}=\underline{v}_{k_s}\underline{M}_{\mathcal{K}_s}=\underline{m}_{k_s}},p_{\underline{Z}_{\mathcal{W}}}\right) \nonumber \\
    & \quad \overset{(b)}{=} \sum_{\underline{m}_{\mathcal{K}_s}} p_{\underline{M}_{\mathcal{K}_s}|\underline{V}_{\mathcal{K}_s}}(\underline{m}_{k_s}|\underline{v}_{k_s}) \cdot \mathbb{V}\left(p_{\underline{Z}_{\mathcal{W}} | \underline{M}_{\mathcal{K}_s}=\underline{m}_{k_s}},p_{\underline{Z}_{\mathcal{W}}}\right) \nonumber \\
    & \quad \overset{(c)}{\leq} \off{\tilde{n} \ell^{-\frac{t}{2}} + }2 \sqrt{2 \ell \tilde{n}  2^{-n^\beta}},
\end{align}
where (a) is from the triangle inequality, (b) is since $\underline{V}_{\mathcal{K}_s} \rightarrow \underline{M}_{\mathcal{K}_s} \rightarrow \underline{Z}_{\mathcal{W}}$ is a Markov chain, and (c) is from \eqref{eq:linear-msg-bound}.
For the information leakage to be negligible, we give an upper\off{ and lower} bound to $\ell$ compared to $n$. \off{$\ell$ is lower bounded by $\omega(\tilde{n}^{\frac{2}{t}})$, s.t. the expression $\tilde{n} \ell^{-\frac{t}{2}}$ is negligible. In addition, }$\ell$ is upper bounded by $o\left(2^{n^\beta}/\tilde{n}\right)$, s.t. the expression $2 \sqrt{2 \ell \tilde{n} 2^{-n^\beta}}$ is negligible. By choosing $\ell$ that upholds\off{both} those bounds, the information leakage to IT-Eve becomes negligible. This holds for any set of $\mathcal{K}_s \subset \mathcal{L}$ source messages, thus the code is $k_s$-IS.

\section{Converse \\ (Proof of Theorem~\ref{Converse})} \label{sec:converse}
In this section, we provide the proof of the converse result as presented in Theorem~\ref{Converse}. Our proof follows similar techniques as presented by Csizar et. al. \cite{csiszar1978broadcast}, Oohama et. al. \cite{watanabe2014optimal}, and Chou et. al. \cite{NegligbleCost}. However, our converse result does not directly follow these known results as it is applied on blocks of individual secure confidential messages and does not rely on a source of randomness shared between Alice and Bob other than the seed.

We consider a multipath communication with $\ell$ links, $w < \ell$ of which are observed by IT-Eve. There exists a coding scheme s.t. Alice delivers $k_s < \ell - w$ source message from the DMS $(\mathcal{V},p_V)$ in the presence of IT-Eve, to Bob in an IS manner, s.t.
\begin{equation} \label{eq:converse-secrecy}    I(\underline{V}_{\mathcal{K}_s};\underline{Z}_{\mathcal{W}}) \leq \epsilon_s \text{, (Secure)}
\end{equation}
\begin{equation} \label{eq:converse-reliability}
    \mathbb{P}(\underline{\hat{V}}_{\mathcal{L}}(\underline{Y}_{\mathcal{L}}) \neq \underline{V}_{\mathcal{L}}) \leq \epsilon_e \text{, (Reliable)}
\end{equation}
\begin{equation} \label{eq:converse-seed}
    \frac{d_n}{n} \leq \epsilon_{d} \text{, (Seed)}
\end{equation}
\off{
\begin{alignat}{1}
    P(\underline{V}_{\mathcal{L}} \neq \hat{\underline{V}}_{\mathcal{L}}) &\leq \epsilon^{'}_{n} \text{ (reliability)} \label{Converse:Reliability} \\
    I(\underline{V}_{K_s};\underline{Z}_{\mathcal{L}}) &\leq \delta_n \text{ ($k_s$-IS)} \label{Converse:IS} \\
    \frac{d_n}{n} &\leq \mu_n \text{ (sub-linear seed)} \label{eq:converse-seed}
\end{alignat}
Where we have that in the asymptotic regime $\lim\limits_{n \rightarrow \infty} \epsilon^{'}_{n} = \lim\limits_{n \rightarrow \infty} \delta_n = \lim\limits_{n \rightarrow \infty} \mu_n = 0$. We also denote $\epsilon_n = \epsilon^{'}_{n} + \frac{1}{n}$.}
where $\lim\limits_{n,\ell \rightarrow \infty} \epsilon_e = \lim\limits_{n,\ell \rightarrow \infty} \epsilon_s = \lim\limits_{n \rightarrow \infty} \epsilon_{d} = 0$. Consequently, we have
\begin{align}\label{TotalSecrecy}
&H(\underline{V}_{\mathcal{K}_w}) + \ell n \epsilon_d \notag \\ 
&\quad \geq H(\underline{V}_{\mathcal{K}_w}) + \ell d_n \notag \\
  &\quad \overset{(a)}{=} H(\underline{V}_{\mathcal{K}_w}) + H(\underline{U}_{d_n}) \notag \\
  &\quad= H(\underline{V}_{\mathcal{K}_w},\underline{U}_{d_n}) \notag \\
  &\quad\overset{(b)}{=} H(\underline{V}_{\mathcal{K}_w},\underline{U}_{d_n}|\underline{V}_{\mathcal{K}_s}) \notag \\
  &\quad\overset{(c)}{\geq} H(\underline{X}_{\mathcal{L}}|\underline{V}_{\mathcal{K}_s}) \notag \\
  &\quad\geq I(\underline{X}_{\mathcal{L}};\underline{Z}_{\mathcal{W}}|\underline{V}_{\mathcal{K}_s}) \notag \\
  &\quad\overset{(d)}{=} H(\underline{Z}_{\mathcal{W}}|\underline{V}_{\mathcal{K}_s}) \notag \\
  &\quad\overset{(e)}{\geq} H(\underline{Z}_{\mathcal{W}}) - \epsilon_s \notag \\
  &\quad= I(\underline{X}_{\mathcal{L}};\underline{Z}_{\mathcal{W}}) - \epsilon_s,
\end{align}
where (a) and (b) are from the independence between the message and the seed. (c) is because $\underline{X}_{\mathcal{L}}$ is a function of $\underline{V}_{\mathcal{L}}$ and $\underline{U}_{d_n}$. \off{(between (c)+(d)) we assume $H(\underline{X}_{\mathcal{L}}|\underline{Z}_{\mathcal{W}},\underline{V}_{\mathcal{K}_s} \approx 0$}(d) is since we consider noiseless channels and (e) holds by~\eqref{eq:converse-secrecy}.\off{ and (e) hold from $\underline{V}_{\mathcal{L}}-\underline{X}_{\mathcal{L}}-\underline{Z}_{\mathcal{L}}$ being  Markov chain.}

We continue with
\begin{align}\label{TotalRandomness}
&H(\underline{V}_{\mathcal{K}_w}) + H(\underline{V}_{\mathcal{K}_s}) \notag \\
&\quad \overset{(a)}{=} H(\underline{V}_{\mathcal{K}_w}\underline{V}_{\mathcal{K}_s}|\underline{U}_{d_n}) \notag \\
&\quad=I(\underline{V}_{\mathcal{K}_w}\underline{V}_{\mathcal{K}_s};\underline{Y}_{\mathcal{L}}|\underline{U}_{d_n}) + \notag \\ 
& \quad\quad\quad H(\underline{V}_{\mathcal{K}_w}\underline{V}_{\mathcal{K}_s}|\underline{Y}_{\mathcal{L}}\underline{U}_{d_n}) \notag \\
& \quad \overset{(b)}{\leq} I(\underline{V}_{\mathcal{K}_w}\underline{V}_{\mathcal{K}_s};\underline{Y}_{\mathcal{L}}|\underline{U}_{d_n}) + \epsilon_e \notag \\
&\quad \leq I(\underline{V}_{\mathcal{K}_w}\underline{V}_{\mathcal{K}_s}\underline{U}_{d_n};\underline{Y}_{\mathcal{L}}) + \epsilon_e \notag \\
&\quad \overset{(c)}{=} I(\underline{X}_{\mathcal{L}};\underline{Y}_{\mathcal{L}}) + \epsilon_e,
\end{align}
where (a) is from the independence between the seed and the source. (b) holds from Fano inequality and \eqref{eq:converse-reliability}. (c) holds since the channels are noiseless.\off{$\underline{V}_{K_w}\underline{V}_{K_s},\underline{U}_{d_n} - \underline{X}_{\mathcal{L}} - \underline{Y}_{\mathcal{L}}$ is Markov chain.inequality between (b) and (c) holds because we assumed $I(\underline{X}_{\mathcal{L}};\underline{U}_{d_n}) \approx = 0$. that is since the we can look at $I(\underline{X}_{\mathcal{L}};\underline{U}_{d_n}) = H(\underline{X}_{\mathcal{L}}) - H(\underline{X}_{\mathcal{L}}|\underline{U}_{d_n})$, the second term is almost equal to the first because if the difference between them was big then it would mean that the codebook is built in such a way that different values of $\underline{V}_{\mathcal{L}}$ along with the same value of $\underline{U}_{d_n}$ result in the same $\underline{X}_{\mathcal{L}}$ which would make the code unreliable. this fact is true because as $H(\underline{X}_{\mathcal{L}}|\underline{U}_{d_n})$ gets smaller it means that $\underline{U}_{d_n}$ holds significant information about $\underline{X}_{\mathcal{L}}$, i.e. a lot $\underline{X}_{\mathcal{L}}$ pointed by that $\underline{U}_{d_n}$ are the same one, thus we have some information about $\underline{X}_{\mathcal{L}}$.}

Finally, we have
\begin{align} \label{ConfidentialRandomness}
    &H(\underline{V}_{\mathcal{K}_s}) \notag \\
    &\quad \overset{(a)}{=} H(\underline{V}_{\mathcal{K}_s}\underline{V}_{\mathcal{K}_w}) - H(\underline{V}_{\mathcal{K}_w}) \notag \\
     &\quad= I(\underline{V}_{\mathcal{K}_s}\underline{V}_{\mathcal{K}_w};\underline{Y}_{\mathcal{L}}\underline{U}_{d_n}) + H(\underline{V}_{\mathcal{K}_s}\underline{V}_{\mathcal{K}_w}|\underline{Y}_{\mathcal{L}}\underline{U}_{d_n}) \notag \\
     & \quad\quad - H(\underline{V}_{\mathcal{K}_w}) \notag \\
     &\quad \overset{(b)}{\leq} I(\underline{V}_{\mathcal{K}_s}\underline{V}_{\mathcal{K}_w};\underline{Y}_{\mathcal{L}}\underline{U}_{d_n}) + \epsilon_e - H(\underline{V}_{\mathcal{K}_w}) \notag \\
     &\quad \overset{(c)}{\leq} I(\underline{V}_{\mathcal{K}_s}\underline{V}_{\mathcal{K}_w};\underline{Y}_{\mathcal{L}}) + I(\underline{V}_{\mathcal{K}_s}\underline{V}_{\mathcal{K}_w};\underline{U}_{d_n}|\underline{Y}_{\mathcal{L}}) \notag \\
     & \quad\quad -I(\underline{V}_{\mathcal{K}_w};\underline{Z}_{\mathcal{W}}|\underline{V}_{\mathcal{K}_s}) + \epsilon_e \notag \\
     &\quad = I(\underline{V}_{\mathcal{K}_s}\underline{V}_{\mathcal{K}_w},\underline{U}_{d_n};\underline{Y}_{\mathcal{L}}) - I(\underline{U}_{d_n};\underline{Y}_{\mathcal{L}}|\underline{V}_{\mathcal{K}_s}\underline{V}_{\mathcal{K}_w}) \notag \\
     & \quad\quad + I(\underline{V}_{\mathcal{K}_s}\underline{V}_{\mathcal{K}_w};\underline{U}_{d_n}|\underline{Y}_{\mathcal{L}}) - I(\underline{V}_{\mathcal{K}_w};\underline{Z}_{\mathcal{W}}|\underline{V}_{\mathcal{K}_s}) + \epsilon_e \notag \\
     & \quad \overset{(d)}{=} I(\underline{X}_{\mathcal{L}};\underline{Y}_{\mathcal{L}}) -  I(\underline{U}_{d_n};\underline{Y}_{\mathcal{L}}|\underline{V}_{\mathcal{K}_s}\underline{V}_{\mathcal{K}_w}) \notag \\
     & \quad\quad + I(\underline{V}_{\mathcal{K}_s}\underline{V}_{\mathcal{K}_w};\underline{U}_{d_n}|\underline{Y}_{\mathcal{L}}) - I(\underline{V}_{\mathcal{K}_w};\underline{Z}_{\mathcal{W}}|\underline{V}_{\mathcal{K}_s}) + \epsilon_e \notag \\
     &\quad\overset{(e)}{\leq} I(\underline{X}_{\mathcal{L}};\underline{Y}_{\mathcal{L}}) -  I(\underline{U}_{d_n};\underline{Y}_{\mathcal{L}}|\underline{V}_{\mathcal{K}_s}\underline{V}_{\mathcal{K}_w}) \notag \\
     &\quad\quad + I(\underline{V}_{\mathcal{K}_s}\underline{V}_{\mathcal{K}_w};\underline{U}_{d_n}|\underline{Y}_{\mathcal{L}}) - I(\underline{V}_{\mathcal{K}_s}\underline{V}_{\mathcal{K}_w};\underline{Z}_{\mathcal{W}}) \notag \\
     & \quad\quad + \epsilon_e + \epsilon_s \notag \\
     & \quad\overset{(f)}{=} I(\underline{X}_{\mathcal{L}};\underline{Y}_{\mathcal{L}}) - I(\underline{X}_{\mathcal{L}};\underline{Z}_{\mathcal{W}}) + \epsilon_e + \epsilon_s \notag \\
     & \quad\quad + I(\underline{U}_{d_n};\underline{Z}_{\mathcal{W}}|\underline{V}_{\mathcal{K}_s}\underline{V}_{\mathcal{K}_w}) - I(\underline{U}_{d_n};\underline{Y}_{\mathcal{L}}|\underline{V}_{\mathcal{K}_s}\underline{V}_{\mathcal{K}_w}) \notag \\
     & \quad\quad + I(\underline{V}_{\mathcal{K}_s}\underline{V}_{\mathcal{K}_w};\underline{U}_{d_n}|\underline{Y}_{\mathcal{L}}) \notag \\
     & \quad = I(\underline{X}_{\mathcal{L}};\underline{Y}_{\mathcal{L}}) - I(\underline{X}_{\mathcal{L}};\underline{Z}_{\mathcal{W}}) + \epsilon_e + \epsilon_s \notag \\
     & \quad\quad + I(\underline{U}_{d_n};\underline{Z}_{\mathcal{W}}|\underline{V}_{\mathcal{K}_s}\underline{V}_{\mathcal{K}_w}) - I(\underline{U}_{d_n};\underline{Y}_{\mathcal{L}}|\underline{V}_{\mathcal{K}_s}\underline{V}_{\mathcal{K}_w}) \notag \\
     & \quad\quad + H(\underline{U}_{d_n}|\underline{Y}_{\mathcal{L}}) - H(\underline{U}_{d_n}|\underline{V}_{\mathcal{K}_s}\underline{V}_{\mathcal{K}_w},\underline{Y}_{\mathcal{L}}) \notag \\
     & \quad \leq I(\underline{X}_{\mathcal{L}};\underline{Y}_{\mathcal{L}}) - I(\underline{X}_{\mathcal{L}};\underline{Z}_{\mathcal{W}}) + \epsilon_e + \epsilon_s \notag \\
     & \quad\quad + H(\underline{U}_{d_n}|\underline{Y}_{\mathcal{L}},\underline{V}_{\mathcal{K}_s}\underline{V}_{\mathcal{K}_w}) - H(\underline{U}_{d_n}|\underline{Z}_{\mathcal{W}},\underline{V}_{\mathcal{K}_s}\underline{V}_{\mathcal{K}_w}) \notag \\
     & \quad\quad + H(\underline{U}_{d_n}) \notag \\
     & \quad \overset{(g)}{\leq} I(\underline{X}_{\mathcal{L}};\underline{Y}_{\mathcal{L}}) - I(\underline{X}_{\mathcal{L}};\underline{Z}_{\mathcal{W}}) \notag \\
     & \quad\quad + \epsilon_e + \epsilon_s + \ell \epsilon_d,
\end{align}
where (a) and (c) are from the independence of the messages. (b) holds by Fano's inequality and \eqref{eq:converse-reliability}. (d) and (f) holds since the channels are noiseless. (e) holds from \eqref{eq:converse-secrecy} and (g) holds from \eqref{eq:converse-seed}. 

\off{The quantities $\ell n \epsilon_e$, $\ell n \epsilon_s$ and $\ell \epsilon_d$ are negligible compared to $I(\underline{X}_{\mathcal{L}};\underline{Y}_{\mathcal{L}})$ and $I(\underline{X}_{\mathcal{L}};\underline{Z}_{\mathcal{W}})$.}

\section{Partial Encryption against Crypto-Eve\\ with Linear Code \\ (Security Proof of Theorem~\ref{IndividualLinear-SS-CCA1})} \label{sec:linear-partial-enc}

In this section, we provide the security proof of Theorem~\ref{IndividualLinear-SS-CCA1}. We aim to show that each column $j \in \{1,...,\lceil \tilde{n} / \mu \rceil\}$ of the message matrix $\underline{M}_{\mathcal{L}}$ is ISS-CCA1. The proof is based on the equivalence between semantic security and indistinguishability \cite{goldwasser2019probabilistic}. We consider the maximal advantage for Crypto-Eve (maximal divination from uniform distribution probability), given the almost uniform messages after the source encoding stage (see Sec.~\ref{sec:source_code}) and prove that each column of the message matrix $\underline{M}_{\mathcal{L}}$ is IIND-CCA1 as given in \cite[Definition 4]{cohen2022partial}, and thus it is ISS-CCA1.

\off{As described in Sec.~\ref{sec:NU-HUNCC}, Alice encodes and encrypts the messages as given in lines $2 - 20$ of Algorithm~\ref{algo:NU-HUNCC}.}\off{ We assume the message matrix is the output of the source encoder from Sec.~\ref{sec:source_code}.} Alice and Crypto-Eve start playing the game as defined in \cite[definition 4]{cohen2022partial}. First, using the security parameter $\mu c$, the public and secret keys $(p_{\mu c},s_{\mu c})$ are created.\off{We note that the security parameter $\mu c$ is a function of the number of encrypted bits $1 \leq \mu c < \ell$.} Crypto-Eve sends a polynomial amount of ciphertexts to Alice and receives back their decryption. At this stage, Crypto-Eve chooses $i^{*} \in \{1,...,k_s\}$ (the case for $i^{*} \in \{k_s+1,...,l\}$ follows analogously), and two possible messages $M_{i^{*},1}$ and $M_{i^{*},2}$, both over the extension field $\mathbb{F}_{2^\mu}$.

We add a step to the game that gives Crypto-Eve an additional advantage over Alice to show a stronger statement than in Definition~\ref{def:individuall-SS-CCA1}. Crypto-Eve is given the symbols in positions $\{1,...,k_s\} \setminus i^{*}$. We show that still, Crypto-Eve is not able to distinguish between $M_{i^{*},1}$ and $M_{i^{*},2}$. Alice draws the symbols in positions $\{k_s+1,...,\ell\}$ from the distribution induced by the source encoder, and chooses $h \in \{1,2\}$ uniformly at random s.t. the symbol in position $i^{*}$ is $M_{i^{*},h}$. The message received is denoted by $M^\ell_h(j) \in \mathbb{F}_{2^\mu}^{\ell}$.

Alice encrypts $M^\ell_h(j)$ using $Crypt^{\dag}_2$~\eqref{eq:linear_encryption_crypt2}. First, Alice employs the linear IS channel code Sec.~\ref{sec:linear-IS-code} to receive the encoded codeword denoted by $X^\ell_h(j) \in \mathbb{F}_{2^\mu}^{\ell}$ which is injectively mapped into a codeword over $\mathbb{F}_{2}^{\ell \times \mu}$. Second, Alice encrypts the first $c$ lines ($\mu c$ bits) from $X^\ell_h(j)$ using $Crypt^{\dag}_1$. For the purpose of this proof, we denote the encrypted ciphertext by $\kappa = Crypt^{\dag}_2(M^\ell_h(j))$.

Upon receiving the ciphertext, $\kappa$, Crypto-Eve tries to guess $h$. $\mu w = \mu (\ell - c)$ of the symbols from $\kappa$ are seen by Crypto-Eve as plaintext\off{ and are injectively mapped back into symbols over $\mathbb{F}_{2^\mu}^{w}$}. The channel code is a linear coset code that maps each message of $\ell$ symbols to a different codeword of size $\ell$. Crypto-Eve can potentially reduce the number of possible codewords from each coset by observing $w$ of the codeword symbols. We denote the set of the remaining possible codewords from coset 1 and coset 2 by $\mathcal{B}_1$ and $\mathcal{B}_2$, respectively. 
\begin{corollary}\label{coro:B1equalsB2}
Consider the setting of Theorem~\ref{IndividualLinear-SS-CCA1} and the sets $\mathcal{B}_1$ and $\mathcal{B}_2$ as defined above. The number of codewords in the sets is equal. i.e.: $|\mathcal{B}_1| = |\mathcal{B}_2|$.
\end{corollary}

{\em Proof:} The proof for Corollary~\ref{coro:B1equalsB2} is provided in Appendix.~\ref{coro:proofB1equalsB2}.

We assume the best possible scenario for Crypto-Eve s.t.: 1) The induced probability of the codewords from coset 1 is as high as possible while the induced probability of the codewords from coset 2 is as low as possible\footnote{In \cite{cohen2022partial} the induced probabilities of each of the codewords were equal since the probability of the messages was uniform.}, and 2) the advantage Crypto-Eve has from $Crypt^{\dag}_1$ is as high as possible.

We start by bounding the maximum and minimum possible probabilities of the codewords in $\mathcal{B}_1$ and $\mathcal{B}_2$. We denote those probabilities by $p_{max}$ and $p_{min}$. First, we consider $j$ s.t. $[j \mu, j (\mu + 1)] \in \mathcal{H}_V$, i.e. the bits that are mapped to the symbol in column $j$ were not padded by the uniform seed, and their entropy is lower bounded by $1 - \delta_n$. The case for $[j \mu, j (\mu + 1)] \in \mathcal{J}_V$ will be addressed later on.

For each column $j$, we recall the set of possible messages from symbols $\{k_s+1,...,\ell\}$, $M_{k_w} \in \mathbb{F}_{2^\mu}^{k_w}$. In each column $j$ of the message matrix, the symbols are independent since they are obtained from independent source messages. Thus, we have
\begin{align*}
    p(M_{k_w} =m_{k_w}) &= \prod_{i=1}^{k_w}p(M_{k_w,i} = m_{k_w,i}) \\
    &= \prod_{i=1}^{k_w} \prod_{j'=1}^{\mu} p(M_{k_w,i,j'} = m_{k_w,i,j'}).
\end{align*}
From \eqref{eq:unreliable-group}, we conclude that
\begin{equation*} \label{eq:bitEntropy}
    \begin{aligned}
        H(m_{k_w,i,j'}) \geq H\left(m_{k_w,i,j'}|m_{k_w,i,j'}^{j\mu + j'-1}\right) \geq 1 - \delta_n,
    \end{aligned}
\end{equation*}
where $m_{k_w,i,j'}^{j \mu +j'-1} = \left(m_{k_w,i,j'}(1),...,m_{k_w,i,j'}(j \mu +j'- 1)\right)$ are the bits in columns $1$ to $j-1$ of the $i$-th row in the message matrix. We denote $\zeta \in (0,\frac{1}{2})$ s.t. $H(\frac{1}{2} - \zeta) = H(\frac{1}{2} + \zeta) = 1 - \delta_n$, and use $\zeta$ to bound $p_{min}$ and $p_{max}$
\begin{equation*}\label{eq:p-bounds}
    \begin{aligned}
        p_{min} \triangleq \left(\frac{1}{2} - \zeta\right)^{\mu k_w} \leq p(\underline{m}_{k_w}) \leq \left(\frac{1}{2} + \zeta\right)^{\mu k_w} \triangleq p_{max}.
    \end{aligned}
\end{equation*}
We bound $\zeta$ using the binary entropy bounds given in \cite{topsoe2001bounds}, $4p(1-p) \leq H_b(p) \leq \left(4p(1-p)\right)^{\ln{4}}$ for $0 < p < 1$. That is, by replacing $p = \frac{1}{2} + \zeta$ we obtain the upper bound by
\begin{equation*} \label{eq:right-side}
    \begin{aligned}
        H_b\left(\frac{1}{2} + \zeta\right) \leq \left(4\left(\frac{1}{2} + \zeta\right)\cdot \left(\frac{1}{2} - \zeta\right)\right)^{\frac{1}{\ln 4}},
    \end{aligned}
\end{equation*}
and the lower bound by
\begin{equation*} \label{eq:linear-left-side}
    \begin{aligned}
        H_b\left(\frac{1}{2} + \zeta\right) \geq 4\left(\frac{1}{2} + \zeta\right)\left(\frac{1}{2} - \zeta\right).
    \end{aligned}
\end{equation*}
From which, we conclude
\begin{align*}
    \frac{1}{2} + \zeta \leq \frac{1}{2} + \frac{1}{2}\cdot \sqrt{1 - (1 - \delta_n)^{\ln 4}},
\end{align*}
and therefore, the upper bound for $\zeta$ is given by
\begin{equation}\label{eq:linear-zeta-bound}
    \begin{aligned}
        \zeta \leq \frac{\sqrt{1 - (1 - \delta_n)^{\ln 4}}}{2}.
    \end{aligned}
\end{equation}

Now, we give the probability for some codeword from the set $\mathcal{B}_1 \cup \mathcal{B}_2$. We denote the codeword by $\alpha \in \mathcal{B}_1 \cup \mathcal{B}_2$ and we denote $|\mathcal{B}_1| = |\mathcal{B}_2| = |\mathcal{B}|$ (see Corollary~\ref{coro:B1equalsB2}). The probability for $\alpha$ is
\begin{equation} \label{eq:linear-one-set}
    \begin{aligned}
        \mathbb{P}[\alpha | \alpha \in \mathcal{B}_1 \cup \mathcal{B}_2] = 
        \begin{cases}
            \frac{p_{max}}{|\mathcal{B}| \cdot (p_{max} + p_{min})} & \text{if } \alpha \in \mathcal{B}_1 \\
            \frac{p_{min}}{|\mathcal{B}| \cdot (p_{max} + p_{min})} & \text{if } \alpha \in \mathcal{B}_2
        \end{cases}.
    \end{aligned}
\end{equation}
From \eqref{eq:linear-one-set}, we conclude that the probability for a codeword $\alpha$ to be from the set $\mathcal{B}_1$ is
\begin{equation} \label{eq:linear-bin1}
    \begin{aligned}
        \mathbb{P}[\alpha \in \mathcal{B}_1 | \alpha \in \mathcal{B}_1 \cup \mathcal{B}_2] = 
            \frac{p_{max}}{p_{max} + p_{min}}.
    \end{aligned}
\end{equation}

Since $Crypt_1^{\dag}$ is SS-CCA1, Crypto-Eve can obtain some negligible information about the original codeword or some function of the codeword. We consider the function $f: \{\mathcal{B}_1 \cup \mathcal{B}_2\} \rightarrow \{1,2\}$, i.e. the function that takes a codeword from the set $\{\mathcal{B}_1 \cup \mathcal{B}_2\}$ and outputs whether it belongs to $\mathcal{B}_1$ or $\mathcal{B}_2$
\begin{equation} \label{eq:linear-f(c*)}
    \begin{aligned}
        f(\alpha \in \{\mathcal{B}_1 \cup \mathcal{B}_2\}) = 
        \begin{cases}
            1 & \text{if }  \alpha \in \mathcal{B}_1\\
            2 & \text{if }  \alpha \in \mathcal{B}_2
        \end{cases}.
    \end{aligned}
\end{equation}
From the definition of SS-CCA1, we conclude
\begin{equation} \label{eq:linear-P(c*)}
    \begin{aligned}
        \mathbb{P}[M_{i^{*},1}] =  \mathbb{P}[\alpha \in \mathcal{B}_1] + \epsilon_{ss-cca1},
    \end{aligned}
\end{equation}
where $\epsilon_{ss-cca1}$ is a negligible function in $c$ s.t. $\epsilon_{ss-cca1} \leq \frac{1}{(\mu c)^{d}}$. Thus, by substituting \eqref{eq:linear-bin1} into \eqref{eq:linear-P(c*)}, we have
\begin{align}
    \mathbb{P}[M_{i^{*},1}] - \frac{1}{2} &=  \mathbb{P}[\alpha \in \mathcal{B}_1] + \epsilon_{ss-cca1} - \frac{1}{2} \nonumber \\
        &= \frac{p_{max} - p_{min}}{p_{max} + p_{min}} + \frac{1}{(\mu c)^{d}} .\label{eq:linear-adv-M}
\end{align}

We bound the term $\frac{p_{max}-p_{min}}{p_{max}+p_{min}}$ from \eqref{eq:linear-adv-M} by
\begin{align}
    \frac{p_{max}-p_{min}}{p_{max}+p_{min}} &= \frac{(\frac{1}{2}+\zeta)^{\mu k_w}-(\frac{1}{2}-\zeta)^{\mu k_w}}{(\frac{1}{2}+\zeta)^{\mu k_w}+(\frac{1}{2}-\zeta)^{\mu k_w}} \nonumber \\
    & = \frac{\sum_{i=0}^{\mu k_w}\binom{\mu k_w}{i}(2\zeta)^i - \sum_{i=0}^{\mu k_w}\binom{\mu k_w}{i}(-2\zeta)^i}{\sum_{i=0}^{\mu k_w}\binom{\mu k_w}{i}(2\zeta)^i + \sum_{i=0}^{\mu k_w}\binom{\mu k_w}{i}(-2\zeta)^i} \nonumber \\
    & = \frac{2\sum_{i=0}^{\lfloor \frac{\mu k_w-1}{2} \rfloor}\binom{\mu k_w}{2i+1}(2\zeta)^{2i+1}}{2\sum_{i=0}^{\lfloor \frac{\mu k_w}{2} \rfloor}\binom{\mu k_w}{2i}(2\zeta)^{2i}} \nonumber \\
    & \overset{(a)}{<} 2\zeta \mu k_w \overset{(b)}{\leq} 2^{\frac{3}{2}}2^{\frac{-n^{\beta}}{2}}\mu k_w, \label{eq:linear-pmax-pmin-bound}
\end{align}
where (a) follows from the upper bound  $\frac{\sum_{i=0}^{\lfloor (\mu k_w-1) / 2 \rfloor}\binom{\mu k_w}{2i+1}(2\zeta)^{2i}}{\sum_{i=0}^{\lfloor \mu k_w / 2 \rfloor}\binom{\mu k_w}{2i}(2\zeta)^{2i}} < \mu k_w$, which can be easily proved, and (b) is from \eqref{eq:linear-zeta-bound}.

By substituting \eqref{eq:linear-pmax-pmin-bound} into \eqref{eq:linear-adv-M}, it can be shown that for every $d'$ Crypto-Eve's advantage can be made smaller than $\frac{1}{(\mu c)^{d'}}$ by choosing an appropriate $d$. Thus, we showed that column $j \in \mathcal{H}_V$ is ISS-CCA1.

For the case of a column $j$ s.t. $j \in \mathcal{J}_V$, ISS-CCA1 follows directly from the SS-CCA1 of $Crypt_1$, considering Crypto-Eve has the maximal advantage in guessing the seed from its ciphertext. Thus, we can conclude the cryptosystem is IIND-CCA1 and thus ISS-CCA1 as well \cite{goldwasser2019probabilistic}.

\subsection{Row-based encryption}
So far, we assumed that $Crypt^{\dag}_1$ was employed on $c$ symbols of each column of the message matrix $\underline{M}_{\mathcal{L}}$. Nevertheless, $Crypt^{\dag}_1$ can be applied on entire rows instead. In this subsection, we discuss the effects of the encryption order (column encryption/row encryption) on the security level of NU-HUNCC against Crypto-Eve. We demonstrate that the scheme maintains its ISS-CCA1 status as long as there are $c$ encrypted links, irrespective of the order in which the encryption is carried out. The proofs provided in this section rely on the results from Sec.~\ref{sec:linear-partial-enc}. First, we show that the encryption of each individual symbol is ISS-CCA1, and we follow by showing that encrypting several symbols of each row at a time is ISS-CCA1 secured as well.

\underline{\textit{Symbol encryption:}} In this encryption scheme, each symbol is encrypted separately as opposed to encrypting $c$ symbols of the same column altogether (Sec~\ref{sec:linear-partial-enc}). Each symbol is composed of $\mu$ bits, thus the encryption is performed on $\mu$ bits instead of $\mu c$ bits. Alice and Crypto-Eve play the game as defined in \cite[Defenition 4]{cohen2022partial}. Crypto-Eve, upon receiving the ciphertext $\kappa$, can reduce the number of possible codewords using the same elimination process as shown in Sec.~\ref{sec:linear-partial-enc} remaining with the sets $\mathcal{B}_1$ and $\mathcal{B}_2$.\off{ Crypto-Eve remains with the possible codewords sets , $\mathcal{B}_1$ and $\mathcal{B}_2$, respectively.} The induced probability of each of the possible codewords from $\mathcal{B}_1$ and $\mathcal{B}_2$ remains the same as in \eqref{eq:linear-one-set} and \eqref{eq:linear-bin1}. However, since each of the $c$ symbols was encrypted separately, the analysis for Crypto-Eve's advantage from $Crypt^{\dag}_1$ is now employed on each encrypted symbol separately as well. For $i \in \{1,...,c\}$ let $\epsilon_{ss-cca1,i}$ be Crypt-Eve's advantage in guessing whether the codeword is from $\mathcal{B}_1$ or $\mathcal{B}_2$. We denote by $\epsilon_{ss-cca1} = \max_{\substack{1 \leq i \leq c}} 
 \epsilon_{ss-cca1,i} \leq \frac{1}{\mu^d}$. By substituting $\epsilon_{ss-cca1}$ into \eqref{eq:linear-P(c*)} we obtain Crypto-Eve's overall advantage against symbol encryption. Crypto-Eve's advantage from the induced probability of the codewords is given by $2^{\frac{3}{2}}2^{\frac{-n^{\beta}}{2}}k_w$, as presented in \eqref{eq:linear-pmax-pmin-bound}. This can be shown to be smaller than $\frac{1}{\mu^d}$ by choosing an appropriate $\tilde{n}$. Thus we conclude that symbol encryption scheme is ISS-CCA1.

\underline{\textit{Row encryption:}} We now show that encrypting several symbols from the same row of the message matrix, $\underline{M}_{\mathcal{L}} \in \mathbb{F}_{2^\mu}^{\lceil \tilde{n} / \mu \rceil}$, at a time is ISS-CCA1. In this encryption scheme, Crypto-Eve observes a ciphertext with information about more than one column of the message matrix. This affects only the information Crypto-Eve can obtain from the encrypted symbols since the analysis for the number of possible codewords per coset remains the same as in Sec.~\ref{sec:linear-partial-enc}. Per column, the advantage Crypto-Eve has in this scenario is $\epsilon_{ss-cca1} = \max_{\substack{1 \leq i \leq c}} 
 \epsilon_{ss-cca1,i} \leq \frac{1}{\mu_e^d}$ where we denoted the number of encrypted bits by $\mu_e = \pi \mu$ for $\pi \in \mathbb{N}$ and $\pi > 1$.\off{ This is true since $Crypt_1$ is SS-CCA1, thus Crypto-Eve's advantage in extracting any information from the ciphertext is negligible. By substituting $\epsilon_{ss-cca1}$ into \eqref{eq:linear-P(c*)} we can obtain Crypto-Eve's overall advantage. Since Crypto-Eve's advantage from the induced probability of the codewords is given by $2^{\frac{3}{2}}2^{\frac{-n^{\beta}}{2}}\mu k_w$ \eqref{eq:linear-pmax-pmin-bound} which can be shown to be smaller than $\frac{1}{\mu_e^d}$ by choosing an appropriate $\tilde{n}$ we conclude that this encryption method is ISS-CCA1 secured.} Thus, by the same arguments as for symbol encryption, we conclude that row encryption is ISS-CCA1 as well.

\section{Conclusions And Future Work} \label{sec:discussion}
In this paper, we introduce a novel PQ individual SS-CCA1 (ISS-CCA1) cryptosystem for non-uniform messages at high data rates. This result is given by using a polar source coding scheme with a sub-linear shared seed, a secure channel coding scheme that mixes the almost uniform information, and encrypting a small amount of the mixed data and the seed by any SS-CCA1 PQ cryptosystem. We have shown the achievability of two proposed codes, linear and non-linear, and provided a tight converse result against IT-Eve. We have introduced a new notion of cryptographic security, Individual SS-CCA1 (ISS-CCA1), and have shown that NU-HUNCC is ISS-CCA1 against Crypto-Eve for both the linear and non-linear codes. Additionally, we provided a numerical analysis of the data rate and complexity of the suggested communication scheme, showing it outperforms previously known PQ-secured cryptosystems in the finite length regime.

This work suggests an efficient and universal scheme for transmitting and storing non-uniform finite-length data in a PQ-secure manner, making it a solution that can be adapted into practical communication and information systems.

Future works include the study of secure linear channel coding schemes for the proposed PQ cryptosystem to guarantee reliable communications over noisy channels \cite{jaggi2007resilient,lun2008coding,koetter2008coding,silva2008rank,silva2011universal}. 
\off{ 
Future works include mixing the data by secure linear coding schemes (e.g., as given in \cite{bhattad2005weakly,silva2009universal,SMSM}) for computationally efficient decoding \cite{silva2009fast}, and a converse analysis for the individual secure source and channel scheme considered. Extensions also include the study of secure linear channel coding schemes for the proposed PQ cryptosystem to guarantee reliable communications over noisy channels \cite{koetter2008coding,silva2008rank,silva2011universal}.}

\bibliographystyle{IEEEtran}
\bibliography{refs}

\begin{thebibliography}{10}
\providecommand{\url}[1]{#1}
\csname url@samestyle\endcsname
\providecommand{\newblock}{\relax}
\providecommand{\bibinfo}[2]{#2}
\providecommand{\BIBentrySTDinterwordspacing}{\spaceskip=0pt\relax}
\providecommand{\BIBentryALTinterwordstretchfactor}{4}
\providecommand{\BIBentryALTinterwordspacing}{\spaceskip=\fontdimen2\font plus
\BIBentryALTinterwordstretchfactor\fontdimen3\font minus
  \fontdimen4\font\relax}
\providecommand{\BIBforeignlanguage}[2]{{%
\expandafter\ifx\csname l@#1\endcsname\relax
\typeout{** WARNING: IEEEtran.bst: No hyphenation pattern has been}%
\typeout{** loaded for the language `#1'. Using the pattern for}%
\typeout{** the default language instead.}%
\else
\language=\csname l@#1\endcsname
\fi
#2}}
\providecommand{\BIBdecl}{\relax}
\BIBdecl

\bibitem{goldwasser2019probabilistic}
S.~Goldwasser and S.~Micali, ``Probabilistic encryption \& how to play mental
  poker keeping secret all partial information,'' in \emph{Providing sound
  foundations for cryptography: on the work of Shafi Goldwasser and Silvio
  Micali}, 2019, pp. 173--201.

\bibitem{forouzan2015cryptography}
B.~A. Forouzan and D.~Mukhopadhyay, \emph{Cryptography and network
  security}.\hskip 1em plus 0.5em minus 0.4em\relax Mc Graw Hill Education
  (India) Private Limited New York, NY, USA:, 2015, vol.~12.

\bibitem{bloch2011physical}
M.~Bloch and J.~Barros, \emph{Physical-layer security: from information theory
  to security engineering}.\hskip 1em plus 0.5em minus 0.4em\relax Cambridge
  University Press, 2011.

\bibitem{bernstein2017post}
D.~J. Bernstein and T.~Lange, ``Post-quantum cryptography,'' \emph{Nature},
  vol. 549, no. 7671, pp. 188--194, 2017.

\bibitem{Shannon1949}
C.~E. Shannon, ``Communication theory of secrecy systems,'' \emph{The Bell
  System Technical Journal}, vol.~28, no.~4, pp. 656--715, 1949.

\bibitem{WiretapWyner}
A.~D. Wyner, ``The wire-tap channel,'' \emph{The Bell System Technical
  Journal}, vol.~54, no.~8, pp. 1355--1387, 1975.

\bibitem{liang2009physical}
Y.~Liang, H.~V. Poor, and S.~Shamai, ``Physical layer security in broadcast
  networks,'' \emph{Sec. and Comm. Net.}, vol.~2, no.~3, pp. 227--238, 2009.

\bibitem{liang2009information}
------, ``Information theoretic security,'' \emph{Foundations and
  Trends{\textregistered} in Comm and Inf. Theory}, vol.~5, no. 4--5, pp.
  355--580, 2009.

\bibitem{zhou2013physical}
X.~Zhou, L.~Song, and Y.~Zhang, \emph{Physical layer security in wireless
  communications}.\hskip 1em plus 0.5em minus 0.4em\relax Crc Press, 2013.

\bibitem{cohen2016wiretap}
A.~Cohen and A.~Cohen, ``Wiretap channel with causal state information and
  secure rate-limited feedback,'' \emph{IEEE Transactions on Communications},
  vol.~64, no.~3, pp. 1192--1203, 2016.

\bibitem{carleial1977note}
A.~Carleial and M.~Hellman, ``A note on wyner's wiretap channel (corresp.),''
  \emph{IEEE Transactions on Information Theory}, vol.~23, no.~3, pp. 387--390,
  1977.

\bibitem{SMSM}
A.~Cohen, A.~Cohen, M.~Médard, and O.~Gurewitz, ``Secure multi-source
  multicast,'' \emph{IEEE Transactions on Communications}, vol.~67, no.~1, pp.
  708--723, 2019.

\bibitem{JointIndividual2014}
A.~S. Mansour, R.~F. Schaefer, and H.~Boche, ``Joint and individual secrecy in
  broadcast channels with receiver side information,'' in \emph{2014 IEEE 15th
  International Workshop on Signal Processing Advances in Wireless
  Communications (SPAWC)}, 2014, pp. 369--373.

\bibitem{IndividualDegradedMU2015}
------, ``The individual secrecy capacity of degraded multi-receiver wiretap
  broadcast channels,'' in \emph{2015 IEEE International Conference on
  Communications (ICC)}, 2015, pp. 4181--4186.

\bibitem{SecrecyBroadcast}
Y.~Chen, O.~O. Koyluoglu, and A.~Sezgin, ``Individual secrecy for the broadcast
  channel,'' \emph{IEEE Transactions on Information Theory}, vol.~63, no.~9,
  pp. 5981--5999, 2017.

\bibitem{tan2019can}
J.~Y. Tan, L.~Ong, and B.~Asadi, ``Can marton coding alone ensure individual
  secrecy?'' in \emph{2019 IEEE Information Theory Workshop (ITW)}.\hskip 1em
  plus 0.5em minus 0.4em\relax IEEE, 2019, pp. 1--5.

\bibitem{cohen2023absolute}
A.~Cohen, R.~G. D'Oliveira, C.-Y. Yeh, H.~Guerboukha, R.~Shrestha, Z.~Fang,
  E.~Knightly, M.~M{\'e}dard, and D.~M. Mittleman, ``Absolute security in
  terahertz wireless links,'' \emph{IEEE Journal of Selected Topics in Signal
  Processing}, 2023.

\bibitem{yeh2023securing}
C.-Y. Yeh, A.~Cohen, R.~G. D’Oliveira, M.~M{\'e}dard, D.~M. Mittleman, and
  E.~W. Knightly, ``Securing angularly dispersive terahertz links with
  coding,'' \emph{IEEE Trans. on Inf. Forensics and Security}, 2023.

\bibitem{kaltz2008introduction}
J.~Kaltz and Y.~Lindell, ``Introduction to modern cryptography: principles and
  protocols,'' \emph{Chapman and Hall}, 2008.

\bibitem{RSA}
D.~Boneh \emph{et~al.}, ``Twenty years of attacks on the {RSA} cryptosystem,''
  \emph{Notices of the AMS}, vol.~46, no.~2, pp. 203--213, 1999.

\bibitem{QuantumRSA}
P.~Shor, ``Algorithms for quantum computation: discrete logarithms and
  factoring,'' in \emph{Proceedings 35th Annual Symposium on Foundations of
  Computer Science}, 1994, pp. 124--134.

\bibitem{sidelnikov1992insecurity}
V.~M. Sidelnikov and S.~O. Shestakov, ``On insecurity of cryptosystems based on
  generalized reed-solomon codes,'' 1992.

\bibitem{sendrier1998concatenated}
N.~Sendrier, ``On the concatenated structure of a linear code,''
  \emph{Applicable Algebra in Engineering, Communication and Computing},
  vol.~9, no.~3, pp. 221--242, 1998.

\bibitem{minder2007cryptanalysis}
L.~Minder and A.~Shokrollahi, ``Cryptanalysis of the sidelnikov cryptosystem,''
  in \emph{Advances in Cryptology-EUROCRYPT 2007: 26th Annual International
  Conference on the Theory and Applications of Cryptographic Techniques,
  Barcelona, Spain, May 20-24, 2007. Proceedings 26}.\hskip 1em plus 0.5em
  minus 0.4em\relax Springer, 2007, pp. 347--360.

\bibitem{monico2000using}
C.~Monico, J.~Rosenthal, and A.~Shokrollahi, ``Using low density parity check
  codes in the mceliece cryptosystem,'' in \emph{2000 IEEE International
  Symposium on Information Theory (Cat. No. 00CH37060)}.\hskip 1em plus 0.5em
  minus 0.4em\relax IEEE, 2000, p. 215.

\bibitem{landais2013efficient}
G.~Landais and J.-P. Tillich, ``An efficient attack of a mceliece cryptosystem
  variant based on convolutional codes,'' in \emph{Post-Quantum Cryptography:
  5th International Workshop, PQCrypto 2013, Limoges, France, June 4-7, 2013.
  Proceedings 5}.\hskip 1em plus 0.5em minus 0.4em\relax Springer, 2013, pp.
  102--117.

\bibitem{mceliece1978public}
R.~J. McEliece, ``A public-key cryptosystem based on algebraic,'' \emph{Coding
  Thv}, vol. 4244, pp. 114--116, 1978.

\bibitem{nojima2008semantic}
R.~Nojima, H.~Imai, K.~Kobara, and K.~Morozov, ``{Semantic security for the
  McEliece cryptosystem without random oracles},'' \emph{Designs, Codes and
  Cryptography}, vol.~49, pp. 289--305, 2008.

\bibitem{dottling2012cca2}
N.~Dottling, R.~Dowsley, J.~Muller-Quade, and A.~C. Nascimento, ``{A CCA2
  secure variant of the McEliece cryptosystem},'' \emph{IEEE Trans. on Inf.
  Theory}, vol.~58, no.~10, pp. 6672--6680, 2012.

\bibitem{aguirre2019ind}
F.~Aguirre~Farro and K.~Morozov, ``{On IND-CCA1 Security of Randomized McEliece
  Encryption in the Standard Model},'' in \emph{Code-Based Cryptography: 7th
  International Workshop, CBC 2019, Darmstadt, Germany, May 18--19, 2019,
  Revised Selected Papers 7}.\hskip 1em plus 0.5em minus 0.4em\relax Springer,
  2019, pp. 137--148.

\bibitem{berlekamp1973goppa}
E.~Berlekamp, ``Goppa codes,'' \emph{IEEE Transactions on Information Theory},
  vol.~19, no.~5, pp. 590--592, 1973.

\bibitem{patterson1975algebraic}
N.~Patterson, ``The algebraic decoding of goppa codes,'' \emph{IEEE
  Transactions on Information Theory}, vol.~21, no.~2, pp. 203--207, 1975.

\bibitem{faugere2013distinguisher}
J.-C. Faugere, V.~Gauthier-Umana, A.~Otmani, L.~Perret, and J.-P. Tillich, ``{A
  distinguisher for high-rate McEliece cryptosystems},'' \emph{IEEE Trans. on
  Inf. Theory}, vol.~59, no.~10, pp. 6830--6844, 2013.

\bibitem{el2007wiretap}
S.~Y. El~Rouayheb and E.~Soljanin, ``On wiretap networks ii,'' in \emph{2007
  IEEE International Symposium on Information Theory}.\hskip 1em plus 0.5em
  minus 0.4em\relax IEEE, 2007, pp. 551--555.

\bibitem{HUNCC}
A.~Cohen, R.~G.~L. D’Oliveira, S.~Salamatian, and M.~Médard, ``Network
  coding-based post-quantum cryptography,'' \emph{IEEE Journal on Selected
  Areas in Information Theory}, vol.~2, no.~1, pp. 49--64, 2021.

\bibitem{d2021post}
R.~G. D'Oliveira, A.~Cohen, J.~Robinson, T.~Stahlbuhk, and M.~M{\'e}dard,
  ``Post-quantum security for ultra-reliable low-latency heterogeneous
  networks,'' in \emph{MILCOM 2021-2021 IEEE Military Communications Conference
  (MILCOM)}.\hskip 1em plus 0.5em minus 0.4em\relax IEEE, 2021, pp. 933--938.

\bibitem{cohen2022partial}
A.~Cohen, R.~G. D’Oliveira, K.~R. Duffy, and M.~M{\'e}dard, ``Partial
  encryption after encoding for security and reliability in data systems,'' in
  \emph{2022 IEEE International Symposium on Information Theory (ISIT)}.\hskip
  1em plus 0.5em minus 0.4em\relax IEEE, 2022, pp. 1779--1784.

\bibitem{woo2023cermet}
J.~Woo, V.~A. Vasudevan, B.~Kim, A.~Cohen, R.~G. D'Oliveira, T.~Stahlbuhk, and
  M.~M{\'e}dard, ``{CERMET}: Coding for energy reduction with multiple
  encryption techniques -- it's easy being green,'' \emph{arXiv preprint
  arXiv:2308.05063}, 2023.

\bibitem{kim2024crypto}
B.~D. Kim, V.~A. Vasudevan, J.~Woo, A.~Cohen, R.~G. D’Oliveira, T.~Stahlbuhk,
  and M.~M{\'e}dard, ``Crypto-mine: Cryptanalysis via mutual information neural
  estimation,'' in \emph{ICASSP 2024-2024 IEEE International Conference on
  Acoustics, Speech and Signal Processing (ICASSP)}.\hskip 1em plus 0.5em minus
  0.4em\relax IEEE, 2024, pp. 4820--4824.

\bibitem{han2005folklore}
T.~S. Han, ``Folklore in source coding: Information-spectrum approach,''
  \emph{IEEE Transactions on Information Theory}, vol.~51, no.~2, pp. 747--753,
  2005.

\bibitem{chou2013data}
R.~A. Chou and M.~R. Bloch, ``Data compression with nearly uniform output,'' in
  \emph{2013 IEEE International Symposium on Information Theory}.\hskip 1em
  plus 0.5em minus 0.4em\relax IEEE, 2013, pp. 1979--1983.

\bibitem{chou2015polar}
R.~A. Chou, M.~R. Bloch, and E.~Abbe, ``Polar coding for secret-key
  generation,'' \emph{IEEE Transactions on Information Theory}, vol.~61,
  no.~11, pp. 6213--6237, 2015.

\bibitem{NegligbleCost}
R.~A. Chou, B.~N. Vellambi, M.~R. Bloch, and J.~Kliewer, ``Coding schemes for
  achieving strong secrecy at negligible cost,'' \emph{IEEE Transactions on
  Information Theory}, vol.~63, no.~3, pp. 1858--1873, 2017.

\bibitem{silva2009universal}
D.~Silva and F.~R. Kschischang, ``Universal weakly secure network coding,'' in
  \emph{2009 IEEE Information Theory Workshop on Networking and Information
  Theory}.\hskip 1em plus 0.5em minus 0.4em\relax IEEE, 2009, pp. 281--285.

\bibitem{silva2011universal}
------, ``Universal secure network coding via rank-metric codes,'' \emph{IEEE
  Transactions on Information Theory}, vol.~57, no.~2, pp. 1124--1135, 2011.

\bibitem{ghemawat2003google}
S.~Ghemawat, H.~Gobioff, and S.-T. Leung, ``The google file system,'' in
  \emph{Proceedings of the nineteenth ACM symposium on Operating systems
  principles}, 2003, pp. 29--43.

\bibitem{decandia2007dynamo}
G.~DeCandia, D.~Hastorun, M.~Jampani, G.~Kakulapati, A.~Lakshman, A.~Pilchin,
  S.~Sivasubramanian, P.~Vosshall, and W.~Vogels, ``Dynamo: Amazon's highly
  available key-value store,'' \emph{ACM SIGOPS operating systems review},
  vol.~41, no.~6, pp. 205--220, 2007.

\bibitem{chang2008bigtable}
F.~Chang, J.~Dean, S.~Ghemawat, W.~C. Hsieh, D.~A. Wallach, M.~Burrows,
  T.~Chandra, A.~Fikes, and R.~E. Gruber, ``Bigtable: A distributed storage
  system for structured data,'' \emph{ACM Transactions on Computer Systems
  (TOCS)}, vol.~26, no.~2, pp. 1--26, 2008.

\bibitem{borthakur2011apache}
D.~Borthakur, J.~Gray, J.~S. Sarma, K.~Muthukkaruppan, N.~Spiegelberg,
  H.~Kuang, K.~Ranganathan, D.~Molkov, A.~Menon, S.~Rash \emph{et~al.},
  ``Apache hadoop goes realtime at facebook,'' in \emph{Proceedings of the 2011
  ACM SIGMOD International Conference on Management of data}, 2011, pp.
  1071--1080.

\bibitem{calder2011windows}
B.~Calder, J.~Wang, A.~Ogus, N.~Nilakantan, A.~Skjolsvold, S.~McKelvie, Y.~Xu,
  S.~Srivastav, J.~Wu, H.~Simitci \emph{et~al.}, ``Windows azure storage: a
  highly available cloud storage service with strong consistency,'' in
  \emph{Proceedings of the Twenty-Third ACM Symposium on Operating Systems
  Principles}, 2011, pp. 143--157.

\bibitem{shannon1948mathematical}
C.~E. Shannon, ``A mathematical theory of communication,'' \emph{The Bell
  system technical journal}, vol.~27, no.~3, pp. 379--423, 1948.

\bibitem{ziv1978compression}
J.~Ziv and A.~Lempel, ``Compression of individual sequences via variable-rate
  coding,'' \emph{IEEE transactions on Information Theory}, vol.~24, no.~5, pp.
  530--536, 1978.

\bibitem{cronie2010lossless}
H.~S. Cronie and S.~B. Korada, ``Lossless source coding with polar codes,'' in
  \emph{2010 IEEE International Symposium on Information Theory}.\hskip 1em
  plus 0.5em minus 0.4em\relax IEEE, 2010, pp. 904--908.

\bibitem{slepian1973noiseless}
D.~Slepian and J.~Wolf, ``Noiseless coding of correlated information sources,''
  \emph{IEEE Trans. on Inf. Theory}, vol.~19, no.~4, pp. 471--480, 1973.

\bibitem{SCMUniform}
D.~Kobayashi, H.~Yamamoto, and T.~Ogawa, ``Secure multiplex coding attaining
  channel capacity in wiretap channels,'' \emph{IEEE Transactions on
  Information Theory}, vol.~59, no.~12, pp. 8131--8143, 2013.

\bibitem{bhattad2005weakly}
K.~Bhattad, K.~R. Narayanan \emph{et~al.}, ``Weakly secure network coding,''
  \emph{NetCod, Apr}, vol. 104, pp. 8--20, 2005.

\bibitem{bellare1998relations}
M.~Bellare, A.~Desai, D.~Pointcheval, and P.~Rogaway, ``Relations among notions
  of security for public-key encryption schemes,'' in \emph{Advances in
  Cryptology—CRYPTO'98: 18th Annual International Cryptology Conference Santa
  Barbara, California, USA August 23--27, 1998 Proceedings 18}.\hskip 1em plus
  0.5em minus 0.4em\relax Springer, 1998, pp. 26--45.

\bibitem{dowsley2009cca2}
R.~Dowsley, J.~M{\"u}ller-Quade, and A.~C. Nascimento, ``{A CCA2 secure public
  key encryption scheme based on the McEliece assumptions in the standard
  model},'' in \emph{{Cryptographers’ Track at the RSA Conference}}.\hskip
  1em plus 0.5em minus 0.4em\relax Springer, 2009, pp. 240--251.

\bibitem{csiszar1978broadcast}
I.~Csisz{\'a}r and J.~Korner, ``Broadcast channels with confidential
  messages,'' \emph{IEEE transactions on information theory}, vol.~24, no.~3,
  pp. 339--348, 1978.

\bibitem{ArikanBase2009}
E.~Arikan, ``Channel polarization: A method for constructing capacity-achieving
  codes for symmetric binary-input memoryless channels,'' \emph{IEEE
  Transactions on Information Theory}, vol.~55, no.~7, pp. 3051--3073, 2009.

\bibitem{matt2013one}
C.~Matt and U.~Maurer, ``The one-time pad revisited,'' in \emph{2013 IEEE
  International Symposium on Information Theory}.\hskip 1em plus 0.5em minus
  0.4em\relax IEEE, 2013, pp. 2706--2710.

\bibitem{gabidulin1985theory}
E.~M. Gabidulin, ``Theory of codes with maximum rank distance,'' \emph{Problemy
  peredachi informatsii}, vol.~21, no.~1, pp. 3--16, 1985.

\bibitem{roth1991maximum}
R.~M. Roth, ``Maximum-rank array codes and their application to crisscross
  error correction,'' \emph{IEEE transactions on Information Theory}, vol.~37,
  no.~2, pp. 328--336, 1991.

\bibitem{niederreiter1986knapsack}
H.~Niederreiter, ``Knapsack-type cryptosystems and algebraic coding theory,''
  \emph{Prob. Contr. Inform. Theory}, vol.~15, no.~2, pp. 157--166, 1986.

\bibitem{sidelnikov1994public}
V.~M. Sidelnikov, ``A public-key cryptosystem based on binary reed-muller
  codes,'' 1994.

\bibitem{londahl2012new}
C.~L{\"o}ndahl and T.~Johansson, ``A new version of mceliece pkc based on
  convolutional codes,'' in \emph{Information and Communications Security: 14th
  International Conference, ICICS 2012, Hong Kong, China, October 29-31, 2012.
  Proceedings 14}.\hskip 1em plus 0.5em minus 0.4em\relax Springer, 2012, pp.
  461--470.

\bibitem{hassani2010scaling}
S.~H. Hassani, K.~Alishahi, and R.~Urbanke, ``On the scaling of polar codes:
  Ii. the behavior of un-polarized channels,'' in \emph{2010 IEEE International
  Symposium on Information Theory}.\hskip 1em plus 0.5em minus 0.4em\relax
  IEEE, 2010, pp. 879--883.

\bibitem{wang2023sub}
H.-P. Wang, T.-C. Lin, A.~Vardy, and R.~Gabrys, ``Sub-4.7 scaling exponent of
  polar codes,'' \emph{IEEE Transactions on Information Theory}, 2023.

\bibitem{wang2021complexity}
H.-P. Wang, ``Complexity and second moment of the mathematical theory of
  communication,'' \emph{arXiv preprint arXiv:2107.06420}, 2021.

\bibitem{korada2010polar}
S.~B. Korada and R.~L. Urbanke, ``Polar codes are optimal for lossy source
  coding,'' \emph{IEEE Transactions on Information Theory}, vol.~56, no.~4, pp.
  1751--1768, 2010.

\bibitem{arikan2009rate}
E.~Arikan and E.~Telatar, ``On the rate of channel polarization,'' in
  \emph{2009 IEEE Int. Sym. on Inf. Theory}.\hskip 1em plus 0.5em minus
  0.4em\relax IEEE, 2009, pp. 1493--1495.

\bibitem{1053968}
S.~Kullback, ``A lower bound for discrimination information in terms of
  variation (corresp.),'' \emph{IEEE Trans. on Inf. Theory}, vol.~13, no.~1,
  pp. 126--127, 1967.

\bibitem{watanabe2014optimal}
S.~Watanabe and Y.~Oohama, ``The optimal use of rate-limited randomness in
  broadcast channels with confidential messages,'' \emph{IEEE Transactions on
  Information Theory}, vol.~61, no.~2, pp. 983--995, 2014.

\bibitem{topsoe2001bounds}
F.~Tops{\o}e, ``Bounds for entropy and divergence for distributions over a
  two-element set,'' \emph{J. Ineq. Pure Appl. Math}, vol.~2, no.~2, 2001.

\bibitem{jaggi2007resilient}
S.~Jaggi, M.~Langberg, S.~Katti, T.~Ho, D.~Katabi, and M.~M{\'e}dard,
  ``Resilient network coding in the presence of byzantine adversaries,'' in
  \emph{IEEE INFOCOM 2007-26th}.\hskip 1em plus 0.5em minus 0.4em\relax IEEE,
  2007, pp. 616--624.

\bibitem{lun2008coding}
D.~S. Lun, M.~M{\'e}dard, R.~Koetter, and M.~Effros, ``On coding for reliable
  communication over packet networks,'' \emph{Physical Communication}, vol.~1,
  no.~1, pp. 3--20, 2008.

\bibitem{koetter2008coding}
R.~Koetter and F.~R. Kschischang, ``Coding for errors and erasures in random
  network coding,'' \emph{IEEE Trans. on Inf. Theory}, vol.~54, no.~8, pp.
  3579--3591, 2008.

\bibitem{silva2008rank}
D.~Silva, F.~R. Kschischang, and R.~Koetter, ``A rank-metric approach to error
  control in random network coding,'' \emph{IEEE Trans. on Inf. Theory},
  vol.~54, no.~9, pp. 3951--3967, 2008.

\bibitem{aldous1983random}
D.~Aldous, ``Random walks on finite groups and rapidly mixing markov chains,''
  in \emph{S{\'e}minaire de Probabilit{\'e}s XVII 1981/82: Proceedings}.\hskip
  1em plus 0.5em minus 0.4em\relax Springer, 1983, pp. 243--297.

\end{thebibliography}

\appendices
\section{Secure Individual Non-Linear code\\ against IT-Eve \\ (Proof of Theorem~\ref{Direct})}\label{sec:IT-Eve} \label{sec:random-code-proof}
We give here the full secrecy and reliability analysis of the $k_s$-IS non-linear coding scheme (proof of Theorem~\ref{Direct}). We assume a noiseless multipath communication system with $\ell$ links and an eavesdropper, IT-Eve, with access to any subset $\mathcal{W} \subset \mathcal{L}$ of the links s.t. $|\mathcal{W}| \triangleq w < \ell$. Alice wants to send $\ell$ confidential messages to Bob while keeping IT-Eve ignorant about any set of $k_s \leq \ell - w - \ell\epsilon$ messages individually. We assume Alice uses the source coding scheme from Sec.~\ref{sec:source_code} and the random channel coding scheme from Sec.~\ref{sec:random-IS-code}. The proof provided in this section heavily relies on the proof given in Sec.~\ref{sec:linear-code-proof}. In this section, we provide the additional steps required for completing the proof of Theorem~\ref{Direct}.\off{we provide in this section the  We provide here only the missing parts that were not required to complete the proof from Sec.~\ref{sec:linear-code-proof} but are necessary for proving Theorem~\ref{Direct}.} We start by showing the code is reliable.

\subsection{Reliability - Non-Linear Code} \label{sec:random-code-reliability}
The reliability of the proposed non-linear end-to-end source-channel coding schemes is a consequence of the reliability of the source and channel decoders over the $\ell$ rows of the source matrix and $\tilde{n}$ columns of the message matrix.\off{ Hence, the proof is based on a similar technique given in \cite{NegligbleCost}, with the additional use of a union bound analyzed herein.}

First, in the scheme proposed, each column of the encoded message matrix is separately decoded by Bob. We bound the decoding error probability of the channel coding scheme. Recall the uniformly distributed message matrix, $\underline{\tilde{M}}_{\mathcal{L}}$, and its corresponding uniform distribution by $p_{U_{\underline{M}_{\mathcal{L}}}}$. We consider the optimal coupling as given in \cite[Lemma 3.6]{aldous1983random}, between $\underline{M}_{\mathcal{L}}$ and $\underline{\tilde{M}}_{\mathcal{L}}$. We denote $\varepsilon \triangleq \{\underline{M}_{\mathcal{L}} \neq \underline{\tilde{M}}_{\mathcal{L}}\}$. From the coupling lemma, we have that $\mathbb{P}[\varepsilon] = \mathbb{V}(p_{\underline{M}_{\mathcal{L}}},p_{U_{\underline{M}_{\mathcal{L}}}})$. The decoded message matrix is denoted by $\underline{\hat{M}}_{\mathcal{L}}$. Thus, the decoding error probability of the message matrix is given by
\begin{align}
    &\mathbb{P}\left[\underline{\hat{M}}_{\mathcal{L}} \neq \underline{M}_{\mathcal{L}} | \underline{M}_{\mathcal{L}} = \underline{m}_{\ell}\right] \nonumber \\
    & \quad= \mathbb{P}\left[\underline{\hat{M}}_{\mathcal{L}} \neq \underline{M}_{\mathcal{L}} | \underline{M}_{\mathcal{L}} = \underline{m}_{\ell},\varepsilon^C\right]\mathbb{P}\left[\varepsilon^C\right] \nonumber \\
    & \quad\quad + \mathbb{P}\left[\underline{\hat{M}}_{\mathcal{L}} \neq  \underline{M}_{\mathcal{L}} | \underline{M}_{\mathcal{L}} = \underline{m}_{\ell},\varepsilon\right]\mathbb{P}\left[\varepsilon\right] \nonumber \\
    & \quad \leq \mathbb{P}\left[\underline{\hat{M}}_{\mathcal{L}} \neq  \underline{M}_{\mathcal{L}} | \underline{M}_{\mathcal{L}} = \underline{m}_{\ell},\varepsilon^C\right] + \mathbb{P}\left[\varepsilon\right] \nonumber \\
    & \quad \overset{(a)}{=} \mathbb{P}\left[\underline{\hat{M}}_{\mathcal{L}} \neq  \underline{M}_{\mathcal{L}} | \underline{M}_{\mathcal{L}} = \underline{m}_{\ell},\varepsilon^C\right] + \mathbb{V}(p_{\underline{M}_{\mathcal{L}}},p_{U_{\underline{M}_{\mathcal{L}}}}) \nonumber \\
    & \quad \overset{(b)}{\leq} \mathbb{P}\left[\underline{\hat{\tilde{M}}}_{\mathcal{L}} \neq \underline{\tilde{M}}_{\mathcal{L}} | \underline{\tilde{M}}_{\mathcal{L}} = \underline{m}_{\ell}\right] + \sqrt{2 \tilde{n} \ell 2^{-n^\beta}}, \label{eq:column_Pe}
\end{align}
where (a) is from the coupling Lemma \cite[Lemma 3.6]{aldous1983random}, and (b) is from \eqref{eq:uniform-bound}. The left term in \eqref{eq:column_Pe} refers to the decoding error probability of the message matrix assuming the message distribution was entirely uniform. The decoding error probability of each column is upper bounded by $O\left(2^{-\ell}\right)$ \cite[Section \uppercase\expandafter{\romannumeral4}]{SMSM}. Thus, the error probability for the entire message matrix is given by the union bound on all the columns of the message matrix
\begin{equation}\label{eq:Column_Pe}
    \begin{aligned}
    \mathbb{P}\left[\underline{\hat{\tilde{M}}}_{\mathcal{L}} \neq \underline{\tilde{M}}_{\mathcal{L}} | \underline{\tilde{M}}_{\mathcal{L}} = \underline{m}_{\ell}\right]  \leq  \tilde{n} \cdot O\left(2^{-\ell}\right).
    \end{aligned}
\end{equation}

After obtaining the decoded message matrix, Bob can decode each row of the matrix separately to recover the source messages as shown in Sec.~\ref{sec:linear-code-reliability}. 
\off{\textcolor{blue}{The analysis for the error probability in the source decoding process remains as given Sec.~\ref{sec:linear-code-reliability}, equation \eqref{eq:UnionError} and thus we obtain}
\begin{equation*} \off{\label{eq:UnionError}}
    \mathbb{P}\left(\bigcup_{i=1}^{\ell} \left\{\underline{\hat{V}}_{\mathcal{L},i} \neq g_{s,n}\left(f_{s,n}(\underline{V}_{\mathcal{L},i},\underline{U}_{d_J,i},\underline{U}_{d_J,i})\right)\right\}\right) \leq \ell2^{-n^{\beta}}.
\end{equation*}}

Finally, we bound the total error probability for the source-channel coding scheme and obtain
\begin{align}
    &\mathbb{P}\left[\underline{V}_{\mathcal{L}} \neq \underline{\hat{V}}_{\mathcal{L}}\right] \nonumber \\
    & \quad \leq \mathbb{P}\left[\underline{V}_{\mathcal{L}} \neq \underline{\hat{V}}_{\mathcal{L}}|\underline{M}_{\mathcal{L}} \neq \underline{\hat{M}}_{\mathcal{L}}\right]\mathbb{P}\left[\underline{M}_{\mathcal{L}} \neq \underline{\hat{M}}_{\mathcal{L}}\right] \nonumber \\
    & \quad\quad\quad + \mathbb{P}\left[\underline
    {V}_{\mathcal{L}} \neq \underline{\hat{V}}_{\mathcal{L}}|\underline{M}_{\mathcal{L}} = \underline{\hat{M}}_{\mathcal{L}}\right]\mathbb{P}\left[\underline{M}_{\mathcal{L}} = \underline{\hat{M}}_{\mathcal{L}}\right] \nonumber \\
    & \quad \leq  \mathbb{P}\left[\underline{M}_{\mathcal{L}} \neq \underline{\hat{M}}_{\mathcal{L}}\right] \nonumber \\
    & \quad\quad\quad + \mathbb{P}\left[\underline{V}_{\mathcal{L}} \neq \underline{\hat{V}}_{\mathcal{L}}|\underline{M}_{\mathcal{L}} = \underline{\hat{M}}_{\mathcal{L}}\right]  \nonumber \\
    & \quad \overset{(a)}{\leq} \tilde{n} O\left(2^{-\ell}\right) +  \sqrt{2 \tilde{n}  \ell  2^{-n^\beta}} + \ell2^{-n^{\beta}}, \nonumber
\end{align}
where (a) is given by \eqref{eq:column_Pe}, \eqref{eq:Column_Pe}, and \eqref{eq:UnionError}. 

By requiring $\ell$ to be lower bounded by $\omega(\log{\tilde{n}})$, the expression $\tilde{n} O\left(2^{-\ell}\right)$ becomes negligible. In addition, requiring $\ell$ to be upper bounded by $ o\left(2^{n^\beta}/\tilde{n}\right)$, the expression $\sqrt{2 \tilde{n} \ell 2^{-n^\beta}} + \ell2^{-n^{\beta}}$ becomes negligible as well. By choosing $\ell$ that upholds both bounds the decoding error probability becomes negligible.

\off{In this section, we provide a security proof sketch for Theorem~\ref{Direct} against IT-Eve. Due to space limitations, we refer to Appendix~\ref{appendix:IT-secrecy} for the full security proof. 
We consider noiseless multipath communication as described in Section~\ref{sec:sys}. To transmit $\ell$ confidential messages, all from the source $(\mathcal{V},p_V)$, Alice employs the polar codes-based source encoder and channel encoder from Section~\ref{sec:main_results}. Here, we show that IT-Eve, observing any subset of $w$ links, $\underline{Z}_{\mathcal{W}}$, can't obtain any significant information about any set of $k_s$ messages individually. We denote a set of $k_s$ messages by $\mathcal{K}_s$, s.t. $|\mathcal{K}_s| \triangleq k_s \leq \ell - w - \ell\epsilon$. The uniformly distributed message matrix is denoted by $\underline{\tilde{M}}_{\mathcal{L}}$, and its distribution is denoted by $p_{U_{\underline{M}_{\mathcal{L}}}}$. The distribution of $\underline{Z}_{\mathcal{W}}$ induced from uniformly distributed messages is denoted by $\tilde{p}_{\underline{Z}_{\mathcal{W}}}$. Thus, using similar techniques given in \cite{NegligbleCost} for strong security in broadcast wiretap channels\footnote{In \cite{NegligbleCost}, it was shown that a polar codes-based source encoder could be used to achieve strong secrecy over a broadcast channel with non-uniform public and confidential messages. Our setting efficiently delivers $\ell$ confidential messages, keeping them $k_s$-IS by using a joint channel coding scheme.}, we have
\vspace{-0.15cm}
\begin{align}&\mathbb{V}\left(p_{\underline{Z}_{\mathcal{W}}|\underline{M}_{\mathcal{K}_s}=\underline{m}_{k_s}},p_{\underline{Z}_{\mathcal{W}}}\right) \leq  \mathbb{V}\left(\tilde{p}_{\underline{Z}_{\mathcal{W}}|\underline{M}_{\mathcal{K}_s}=\underline{m}_{k_s}},\tilde{p}_{\underline{Z}_{\mathcal{W}}}\right) \nonumber \\ 
    & + \mathbb{V}\left(p_{\underline{Z}_{\mathcal{W}}|\underline{M}_{\mathcal{K}_s}=\underline{m}_{k_s}},\tilde{p}_{\underline{Z}_{\mathcal{W}}|\underline{M}_{\mathcal{K}_s}=\underline{m}_{k_s}}\right) + \mathbb{V}\left(\tilde{p}_{\underline{Z}_{\mathcal{W}}},p_{\underline{Z}_{\mathcal{W}}}\right), \label{eq:k_s_ind}
\vspace{-0.1cm}    
\end{align}
by using the triangle inequality. Now, we consider each expression separately. The first expression in \eqref{eq:k_s_ind}\off{, $\mathbb{V}(\tilde{p}_{\underline{Z}_{\mathcal{L}}|\underline{M}_{\mathcal{K}_s}=\underline{m}_{k_s}},\tilde{p}_{\underline{Z}_{\mathcal{W}}})$,} is bounded using \cite[Telescoping Expansion]{korada2010polar}, thus we have 
\vspace{-0.15cm}
\begin{align*}
\mathbb{V}\left(\tilde{p}_{\underline{Z}_{\mathcal{W}}|\underline{M}_{\mathcal{K}_s}=\underline{m}_{k_s}},\tilde{p}_{\underline{Z}_{\mathcal{W}}}\right) \leq \sum_{j=1}^{\tilde{n}} \mathbb{V}\left(\tilde{p}_{\underline{Z}_{\mathcal{W}}^{(j)}|\underline{M}_{\mathcal{K}_s}^{(j)}=\underline{m}_{k_s}^{(j)}},\tilde{p}_{\underline{Z}_{\mathcal{W}}}^{(j)}\right).
\vspace{-0.1cm}
\end{align*}
By applying \cite[Theorem 1]{SMSM}, it can be shown that for every $1 \leq j \leq \tilde{n}$, $\mathbb{V}(\tilde{p}_{\underline{Z}_{\mathcal{W}}^{(j)}|\underline{M}_{\mathcal{K}_s}^{(j)}=\underline{m}_{k_s}^{(j)}},\tilde{p}_{\underline{Z}_{\mathcal{W}}}^{(j)}) \leq \ell^{-\frac{t}{2}}$ for some $t \geq 1$ s.t. $\ell\epsilon =  \lceil t\log{\ell}\rceil$. 
We note, that as $t$ grows, the number $k_s$ of IS messages decreases. However, the information leakage decreases significantly as well. Now, we bound the third expression in \eqref{eq:k_s_ind}. As demonstrated in Appendix~\ref{appendix:IT-secrecy}, the same bound applies to the second expression in \eqref{eq:k_s_ind}. 
By using the independence between the messages, and applying the law of total probability to the distribution of IT-Eve's observations it can be shown that 
\begin{equation}\label{eq:bound_z}
\begin{aligned}
   \mathbb{V}\left(\tilde{p}_{\underline{Z}_{\mathcal{W}}},p_{\underline{Z}_{\mathcal{W}}}\right) \leq \mathbb{V}\left(p_{\underline{M}_{\mathcal{L}}},p_{U_{\underline{M}_{\mathcal{L}}}}\right). 
\end{aligned}
\end{equation}
We denote the KL-divergence between two distributions by $\mathbb{D}(\cdot||\cdot)$. Since the entropy of each bit in $\underline{M}_{\mathcal{L}}$ is at least $1-\delta_n$, it can be shown that $\mathbb{D}(p_{\underline{M}_{\mathcal{L}}}||p_{U_{\underline{M}_{\mathcal{L}}}}) \leq \ell \tilde{n} 2^{-n^\beta}$. By invoking the Pinsker inequality \cite{1053968}, we have that \eqref{eq:bound_z} is bounded by $\sqrt{2 \ell \tilde{n} 2^{-n^\beta}}$. Now, by using the triangle inequality and the fact that $\underline{V}_{\mathcal{K}_s} \rightarrow \underline{M}_{\mathcal{K}_s} \rightarrow \underline{Z}_{\mathcal{W}}$ is a Markov chain, we finally have: $\mathbb{V}(p_{\underline{Z}_{\mathcal{W}}|{\underline{V}_{\mathcal{K}_s}=\underline{v}_{k_s}}},p_{\underline{Z}_{\mathcal{W}}}) \leq 2\sqrt{2\ell \tilde{n} 2^{-n^\beta}} + \tilde{n} \ell^{-\frac{t}{2}}$. \off{Thus, for the information leakage to be negligible, we give lower and upper bounds on the size of $\ell$ compared to $n$.}The lower bound on $\ell$ is given by $\omega(\tilde{n}^{\frac{2}{t}})$, s.t. $\tilde{n} \ell^{-\frac{t}{2}}$ is negligible. The upper bound on $\ell$ is given by $o\left(2^{n^\beta}/\tilde{n}\right)$, s.t. $2\sqrt{2\ell \tilde{n} 2^{-n^\beta}}$ is negligible.} 

\subsection{Information Leakage - Non-Linear Code}

We start by recalling equations~\eqref{FullSecrecy}-\eqref{eq:FullZ} which hold regardless of the underlined channel code. The upper bound for expressions~\eqref{eq:CondSec} and~\eqref{eq:FullZ} that was given in Sec.~\ref{sec:linear-leakage-proof} holds for the non-linear channel code as it does not depend on the structure of the code. However,~\eqref{eq:UnifSec}, represents exactly the information leakage of the non-linear code given uniformaly distributed input messages. Thus, the bound provided for this expression in Sec.~\ref{sec:linear-leakage-proof} does not hold for the non-linear code. We now bound~\eqref{eq:UnifSec}, thus we have

\off{Recall the set $\mathcal{K}_s \subset \mathcal{L}$ where. $|\mathcal{K}_s| \triangleq k_s$, and the set  $\mathcal{K}_w \triangleq \mathcal{L} \setminus \mathcal{K}_s$. We denote by $\underline{M}_{\mathcal{K}_s} \subset \underline{M}_{\mathcal{L}}$ the subset of the secured messages and by $\underline{M}_{\mathcal{K}_w} \subset \underline{M}_{\mathcal{L}} \setminus \underline{M}_{\mathcal{K}_s}$ the rest of the messages. The distribution of $\underline{Z}_{\mathcal{W}}$ induced by the uniform message matrix is denoted by $\tilde{p}_{\underline{Z}_{\mathcal{W}}}$ or $\tilde{p}_{\underline{Z}_{\mathcal{W}}|\underline{M}_{\mathcal{K}_s}=\underline{m}_{k_s}}$.
\textcolor{blue}{We start the proof similarly as in Sec.~\ref{sec:linear-leakage-proof}.} For any $\underline{m}_{k_s} \in \underline{\mathcal{M}}_{k_s}$
\begin{alignat}{1}
    &\mathbb{V}\left(p_{\underline{Z}_{\mathcal{W}}|\underline{M}_{\mathcal{K}_s}=\underline{m}_{k_s}},p_{\underline{Z}_{\mathcal{W}}}\right) \label{eq:var_dis_random_IS}\\
    & \quad \overset{(a)}{\leq} \mathbb{V}\left(p_{\underline{Z}_{\mathcal{W}}|\underline{M}_{\mathcal{K}_s}=\underline{m}_{k_s}},\tilde{p}_{\underline{Z}_{\mathcal{W}}|\underline{M}_{\mathcal{K}_s}=\underline{m}_{k_s}}\right) \label{eq:var_dis_cond_random_IS} \\
    & \quad\quad + \mathbb{V}\left(\tilde{p}_{\underline{Z}_{\mathcal{W}}|\underline{M}_{\mathcal{K}_s}=\underline{m}_{k_s}},\tilde{p}_{\underline{Z}_{\mathcal{W}}}\right) \label{eq:var_dis_unif_random_IS} \\
    & \quad\quad + \mathbb{V}\left(\tilde{p}_{\underline{Z}_{\mathcal{W}}},p_{\underline{Z}_{\mathcal{W}}}\right), \label{eq:var_dis_obs_random_IS}
\end{alignat}
where (a) is from the triangle inequality. Since this is true for all  $\underline{m}_{k_s} \in \underline{\mathcal{M}}_{k_s}$, from now on we omit the equality $\underline{M}_{\mathcal{K}_s} = \underline{m}_{k_s}$ for ease of notation.

\textcolor{blue}{We provided a bound to expressions \eqref{eq:var_dis_cond_random_IS} and \eqref{eq:var_dis_obs_random_IS} in Sec.~\ref{sec:linear-leakage-proof} which also holds directly for the proposed random coding scheme. Thus, we focus on bounding \eqref{eq:var_dis_unif_random_IS}.}}
\begin{align} \label{eq:HelpLemmaBound}
    &\mathbb{V}\left(\tilde{p}_{\underline{Z}_{\mathcal{W}}|\underline{M}_{\mathcal{K}_s}},\tilde{p}_{\underline{Z}_{\mathcal{W}}}\right) \nonumber \\
    & \quad = \sum_{\underline{z}_w}\left|\tilde{p}_{\underline{Z}_{\mathcal{W}}|\underline{M}_{\mathcal{K}_s}}(\underline{z}_w|\underline{m}_{k_s})-\tilde{p}_{\underline{Z}_{\mathcal{W}}}(\underline{z}_w)\right| \nonumber \\
    & \quad \overset{(a)}{=} \sum_{\underline{z}_\mathcal{W}}\left|\prod_{j=1}^{\tilde{n}}\tilde{p}_{\underline{Z}_{\mathcal{W}}^{(j)}|\underline{M}_{\mathcal{K}_s}^{(j)}}(\underline{z}^{(j)}|\underline{m}_{k_s}^{(j)}) - \prod_{j=1}^{\tilde{n}}\tilde{p}_{\underline{Z}_{\mathcal{W}}^{(j)}}(\underline{z}^{(j)})\right| \nonumber \\
    & \quad \overset{(b)}{\leq} \sum_{\underline{z}_\mathcal{W}} \sum_{j=1}^{\tilde{n}} \left|\tilde{p}_{\underline{Z}_{\mathcal{W}}^{(j)}|\underline{M}_{\mathcal{K}_s}^{(j)}}(\underline{z}^{(j)}|\underline{m}_{k_s}^{(j)})-\tilde{p}_{\underline{Z}_{\mathcal{W}}^{(j)}}(\underline{z}^{(j)})\right| \cdot \nonumber \\
    & \quad\quad\quad\quad \prod_{q=1}^{j-1}\tilde{p}_{\underline{Z}_{\mathcal{W}}^{(q)}|\underline{M}_{\mathcal{K}_s}^{(q)}}(\underline{z}^{(q)}|\underline{m}_{k_s}^{(q)}) \cdot \prod_{k=j+1}^{\tilde{n}}\tilde{p}_{\underline{Z}_{\mathcal{W}}^{(k)}}(\underline{z}^{(k)}) \nonumber \\
    & \quad \overset{(c)}{=} \sum_{j=1}^{\tilde{n}} \sum_{\underline{z}^{(j)}}\left|\tilde{p}_{\underline{Z}_{\mathcal{W}}^{(j)}|\underline{M}_{\mathcal{K}_s}^{(j)}}(\underline{z}^{(j)}|\underline{m}_{k_s}^{(j)})-\tilde{p}_{\underline{Z}_{\mathcal{W}}^{(j)}}(\underline{z}^{(j)})\right|\cdot  \nonumber \\
    & \quad\quad\quad \sum_{\underline{\hat{z}}^{(j)}}\prod_{q=1}^{j-1}\tilde{p}_{\underline{Z}_{\mathcal{W}}^{(q)}|\underline{M}_{\mathcal{K}_s}^{(q)}}(\underline{z}^{(q)}|\underline{m}_{k_s}^{(q)}) \cdot \prod_{k=i+1}^{\tilde{n}}\tilde{p}_{\underline{Z}_{\mathcal{W}}^{(k)}}(\underline{z}^{(k)}) \nonumber \\
    & \quad = \sum_{j=1}^{\tilde{n}} \sum_{\underline{z}^{(j)}}\left|\tilde{p}_{\underline{Z}_{\mathcal{W}}^{(j)}|\underline{M}_{\mathcal{K}_s}^{(j)}}(\underline{z}^{(j)}|\underline{m}_{k_s}^{(j)})-\tilde{p}_{\underline{Z}_{\mathcal{W}}^{(j)}}(\underline{z}^{(j)})\right| \overset{(d)}{\leq} \tilde{n} \ell^{-\frac{t}{2}},
\end{align} 
where (a) holds since $\tilde{p}_{\underline{Z}_{\mathcal{W}}|\underline{M}_{\mathcal{K}_s}}$ and $\tilde{p}_{\underline{Z}_{\mathcal{W}}}$ are induced from a completely uniform distribution, and (b) holds from \cite[Telescoping Expansion]{korada2010polar}. (c) is from denoting $\underline{\hat{z}}^{(j)}=\left(\underline{z}^{(1)},...,\underline{z}^{(j-1)},\underline{z}^{(j+1)},...,\underline{z}^{(\tilde{n})}\right)$. (d) holds according to \cite[Theorem 1]{SMSM}, by choosing $k_s \leq \ell - w - \ell\epsilon$, and $\ell\epsilon = \lceil t \log{\ell}\rceil$ for $t \geq 1$.

Now, we return to \eqref{FullSecrecy}. By substituting \eqref{eq:CondSec}-\eqref{eq:FullZ} with \eqref{eq:uniform-bound} and \eqref{eq:HelpLemmaBound}, we show that
\begin{equation} \label{eq:random_msg_leakage}
    \mathbb{V}\left(p_{\underline{Z}_{\mathcal{W}} | \underline{M}_{\mathcal{K}_s}=\underline{m}_{k_s}},p_{\underline{Z}_{\mathcal{W}}}\right) \leq  \tilde{n} \ell^{-\frac{t}{2}} + 2 \sqrt{2 \ell \tilde{n} 2^{-n^\beta}}.
\end{equation}
\off{\[
\mathbb{V}\left(p_{\underline{Z}_{\mathcal{W}} | \underline{M}_{\mathcal{K}_s}=\underline{m}_{k_s}},p_{\underline{Z}_{\mathcal{W}}}\right) \leq  \tilde{n} \ell^{-\frac{t}{2}} + 2 \sqrt{2 \ell \tilde{n} 2^{-n^\beta}}.
\]
\begin{multline*}
\begin{aligned}
    &\mathbb{V}\left(p_{\underline{Z}_{\mathcal{W}} | \underline{M}_{K_s}=\underline{m}_{k_s}},p_{\underline{Z}_{\mathcal{W}}}\right) \\
    & \quad \leq \mathbb{V}\left(p_{\underline{Z}_{\mathcal{W}}|\underline{M}_{K_s}},\tilde{p}_{\underline{Z}_{\mathcal{W}}|\underline{M}_{K_s}}\right) +\\
    & \quad\quad\quad  \mathbb{V}\left(\tilde{p}_{\underline{Z}_{\mathcal{W}}|\underline{M}_{K_s}=\underline{m}_{k_s}},\tilde{p}_{\underline{Z}_{\mathcal{W}}}\right) + \mathbb{V}\left(\tilde{p}_{\underline{Z}_{\mathcal{W}}},p_{\underline{Z}_{\mathcal{W}}}\right) \\
    & \quad \leq \tilde{n} \cdot \ell^{-\frac{t}{2}} + 2\cdot \sqrt{2\cdot \ell \cdot \tilde{n} \cdot 2^{-n^\beta}}.
\end{aligned}
\end{multline*}}

We have shown that the non-linear channel coding scheme employed on $\underline{M}_{\mathcal{L}}$ is $k_s$-IS.\off{ To conclude the leakage proof against IT-Eve, we show that the scheme remains $k_s$-IS secured even when using the source coding scheme, by bounding $\mathbb{V}\left(p_{\underline{Z}_{\mathcal{W}} | \underline{V}_{\mathcal{K}_s}=\underline{v}_{k_s}},p_{\underline{Z}_{\mathcal{W}}}\right)$, and showing that for any $\underline{v}_{k_s} \in \underline{\mathcal{V}}_{\mathcal{K}_s}$ the
information leakage becomes negligible.} To conclude the proof we bound $\mathbb{V}\left(p_{\underline{Z}_{\mathcal{W}} | \underline{V}_{\mathcal{K}_s}=\underline{v}_{k_s}},p_{\underline{Z}_{\mathcal{W}}}\right)$ similarly as in \eqref{eq:final_leakage} and obtain
\begin{align*}
    &\mathbb{V}\left(p_{\underline{Z}_{\mathcal{W}} | \underline{V}_{\mathcal{K}_s}=\underline{v}_{k_s}},p_{\underline{Z}_{\mathcal{W}}}\right) \nonumber \\
    & \quad = \sum_{\underline{m}_{\mathcal{K}_s}} p_{\underline{M}_{\mathcal{K}_s}|\underline{V}_{\mathcal{K}_s}}(\underline{m}_{k_s}|\underline{v}_{k_s}) \cdot \mathbb{V}\left(p_{\underline{Z}_{\mathcal{W}} | \underline{M}_{\mathcal{K}_s}=\underline{m}_{k_s}},p_{\underline{Z}_{\mathcal{W}}}\right) \\
    & \quad \overset{(a)}{\leq} \tilde{n} \ell^{-\frac{t}{2}} + 2 \sqrt{2 \ell \tilde{n}  2^{-n^\beta}},
\end{align*}
where (a) is from \eqref{eq:random_msg_leakage}. For the information leakage to be negligible, we give an upper and lower bound to $\ell$ compared to $n$. $\ell$ is lower bounded by $\omega(\tilde{n}^{\frac{2}{t}})$, s.t. the expression $\tilde{n} \ell^{-\frac{t}{2}}$ is negligible. In addition, $\ell$ is upper bounded by $o\left(2^{n^\beta}/\tilde{n}\right)$, s.t. the expression $2 \sqrt{2 \ell \tilde{n} 2^{-n^\beta}}$ is negligible. By choosing $\ell$ upholding both bounds, the information leakage to IT-Eve becomes negligible. This holds for any set of $\mathcal{K}_s \subset \mathcal{L}$  source messages, thus the non-linear code is $k_s$-IS.

\section{Partial Encryption against Crypto-Eve\\ with Non-Linear Code \\ (Proof of Theorem~\ref{Individual-SS-CCA1})}\label{sec:Crypto-Eve} \label{sec:random-partial-enc}
We give here the full security proof of Theorem~\ref{Individual-SS-CCA1}. We aim to show that each column $j \in \{1,...,\tilde{n}\}$ of the message matrix $\underline{M}_{\mathcal{L}}$ is ISS-CCA1. The proof given in this section follows similar techniques and arguments as the proof showed in Sec.~\ref{sec:linear-partial-enc}. However, since the IS channel code in this section is changed to a non-linear code, there are some differences in the proof that require revision. In this section, we will pinpoint those differences and update the proof according to the characteristics of the non-linear code.\off{ Thus, we will 
The proof is based on the equivalence between semantic security and indistinguishability \cite{goldwasser2019probabilistic}. We consider the maximal advantage for Crypto-Eve (maximal divination from uniform distribution probability), given the almost uniform messages after the source encoding stage (see Sec.~\ref{sec:source_code}) and prove that each column of the message matrix $\underline{M}_{\mathcal{L}}$ is IIND-CCA1 as given in \cite[Definition 4]{cohen2022partial}, and thus it is ISS-CCA1. The proof provided in this section follows the same steps as in Sec.~\ref{sec:linear-partial-enc}. The main differences between the two proofs are: 1) The order of the encryption and 2) The number of possible codewords remaining from each bin after Crypto-Eve being able to see part of the plaintext.} 

\off{We assume the message matrix, $\underline{M}_{\mathcal{L}}$, is the output of the source encoder from Sec.~\ref{sec:source_code} with almost uniform messages. }Alice and Crypto-Eve start playing the game as defined in \cite[definition 4]{cohen2022partial}. First, using the security parameter $c$, the public and secret keys are created $(p_c,s_c)$.\off{ We note that the security parameter $c$ is a function of the number of encrypted bits $1 \leq c < \ell$.} 
The game is played as described in Sec.~\ref{sec:linear-code-proof} expect: 1) The messages $M_{i^{*},1}$ and $M_{i^{*},2}$ chosen by Alice are from $\mathbb{F}_{2}$. 2) The encryption alghorithms used by Alice are $Crypt_1$ and $Crypt_2$ given in~\eqref{eq:encryption} and~\eqref{eq:encryption_crypt2} respectively.\off{ Crypto-Eve sends a polynomial amount of ciphertexts to Alice and receives back their decryption. At this stage, Crypto-Eve chooses $i^{*} \in \{1,...,k_s\}$ (the case for $i^{*} \in \{k_s+1,...,l\}$ follows analogously), and two possible messages $M_{i^{*},1}$ and $M_{i^{*},2}$.}

\off{We add a step to the game that gives Crypto-Eve an additional advantage over Alice to show a stronger statement than in Definition~\ref{def:individuall-SS-CCA1}. Crypto-Eve is given the bits in positions $\{1,...,k_s\} \setminus i^{*}$. We show that still, Crypto-Eve is not able to distinguish between $M_{i^{*},1}$ and $M_{i^{*},2}$. Alice draws the bits in positions $\{k_s+1,...,\ell\}$ from the distribution induced by the source encoder, and chooses $h \in \{1,2\}$ uniformly at random s.t. the bit in position $i^{*}$ is $M_{i^{*},h}$. The message received is denoted by $M^\ell_h(j) \in \mathbb{F}_{2}^{\ell}$.}

\off{Alice encrypts $M^\ell_h(j)$ using $Crypt_2$ from \eqref{eq:encryption_crypt2}. First, Alice employs the IS channel code from Sec.~\ref{sec:random-IS-code} to receive the encoded codeword denoted by $X^\ell_h(j) \in \mathbb{F}_{2}^{\ell}$. Second, Alice encrypts the first $c$ bits from $X^\ell_h(j)$ using $Crypt_1$. For the purpose of this proof, we denote the encrypted ciphertext by $\kappa = Crypt_2(M^\ell_h(j))$.}

Upon receiving the ciphertext, $\kappa = Crypt_2(M^\ell_h(j))$, Crypto-Eve tries to guess $h$, which represents the bin in the codebook where the codeword resides. Recall the sets $\mathcal{B}_1$ and $\mathcal{B}_2$ from Sec.~\ref{sec:linear-partial-enc}.\off{ First, $w = \ell - c$ of the bits from $\kappa$ are seen by Crypto-Eve as plaintext. Thus, Crypto-Eve can potentially reduce the number of possible codewords in each of the bins. \off{The average number of codewords that remain per bin after Crypto-Eve observes $w$ of the bits is $2^{\ell\epsilon}$ \cite{SMSM}, where $k_s = \ell - w - \ell\epsilon = c - \ell\epsilon$.} We denote the set of the remaining possible codewords from bin 1 and bin 2 by $\mathcal{B}_1$ and $\mathcal{B}_2$, respectively.} It was shown in \cite[Sec. \uppercase\expandafter{\romannumeral4}]{SMSM} that with high probability, the number of remaining possible codewords per bin deviates between $(1-\epsilon') 2^{\ell\epsilon} \leq |\mathcal{B}_1|,|\mathcal{B}_2| \leq (1+\epsilon') 2^{\ell\epsilon}$
for some $\epsilon' > 0$ and $\ell\epsilon$ s.t. $k_s = \ell - w -\ell\epsilon = c - \ell\epsilon$. We note that for the non-linear code, the sizes of the sets $\mathcal{B}_1$ and $\mathcal{B}_2$ are not necessarily equal as they are for the linear code.

We assume the best possible scenario for Crypt-Eve s.t.: 1) The number of possible codewords remaining in bin 1 is as high as possible where the number of possible codewords remaining in bin 2 is as low as possible, $|\mathcal{B}_1| \geq |\mathcal{B}_2|$, 2) the induced probability of the codewords from bin 1 is as high as possible while the induced probability of the codewords from bin 2 is as low as possible, and 3) the advantage Crypto-Eve has from $Crypt_1$ is the highest possible.

We start by bounding the maximum and minimum possible probabilities of the codewords in $\mathcal{B}_1$ and $\mathcal{B}_2$. We recall the probabilities $p_{max}$ and $p_{min}$ which were introduced in Sec.~\ref{sec:linear-partial-enc}. Since the non-linear code operates over $\mathbb{F}_2$, using the notations from Sec.~\ref{sec:linear-partial-enc}, we have \off{ First, we consider $j \in \mathcal{H}_V$, i.e. the bits in column $j$ were not padded by the uniform seed, and their entropy is lower bounded by $1 - \delta_n$. The case for $j \in \mathcal{J}_V$ will be addressed later on.

For each column $j$, we denote by $M_{k_w} \in \mathbb{F}_{2}^{k_w}$ the set of possible messages from bits $\{k_s+1,...,\ell\}$. In each column $j$ of the message matrix, the bits are independent since they are obtained from independent source messages. Thus, we have}
\begin{align*}
    p(M_{k_w} =m_{k_w}) = \prod_{i=1}^{k_w}p(M_{k_w,i} = m_{k_w,i}),
\end{align*}
From \eqref{eq:unreliable-group}, we conclude that
\begin{equation*} \off{\label{eq:bitEntropy}}
    \begin{aligned}
        H(m_{k_w,i}) \geq H\left(m_{k_w,i}|m_{k_w,i}^{j-1}\right) \geq 1 - \delta_n,
    \end{aligned}
\end{equation*}
where $m_{k_w,i}^{j-1} = \left(m_{k_w,i}(1),...,m_{k_w,i}(j-1)\right)$ are the bits in columns $1$ to $j-1$ of the $i$-th row in the message matrix. By taking $\zeta \in (0,\frac{1}{2})$ s.t. $H(\frac{1}{2} - \zeta) = H(\frac{1}{2} + \zeta) = 1 - \delta_n$, and using \eqref{eq:linear-zeta-bound} we bound $p_{min}$ and $p_{max}$
\begin{equation*}\off{\label{eq:p-bounds}}
    \begin{aligned}
        p_{min} \triangleq \left(\frac{1}{2} - \zeta\right)^{k_w} \leq p(\underline{m}_{k_w}) \leq \left(\frac{1}{2} + \zeta\right)^{k_w} \triangleq p_{max}.
    \end{aligned}
\end{equation*}
\off{$\zeta$ is the same as in Sec.~\ref{sec:linear-partial-enc}, thus we bound using \eqref{eq:linear-zeta-bound}. }\off{the binary entropy bounds given in \cite{topsoe2001bounds}, $4p(1-p) \leq H_b(p) \leq \left(4p(1-p)\right)^{\ln{4}}$ for $0 < p < 1$. That is, by replacing $p = \frac{1}{2} + \zeta$ we obtain the upper bound by
\begin{equation*} \label{eq:right-side}
    \begin{aligned}
        H_b\left(\frac{1}{2} + \zeta\right) \leq \left(4\left(\frac{1}{2} + \zeta\right)\cdot \left(\frac{1}{2} - \zeta\right)\right)^{\frac{1}{\ln 4}},
    \end{aligned}
\end{equation*}
and the lower bound by
\begin{equation*} \label{eq:left-side}
    \begin{aligned}
        H_b\left(\frac{1}{2} + \zeta\right) \geq 4\left(\frac{1}{2} + \zeta\right)\left(\frac{1}{2} - \zeta\right).
    \end{aligned}
\end{equation*}
From which, we conclude
\begin{align*}
    \frac{1}{2} + \zeta \leq \frac{1}{2} + \frac{1}{2}\cdot \sqrt{1 - (1 - \delta_n)^{\ln 4}},
\end{align*}
and therefore, the upper bound for $\zeta$ is given by
\begin{equation}\label{eq:zeta-bound}
    \begin{aligned}
        \zeta \leq \frac{\sqrt{1 - (1 - \delta_n)^{\ln 4}}}{2}.
    \end{aligned}
\end{equation}}

Now, we give the probability for some codeword from the set $\mathcal{B}_1 \cup \mathcal{B}_2$. We denote the codeword by $\alpha \in \mathcal{B}_1 \cup \mathcal{B}_2$. The probability for $\alpha$ is
\begin{equation} \label{eq:one-set}
    \begin{aligned}
        \mathbb{P}[\alpha | \alpha \in \mathcal{B}_1 \cup \mathcal{B}_2] = 
        \begin{cases}
            \frac{p_{max}}{|\mathcal{B}_1| \cdot p_{max} + |\mathcal{B}_2| \cdot p_{min}} & \text{if } \alpha \in \mathcal{B}_1 \\
            \frac{p_{min}}{|\mathcal{B}_1| \cdot p_{max} + |\mathcal{B}_2| \cdot p_{min}} & \text{if } \alpha \in \mathcal{B}_2
        \end{cases}.
    \end{aligned}
\end{equation}
From \eqref{eq:one-set}, we conclude that the probability for a codeword $\alpha$ to be from the set $\mathcal{B}_1$ is
\begin{equation} \label{eq:bin1}
    \begin{aligned}
        \mathbb{P}[\alpha \in \mathcal{B}_1 | \alpha \in \mathcal{B}_1 \cup \mathcal{B}_2] = 
            \frac{|\mathcal{B}_1| \cdot p_{max}}{|\mathcal{B}_1| \cdot p_{max} + |\mathcal{B}_2| \cdot p_{min}}.
    \end{aligned}
\end{equation}

We consider the function $f: \{\mathcal{B}_1 \cup \mathcal{B}_2\} \rightarrow \{1,2\}$, as given in \eqref{eq:linear-f(c*)}. Since $Crypt_1$ is SS-CCA1, we conclude\off{ Since $Crypt_1$ is SS-CCA1, Crypto-Eve can obtain some negligible information about the original codeword or some function of the codeword. }\off{i.e. the function that takes a codeword from the set $\{\mathcal{B}_1 \cup \mathcal{B}_2\}$ and outputs whether it belongs to $\mathcal{B}_1$ or $\mathcal{B}_2$
\begin{equation*} \label{eq:f(c*)}
    \begin{aligned}
        f(\alpha \in \{\mathcal{B}_1 \cup \mathcal{B}_2\}) = 
        \begin{cases}
            1 & \text{if }  \alpha \in \mathcal{B}_1\\
            2 & \text{if }  \alpha \in \mathcal{B}_2
        \end{cases}
    \end{aligned}
\end{equation*}}
\off{From the definition of SS-CCA1, we conclude}
\begin{equation} \label{eq:P(c*)}
    \begin{aligned}
        \mathbb{P}[M_{i^{*},1}] =  \mathbb{P}[\alpha \in \mathcal{B}_1] + \epsilon_{ss-cca1},
    \end{aligned}
\end{equation}
where $\epsilon_{ss-cca1}$ is a negligible function in $c$ s.t. $\epsilon_{ss-cca1} \leq \frac{1}{c^{d}}$. Thus, by substituting \eqref{eq:bin1} into \eqref{eq:P(c*)}, we have
\begin{align}
    \mathbb{P}[M_{i^{*},1}] - \frac{1}{2} &=  \mathbb{P}[\alpha \in \mathcal{B}_1] + \epsilon_{ss-cca1} - \frac{1}{2} \nonumber \\
        \off{&= \frac{|\mathcal{B}|_1 \cdot p_{max}}{|\mathcal{B}_1| \cdot p_{max} + |\mathcal{B}_2| \cdot p_{min}} - \frac{1}{2} + \epsilon_{ss-cca1} \nonumber \\}
        &= \frac{|\mathcal{B}_1| \cdot p_{max} - |\mathcal{B}_2| \cdot p_{min}}{|\mathcal{B}_1| \cdot p_{max} + |\mathcal{B}_2| \cdot p_{min}} + \frac{1}{c^{d}} .\label{eq:adv-M}
\end{align}
We focus on bounding the left term of \eqref{eq:adv-M}
\begin{align}
    &\frac{|\mathcal{B}_1| \cdot p_{max} - |\mathcal{B}_2| \cdot p_{min}}{|\mathcal{B}_1| \cdot p_{max} + |\mathcal{B}_2| \cdot p_{min}} \nonumber \\
        &\overset{(a)}{=} \frac{(1+\epsilon') 2^{\ell\epsilon} p_{max} - (1-\epsilon') 2^{\ell\epsilon} p_{min}}{(1+\epsilon') 2^{\ell\epsilon} p_{max} + (1-\epsilon') 2^{\ell\epsilon} p_{min}} \nonumber \\
        &= \frac{1}{\frac{p_{max}+p_{min}}{p_{max}-p_{min}} + \epsilon'} + \frac{1}{\frac{p_{max}-p_{min}}{p_{max}+p_{min}} + \frac{1}{\epsilon'}}, \label{eq:p_max-p_min}
\end{align}
where (a) is from the assumption that $|\mathcal{B}_1| = (1+\epsilon')\cdot 2^{\ell\epsilon}$ and $|\mathcal{B}_2| = (1-\epsilon')\cdot 2^{\ell\epsilon}$. \off{By choosing $\epsilon' = \ell^{-t}$ s.t. $\ell\epsilon = \lceil t \log{\ell} \rceil$, and $t \geq 1$ \cite{SMSM},} To bound the term $\frac{p_{max}-p_{min}}{p_{max}+p_{min}}$ from \eqref{eq:p_max-p_min} we use the result in \eqref{eq:linear-pmax-pmin-bound} and obtain
\begin{align}
     \frac{p_{max}-p_{min}}{p_{max}+p_{min}} < 2^{\frac{3}{2}}2^{\frac{-n^{\beta}}{2}}k_w. \label{eq:pmax-pmin-bound}
\end{align}
\off{
\begin{align}
    \frac{p_{max}-p_{min}}{p_{max}+p_{min}} &= \frac{(\frac{1}{2}+\zeta)^{k_w}-(\frac{1}{2}-\zeta)^{k_w}}{(\frac{1}{2}+\zeta)^{k_w}+(\frac{1}{2}-\zeta)^{k_w}} \nonumber \\
    & = \frac{\sum_{i=0}^{k_w}\binom{k_w}{i}(2\zeta)^i - \sum_{i=0}^{k_w}\binom{k_w}{i}(-2\zeta)^i}{\sum_{i=0}^{k_w}\binom{k_w}{i}(2\zeta)^i + \sum_{i=0}^{k_w}\binom{k_w}{i}(-2\zeta)^i} \nonumber \\
    & = \frac{2\sum_{i=0}^{\lfloor \frac{k_w-1}{2} \rfloor}\binom{k_w}{2i+1}(2\zeta)^{2i+1}}{2\sum_{i=0}^{\lfloor \frac{k_w}{2} \rfloor}\binom{k_w}{2i}(2\zeta)^{2i}} \nonumber \\
    & \overset{(a)}{<} 2\zeta k_w \overset{(b)}{\leq} 2^{\frac{3}{2}}2^{\frac{-n^{\beta}}{2}}k_w, \label{eq:pmax-pmin-bound}
\end{align}
where (a) follows from the upper bound  $\frac{\sum_{i=0}^{\lfloor (k_w-1) / 2 \rfloor}\binom{k_w}{2i+1}(2\zeta)^{2i}}{\sum_{i=0}^{\lfloor k_w / 2 \rfloor}\binom{k_w}{2i}(2\zeta)^{2i}} < k_w$, which can be easily proved, and (b) is from \eqref{eq:linear-zeta-bound}. }By choosing $l \leq 2^{\frac{n^\beta}{2(t+1)}}$ s.t. $\ell\epsilon = \lceil t \log{\ell} \rceil$, for any $t \geq 1$, we have that $2^{\frac{3}{2}}2^{\frac{-n^{\beta}}{2}}k_w < 2^{\frac{3}{2}} \ell^{-t}$.

By substituting \eqref{eq:pmax-pmin-bound} into \eqref{eq:p_max-p_min} and taking $\epsilon' = \ell^{-t}$ \cite{SMSM}, it can be shown that for every $d'$ Crypto-Eve's advantage can be made smaller than $\frac{1}{c^{d'}}$ by choosing an appropriate $d$, large enough $\ell$ s.t $\ell \leq 2^{\frac{n^\beta}{2(t+1)}}$ and 
taking $t=\log{\ell}$. Thus, we conclude the proof for Theorem~\ref{Individual-SS-CCA1}. \off{Thus, we showed that column $j \in \mathcal{H}_V$ is ISS-CCA1.

For the case of a column $j$ s.t. $j \in \mathcal{J}_V$, ISS-CCA1 follows directly from the SS-CCA1 of $Crypt_1$, considering Crypto-Eve has the maximal advantage in guessing the seed from its ciphertext. Thus, we can conclude the cryptosystem is IIND-CCA1 and thus ISS-CCA1 as well \cite{goldwasser2019probabilistic}.}

\section{Proof of Corollary~\ref{coro:B1equalsB2}}\label{coro:proofB1equalsB2}
In this section, we provide a proof for Corollary~\ref{coro:B1equalsB2}. From Crypto-Eve's observation of the ciphertext $\kappa$, she can reduce the number of possible codewords to $2^{\mu(\ell-w)}$ over all the cosets. The process of reducing the codewords depends only on Crypto-Eve's observations and is independent of the distribution of the messages. Thus, we consider a uniform distribution of the messages to count the number of possible remaining codewords per coset. Now, it remains to show the number of codewords per coset is equal for all cosets. In \cite[Sec. \uppercase\expandafter{\romannumeral6}]{SMSM} it was shown that the information leakage of the linear IS code is 0, i.e., $I(\underline{M}^{(j)}_{\mathcal{K}_s};\underline{Z}^{(j)}_{\mathcal{W}}) = 0$ for all $1 \leq j \leq \lceil \tilde{n} / \mu \rceil$. It can also be shown that
\begin{equation}\label{eq:coro1}
\begin{aligned}
    &I(\underline{M}^{(j)}_{\mathcal{K}_s};\underline{Z}^{(j)}_{\mathcal{W}}) = I(\underline{M}^{(j)}_{\mathcal{K}_s},\underline{X}^{(j)}_{\mathcal{L}};\underline{Z}^{(j)}_{\mathcal{W}}) -  I(\underline{X}^{(j)}_{\mathcal{L}};\underline{Z}^{(j)}_{\mathcal{W}}|\underline{M}^{(j)}_{\mathcal{K}_s}) \\
    & \quad\quad \overset{(a)}{=} H(\underline{Z}^{(j)}_{\mathcal{W}}) - H(\underline{X}^{(j)}_{\mathcal{L}}|\underline{M}^{(j)}_{\mathcal{K}_s}) + H(\underline{X}^{(j)}_{\mathcal{L}}|\underline{M}^{(j)}_{\mathcal{K}_s},\underline{Z}^{(j)}_{\mathcal{W}}),
\end{aligned}
\end{equation}
where (a) is since $H(\underline{Z}^{(j)}_{\mathcal{W}}|\underline{X}^{(j)}_{\mathcal{L}},\underline{M}^{(j)}_{\mathcal{K}_s}) = 0$, as $\underline{Z}^{(j)}_{\mathcal{W}}$ is a subset of $\underline{X}^{(j)}_{\mathcal{L}}$ and the channel is noiseless. From \eqref{eq:coro1} we conclude that
\begin{equation}\label{eq:coro2}
\begin{aligned}
    H(\underline{X}^{(j)}_{\mathcal{L}}|\underline{M}^{(j)}_{\mathcal{K}_s},\underline{Z}^{(j)}_{\mathcal{W}}) = H(\underline{X}^{(j)}_{\mathcal{L}}|\underline{M}^{(j)}_{\mathcal{K}_s}) - H(\underline{Z}^{(j)}_{\mathcal{W}}).
\end{aligned}
\end{equation}
The linear IS code is a one-to-one mapping between messages of size $\ell$ over the extension field $\mathbb{F}_{2^\mu}$, thus we have that $H(\underline{Z}^{(j)}) = \mu w$. We note that this result holds since the codebook construction is independent of the message distribution. Additionally,
\begin{equation}\label{eq:coro3}
\begin{aligned}
    &H(\underline{X}^{(j)}_{\mathcal{L}}|\underline{M}^{(j)}_{\mathcal{K}_s}) \\
    & \quad= H(\underline{M}^{(j)}_{\mathcal{W}}|\underline{M}^{(j)}_{\mathcal{K}_s}) + H(\underline{X}^{(j)}_{\mathcal{L}}|\underline{M}^{(j)}_{\mathcal{K}_s},\underline{M}^{(j)}_{\mathcal{W}}) \\
    & \quad\quad\quad\quad\quad\quad\quad\quad\quad\quad\quad- H(\underline{M}^{(j)}_{\mathcal{W}}|\underline{X}^{(j)}_{\mathcal{L}},\underline{M}^{(j)}_{\mathcal{K}_s}) \\
    & \quad \overset{(a)}{=} \mu w + H(\underline{X}^{(j)}_{\mathcal{L}}|\underline{M}^{(j)}_{\mathcal{K}_s},\underline{M}^{(j)}_{\mathcal{W}}) - H(\underline{M}^{(j)}_{\mathcal{W}}|\underline{X}^{(j)}_{\mathcal{L}},\underline{M}^{(j)}_{\mathcal{K}_s}) \\
    & \quad \overset{(b)}{=} \mu w + \mu (\ell - w - k_s) -  H(\underline{M}^{(j)}_{\mathcal{W}}|\underline{X}^{(j)}_{\mathcal{L}},\underline{M}^{(j)}_{\mathcal{K}_s}) \\
    & \quad \overset{(c)}{=} \mu (\ell -k_s),
\end{aligned}
\end{equation}
where (a) is since $\underline{M}^{(j)}_{\mathcal{W}}$ is independent of $\underline{M}^{(j)}_{\mathcal{K}_s}$. (b) and (c) are since there is a one-to-one mapping between the entire message, $\underline{M}^{(j)}_{\mathcal{L}}$, to the codeword $\underline{X}^{(j)}_{\mathcal{L}}$. Thus, by substituting \eqref{eq:coro3} into \eqref{eq:coro2} we conclude
\begin{equation*}\label{eq:coro4}
\begin{aligned}
H(\underline{X}^{(j)}_{\mathcal{L}}|\underline{M}^{(j)}_{\mathcal{K}_s},\underline{Z}^{(j)}_{\mathcal{W}}) = \mu (\ell - w - k_s).
\end{aligned}
\end{equation*}
{Namely, if Crypto-Eve knows the number of the coset where the codeword resides, $\underline{M}^{(j)}_{\mathcal{K}_s}$ then the number of possible codewords inside this coset is $2^{\mu(\ell -w - k_s)}$. This is true for all values of $\underline{M}^{(j)}_{\mathcal{K}_s}$, thus the number of possible codewords per coset is the same and it is $2^{\mu(\ell -w - k_s)}$. From this we conclude that $|\mathcal{B}_1| = |\mathcal{B}_2|$.}
We note that by taking $k_s = \ell - w$, we obtain $|\mathcal{B}_1| = |\mathcal{B}_2| = 1$.

\off{\section{Source and IS Channel Coding Schemes} 
For completeness, first, we provide here the source coding construction as presented in \cite[Proposition 4]{NegligbleCost} and the IS random channel coding construction as presented in \cite[Section \uppercase\expandafter{\romannumeral4}]{SMSM}, we consider for the proposed NU-HUNCC scheme. Then, we analyze the reliability of NU-HUNCC coding scheme suggested in Section~\ref{sec:main_results}, using the source and IS channel coding schemes on $\ell$ non-uniform messages of size $n$.

\subsection{Polar Source Code}\label{appendix:src_code}
Let $m \in \mathbb{N}$ and the blocklength $n = 2^m$. We denote by $\underline{G}_n$, the polarization transform as defined in \cite{ArikanBase2009} s.t. $\underline{G}_n = \underline{P}_n\begin{bmatrix} 1 & 0 \\ 1 & 1 \end{bmatrix}^{\otimes m}$, where $\otimes$ is the Kronecker product, and $\underline{P}_n$ is the bit reversal matrix. In our proposed scheme, each row of the source message matrix $\underline{V}_{\mathcal{L}}$ is separately encoded by the polarization transformation $\underline{G}_n$: $\underline{A}_n = 
 \underline{V}_\mathcal{L} \cdot \underline{G}_n$. Let $\delta_n \triangleq 2^{-n^\beta}$, for $\beta \in [0,\frac{1}{2})$. For each row $A \subset \underline{A}_n$, the bits are divided into three groups:
\begin{equation} \label{eq:unreliable-group}
    \begin{aligned}
        1) \quad \mathcal{H}_V \triangleq \left\{ j \in [1,n]: H\left(A^{(j)}|A^{j-1}\right) > 1 - \delta_n \right\},
    \end{aligned}
\end{equation}
\begin{equation*}
    \begin{aligned}
     \hspace{-1.05cm}   2) \quad \mathcal{U}_V \triangleq \left\{ j \in [1,n]: H\left(A^{(j)}|A^{j-1}\right) < \delta_n \right\},
    \end{aligned}
\end{equation*}
\begin{equation*} \label{eq:seed-group}
    \begin{aligned}
     \hspace{-4.00cm}  3) \quad  \mathcal{J}_V \triangleq (\mathcal{U}_V \cup \mathcal{H}_V)^{C}.
    \end{aligned}
\end{equation*}
Traditionally, the compressed message in polar codes-based source coding is obtained from the concatenation of groups, $\mathcal{H}_V$, and $\mathcal{J}_V$, the bits with non-negligible entropy. The entropy of the bits from group $\mathcal{J}_V$ is not necessarily high, and they are not uniform.
To ensure the uniformity of those bits, they undergo one-time padding with the uniform seed. We denote by $U_{d_{J}}$, the uniform seed of size $|\mathcal{J}_{V}| \triangleq d_{J}$. A different seed is used for each row, thus we denote the seed matrix by $\underline{U}_{d_{J}} \in \mathbb{F}_{2}^{\ell \times d_{J}}$. Finally, the $i$-th row of the message matrix is given by
\begin{equation} \label{eq:output_src}
    \begin{aligned}
        \underline{M}_{\mathcal{L},i} \triangleq \left[A_i[\mathcal{H}_V],A_i[\mathcal{J}_V] \oplus \underline{U}_{d_{J},i}\right].
    \end{aligned}
\end{equation}
Here $\oplus$ denotes the xor operation over $\mathbb{F}_2$.  
The size of each almost uniform message obtained at the outcome of the source coding scheme is $\tilde{n} = |\mathcal{H}_V| + d_{J}$.

The decoding process is divided into two parts. First, Bob one-time pads the bits from the group $\mathcal{J}_V$ using the seed shared with him by Alice. Second, Bob uses a successive cancellation decoder to reliably decode the original message \cite{ArikanBase2009,cronie2010lossless}. In the suggested scheme, this decoding process is performed separately on each row of the message matrix.

\subsection{IS Random Channel Coding}\label{appendix:msg_encoder}
The IS random channel coding scheme is applied separately on each column of the message matrix $\underline{M}_{\mathcal{L}}$. Let $w < \ell$ be the number of links observed by IT-Eve, and $k_s \leq \ell - w - \ell\epsilon$ be the number of messages kept secured from IT-Eve s.t. $\ell\epsilon$ is an integer.\off{At the end of the secrecy proof, we discuss the effect $\ell\epsilon$ has on the information leakage to IT-Eve.} We denote $k_w \triangleq w + \ell\epsilon$. Each column of the message matrix is divided into two parts of sizes $k_s$ and $k_w$. The first part is denoted by $M_{k_s} \in \mathbb{F}_2^{k_s}$, and the second is denoted by $M_{k_w} \in \mathbb{F}_2^{k_w}$. We now give the detailed code construction for the IS random channel code applied by Alice on each $j$-th column.

\textit{\underline{Codebook Generation}}: Let $P(x) \sim Bernouli(\frac{1}{2})$. There are $2^{k_s}$ possible messages for $M_{k_s}$ and $2^{k_w}$ possible messages for $M_{k_w}$. For each possible message $M_{k_s}$, generate $2^{k_w}$ independent codewords $x^{\ell}(e)$, $1 \leq e \leq 2^{k_w}$, using the distribution $P(X^{\ell}) = \prod_{j=1}^{\ell}P(X_j)$. Thus, we have $2^{k_s}$ bins, each having $2^{k_w}$ possible codewords. The length of the codeword remains the same as the length of the message since the links are noiseless \cite{SMSM} and error correction is not required.\off{, e.g., as given in \cite{cohen2022partial}.}

\textit{\underline{Encoding}}: For each column in the message matrix, $k_s$ bits are used to choose the bin, while the remaining $k_w$ bits are used to select the codeword inside the bin.

\textit{\underline{Decoding}}: Bob decodes each column separately. \off{by finding the received codeword of each column in the codebook.} If the codeword appears only once in the codebook, then $M_{k_s}$ is the index of the bin in which the codeword was found, and $M_{k_w}$ is the index of the codeword inside the bin. However, if the codeword appears more than once, a decoding error occurs. The probability of a codeword appearing more than once in a codebook exponentially decreases as a function of $\ell$. The reliability analysis of the code is given in \cite[Section \uppercase\expandafter{\romannumeral4}]{SMSM}.

\subsection{Reliability} \label{appendix:reliability}
The reliability of the proposed end-to-end source-channel coding schemes is a consequence of the reliability of the source and channel decoders over the $\ell$ rows of the source matrix and $\tilde{n}$ columns of the message matrix. Hence, the proof is based on a similar technique given in \cite{NegligbleCost}, with the additional use of a union bound analyzed herein. \off{We denote the source encoder and decoder pair for each row by $f_{s,n}$ and $g_{s,n}$, respectively, and the channel encoder and decoder for each column by $f_{c,\ell}$ and $g_{c,\ell}$, respectively.}

First, each column of the encoded message matrix is separately decoded. We bound the decoding error probability of the channel coding scheme. We denote the uniformly distributed message matrix by $\underline{\tilde{M}}_{\mathcal{L}}$ and its corresponding uniform distribution by $p_{U_{\underline{M}_{\mathcal{L}}}}$. We consider the optimal coupling as given in \cite[Lemma 3.6]{aldous1983random}, between $\underline{M}_{\mathcal{L}}$ and $\underline{\tilde{M}}_{\mathcal{L}}$. We denote $\varepsilon \triangleq \{\underline{M}_{\mathcal{L}} \neq \underline{\tilde{M}}_{\mathcal{L}}\}$. From the coupling lemma, we have that $\mathbb{P}[\varepsilon] = \mathbb{V}(p_{\underline{M}_{\mathcal{L}}},p_{U_{\underline{M}_{\mathcal{L}}}})$. The decoded message matrix is denoted by $\underline{\hat{M}}_{\mathcal{L}}$. Thus, the decoding error probability of the message matrix is given by
\begin{align}
    &\mathbb{P}\left[\underline{\hat{M}}_{\mathcal{L}} \neq \underline{M}_{\mathcal{L}} | \underline{M}_{\mathcal{L}} = \underline{m}_{\ell}\right] \nonumber \\
    & \quad= \mathbb{P}\left[\underline{\hat{M}}_{\mathcal{L}} \neq \underline{M}_{\mathcal{L}} | \underline{M}_{\mathcal{L}} = \underline{m}_{\ell},\varepsilon^C\right]\mathbb{P}\left[\varepsilon^C\right] \nonumber \\
    & \quad\quad + \mathbb{P}\left[\underline{\hat{M}}_{\mathcal{L}} \neq  \underline{M}_{\mathcal{L}} | \underline{M}_{\mathcal{L}} = \underline{m}_{\ell},\varepsilon\right]\mathbb{P}\left[\varepsilon\right] \nonumber \\
    & \quad \leq \mathbb{P}\left[\underline{\hat{M}}_{\mathcal{L}} \neq  \underline{M}_{\mathcal{L}} | \underline{M}_{\mathcal{L}} = \underline{m}_{\ell},\varepsilon^C\right] + \mathbb{P}\left[\varepsilon\right] \nonumber \\
    & \quad \overset{(a)}{=} \mathbb{P}\left[\underline{\hat{M}}_{\mathcal{L}} \neq  \underline{M}_{\mathcal{L}} | \underline{M}_{\mathcal{L}} = \underline{m}_{\ell},\varepsilon^C\right] + \mathbb{V}(p_{\underline{M}_{\mathcal{L}}},p_{U_{\underline{M}_{\mathcal{L}}}}) \nonumber \\
    & \quad \overset{(b)}{\leq} \mathbb{P}\left[\underline{\hat{\tilde{M}}}_{\mathcal{L}} \neq \underline{\tilde{M}}_{\mathcal{L}} | \underline{\tilde{M}}_{\mathcal{L}} = \underline{m}_{\ell}\right] + \sqrt{2 \tilde{n} \ell 2^{-n^\beta}}, \label{eq:column_Pe}
\end{align}
where (a) is from the coupling Lemma \cite[Lemma 3.6]{aldous1983random}, and (b) is from \eqref{eq:uniform-bound}. The left term in \eqref{eq:column_Pe} refers to the decoding error probability of the message matrix assuming the message distribution was entirely uniform. The decoding error probability of each column is upper bounded by $O\left(2^{-\ell}\right)$ \cite[Section \uppercase\expandafter{\romannumeral4}]{SMSM}. Thus, the error probability for the entire message matrix is given by the union bound on all the columns of the message matrix
\begin{equation}\label{eq:Column_Pe}
    \begin{aligned}
    \mathbb{P}\left[\underline{\hat{\tilde{M}}}_{\mathcal{L}} \neq \underline{\tilde{M}}_{\mathcal{L}} | \underline{\tilde{M}}_{\mathcal{L}} = \underline{m}_{\ell}\right]  \leq  \tilde{n} \cdot O\left(2^{-\ell}\right).
    \end{aligned}
\end{equation}
\off{
First, each column of the encoded message matrix is separately decoded. We start by showing each column is reliably decoded. For ease of notation, we denote the $j$-th column of size $\ell$ from the message matrix by $M(j)$. For each column, $j$, we consider the optimal coupling \cite[Lemma 3.6]{aldous1983random} between $M(j)$ and a uniformly distributed message $\tilde{M}(j)$. We denote $\varepsilon \triangleq \{M(j) \neq \tilde{M}(j)\}$. From the coupling lemma, we have that $\mathbb{P}[\varepsilon] = \mathbb{V}(p_{M(j)},p_{\tilde{M}(j)})$. The decoded message is denoted by $\hat{M}(j)$. Thus, the decoding error probability of the $j$-th column is given by
\begin{equation}\label{eq:column_Pe}
    \begin{split}
    &\mathbb{P}[\hat{M}(j) \neq M(j) | M(j) = m(j)] \\
    & \quad= \mathbb{P}[\hat{M}(j) \neq M(j) | M(j) = m(j),\varepsilon^C]\mathbb{P}[\varepsilon^C] \\
    & \quad\quad + \mathbb{P}[\hat{M}(j) \neq  M(j) | M(j) = m(j),\varepsilon]\mathbb{P}[\varepsilon] \\
    & \quad \leq \mathbb{P}[\hat{M}(j) \neq  M(j) | M(j) = m(j),\varepsilon^C] + \mathbb{P}[\varepsilon] \\
    & \quad = \mathbb{P}[\hat{M}(j) \neq  M(j) | M(j) = m(j),\varepsilon^C] + \mathbb{V}(p_{M(j)},p_{\tilde{M}(j)}) \\
    & \quad \overset{(a)}{\leq} \mathbb{P}[\hat{\tilde{M}}(j) \neq \tilde{M}(j) | \tilde{M}(j) = m(j)] + \sqrt{2\cdot \tilde{n} \ell2^{-n^\beta}} \\
    & \quad \overset{(b)}{\leq} O(2^{-\ell}) + \sqrt{2\cdot \ell2^{-n^\beta}}, 
    \end{split}
\end{equation}

where (a) is from \eqref{eq:uniform-bound} and (b) is from the reliability of the random coding scheme for uniform messages\cite[Section \uppercase\expandafter{\romannumeral4}]{SMSM}. \off{The reliability of the random coding scheme for uniform messages was discussed in \cite{SMSM} and is a direct consequence of traditional random coding reliability analysis \cite[section 3.4]{bloch2011physical}.} By using the union bound on all the columns and \eqref{eq:column_Pe}, we conclude the error probability for the entire message matrix is given by
\begin{equation}\label{eq:Column_Pe}
    \begin{aligned}
    \mathbb{P}[\hat{\underline{M}}_{\mathcal{L}} \neq \underline{M}_{\mathcal{L}} | \underline{M}_{\mathcal{L}} = \underline{m}_{\ell}] \leq  \tilde{n} \cdot O(2^{-\ell}) + \tilde{n} \cdot \sqrt{2\cdot \ell2^{-n^\beta}}
    \end{aligned}.
\end{equation}
}
After obtaining the decoded message matrix, Bob can decode each row of the matrix separately to recover the source messages. Since Bob has the shared seed, he first one-time pads the bits from group $\mathcal{J}_V$ as described in Appendix~\ref{appendix:src_code}. Bob continues by employing successive cancellation decoding \cite{cronie2010lossless}\cite{arikan2009rate}. Such that, the error probability for the $i$-th row is given by
\begin{equation} \label{eq:single_row_pe}
\mathbb{P}\left[\underline{\hat{V}}_{\mathcal{L},i} \neq g_{s,n}\left(f_{s,n}(\underline{V}_{\mathcal{L},i},\underline{U}_{d_J,i},\underline{U}_{d_J,i})\right)\right] \leq 2^{-n^{\beta}},
\end{equation}
for $\beta \in [0,\frac{1}{2})$ as have been chosen in Appendix~\ref{appendix:src_code}. 

Now, using the union bound on all the rows in the recovered message matrix in \eqref{eq:single_row_pe}, we have
\begin{equation} \label{eq:UnionError}
    \mathbb{P}\left(\bigcup_{i=1}^{\ell} \left\{\underline{\hat{V}}_{\mathcal{L},i} \neq g_{s,n}\left(f_{s,n}(\underline{V}_{\mathcal{L},i},\underline{U}_{d_J,i},\underline{U}_{d_J,i})\right)\right\}\right) \leq \ell2^{-n^{\beta}}.
\end{equation}

Finally, we bound the total error probability for the source and channel coding schemed together obtaining
\begin{equation*}\label{eq:tot-Pe}
\begin{aligned}
    &\mathbb{P}\left[V_{\mathcal{L}} \neq \underline{\hat{V}}_{\mathcal{L}}\right] \\
    & \quad \leq \mathbb{P}\left[V_{\mathcal{L}} \neq \underline{\hat{V}}_{\mathcal{L}}|\underline{M}_{\mathcal{L}} \neq \underline{\hat{M}}_{\mathcal{L}}\right]\mathbb{P}\left[\underline{M}_{\mathcal{L}} \neq \underline{\hat{M}}_{\mathcal{L}}\right] \\
    & \quad\quad\quad + \mathbb{P}\left[V_{\mathcal{L}} \neq \underline{\hat{V}}_{\mathcal{L}}|\underline{M}_{\mathcal{L}} = \underline{\hat{M}}_{\mathcal{L}}\right]\mathbb{P}\left[\underline{M}_{\mathcal{L}} = \underline{\hat{M}}_{\mathcal{L}}\right] \\
    & \quad \leq  \mathbb{P}\left[\underline{M}_{\mathcal{L}} \neq \underline{\hat{M}}_{\mathcal{L}}\right] \\
    & \quad\quad\quad + \mathbb{P}\left[V_{\mathcal{L}} \neq \underline{\hat{V}}_{\mathcal{L}}|\underline{M}_{\mathcal{L}} = \underline{\hat{M}}_{\mathcal{L}}\right] \\
    & \quad \overset{(a)}{\leq} \tilde{n} O\left(2^{-\ell}\right) +  \sqrt{2 \tilde{n}  \ell  2^{-n^\beta}} + \ell2^{-n^{\beta}},
\end{aligned}
\end{equation*}
where (a) is given by \eqref{eq:column_Pe}, \eqref{eq:Column_Pe}, and \eqref{eq:UnionError}. 

By requiring $\ell$ to be lower bounded by $\omega(\log{\tilde{n}})$, the expression $\tilde{n} O\left(2^{-\ell}\right)$ becomes negligible. In addition, requiring $\ell$ to be upper bounded by $ o\left(2^{n^\beta}/\tilde{n}\right)$, the expression $\sqrt{2 \tilde{n} \ell 2^{-n^\beta}} + \ell2^{-n^{\beta}}$ becomes negligible as well. By choosing $\ell$ that upholds both bounds the decoding error probability becomes negligible.
\off{For the error probability to be negligible we give $\ell$ upper and lower bounds. The lower bound of $\ell$ is required for $\tilde{n} O\left(2^{-\ell}\right)$ to be negligible. Thus, we require that $\ell = \omega(\log{\tilde{n}})$. The upper bound is required for the expression $\sqrt{2 \tilde{n} \ell 2^{-n^\beta}} + \ell2^{-n^{\beta}}$ to be negligible. This is obtained by choosing $\ell$ s.t. $\ell = o\left(2^{n^\beta}/\tilde{n}\right)$. By choosing $\ell$ that upholds both bounds the decoding error probability becomes negligible.}

\section{$k_s$-IS Information Leakage against IT-Eve}\label{appendix:IT-secrecy}
We give here the full secrecy analysis of the $k_s$-IS coding scheme (security proof of Theorem~\ref{Direct}). We assume a noiseless multipath communication system with $\ell$ links and an eavesdropper, IT-Eve, with access to any subset $\mathcal{W} \subset \mathcal{L}$ of the links s.t. $|\mathcal{W}| \triangleq w < \ell$. Alice wants to send $\ell$ confidential messages to Bob while keeping IT-Eve ignorant about any set of $k_s \leq \ell - w - \ell\epsilon$ messages individually. We assume Alice uses the source coding scheme from Appendix~\ref{appendix:src_code} and the channel coding scheme from Appendix~\ref{appendix:msg_encoder}.

Alice encodes the source message matrix $\underline{V}_{\mathcal{L}}$ using the polar codes-based source encoder (Appendix~\ref{appendix:src_code}). The almost uniform message at the output of the source encoder is of size $\tilde{n} = |\mathcal{H}_V| + d_{J}$, where $d_{J} = |\mathcal{J}_V|$. Thus, we denote the output of the source encoder as given in \eqref{eq:output_src} by $\underline{M}_\mathcal{L} \in \mathbb{F}_2^{\ell \times \tilde{n}}$.

We denote the set $\mathcal{K}_s \subset \mathcal{L}$ s.t. $|\mathcal{K}_s| \triangleq k_s$, and the set  $\mathcal{K}_w \triangleq \mathcal{L} \setminus \mathcal{K}_s$. We start by showing that the channel scheme is $k_s$-IS. We denote by $\underline{M}_{\mathcal{K}_s} \subset \underline{M}_{\mathcal{L}}$ the subset of the secured messages and by $\underline{M}_{\mathcal{K}_w} \subset \underline{M}_{\mathcal{L}} \setminus \underline{M}_{\mathcal{K}_s}$ the rest of the messages. The distribution of $\underline{Z}_{\mathcal{W}}$ induced by the uniform message matrix is denoted by $\tilde{p}_{\underline{Z}_{\mathcal{W}}}$ or $\tilde{p}_{\underline{Z}_{\mathcal{W}}|\underline{M}_{\mathcal{K}_s}=\underline{m}_{k_s}}$. For any $\underline{m}_{k_s} \in \underline{\mathcal{M}}_{k_s}$
\begin{alignat}{1}
    &\mathbb{V}\left(p_{\underline{Z}_{\mathcal{W}}|\underline{M}_{\mathcal{K}_s}=\underline{m}_{k_s}},p_{\underline{Z}_{\mathcal{W}}}\right) \label{FullSecrecy}\\
    & \quad \overset{(a)}{\leq} \mathbb{V}\left(p_{\underline{Z}_{\mathcal{W}}|\underline{M}_{\mathcal{K}_s}=\underline{m}_{k_s}},\tilde{p}_{\underline{Z}_{\mathcal{W}}|\underline{M}_{\mathcal{K}_s}=\underline{m}_{k_s}}\right) \label{eq:CondSec} \\
    & \quad\quad + \mathbb{V}\left(\tilde{p}_{\underline{Z}_{\mathcal{W}}|\underline{M}_{\mathcal{K}_s}=\underline{m}_{k_s}},\tilde{p}_{\underline{Z}_{\mathcal{W}}}\right) \label{eq:UnifSec} \\
    & \quad\quad + \mathbb{V}\left(\tilde{p}_{\underline{Z}_{\mathcal{W}}},p_{\underline{Z}_{\mathcal{W}}}\right), \label{eq:FullZ}
\end{alignat}
where (a) is from the triangle inequality. Since this is true for all  $\underline{m}_{k_s} \in \underline{\mathcal{M}}_{k_s}$, from now on we omit the equality $\underline{M}_{\mathcal{K}_s} = \underline{m}_{k_s}$ for ease of notation.

Now, we bound each of the expressions \eqref{eq:CondSec}-\eqref{eq:FullZ}, starting with \eqref{eq:CondSec}
\begin{align*}
    & \mathbb{V}\left(p_{\underline{Z}_{\mathcal{W}}|\underline{M}_{\mathcal{K}_s}},\tilde{p}_{\underline{Z}_{\mathcal{W}}|\underline{M}_{\mathcal{K}_s}}\right) \\
    & \quad = \sum_{\underline{z}_{w}} \left|p_{\underline{Z}_{\mathcal{W}}|\underline{M}_{\mathcal{K}_s}}(\underline{z}_{w}|\underline{m}_{k_s})-\tilde{p}_{\underline{Z}_{\mathcal{W}}|\underline{M}_{\mathcal{K}_s}}(\underline{z}_{w}|\underline{m}_{k_s})\right| \\
    & \quad = \sum_{\underline{z}_{w}}\left|\sum_{\underline{m}_{k_w}}\left(p_{\underline{Z}_{\mathcal{W}}\underline{M}_{\mathcal{K}_w}|\underline{M}_{\mathcal{K}_s}}(\underline{z}_{w},\underline{m}_{k_w}|\underline{m}_{k_s}) \right. \right. \\
    & \quad\quad\quad\quad\quad\quad\quad \left. \left. - \tilde{p}_{\underline{Z}_{\mathcal{W}}\underline{M}_{\mathcal{K}_w}|\underline{M}_{\mathcal{K}_s}}(\underline{z}_{w},\underline{m}_{k_w}|\underline{m}_{k_s})\right)\right| \\
    & \quad = \sum_{\underline{z}_{w}}\left|\sum_{\underline{m}_{k_w}}\left(p_{\underline{Z}_{\mathcal{W}}|\underline{M}_{\mathcal{K}_w}\underline{M}_{\mathcal{K}_s}}(\underline{z}_{w}|\underline{m}_{k_w},\underline{m}_{k_s}) \right. \right. \\
    & \quad\quad\quad\quad\quad\quad\quad\quad\quad\quad\quad \left. \left.\cdot p_{\underline{M}_{\mathcal{K}_w}|\underline{M}_{\mathcal{K}_s}}(\underline{m}_{k_w}|\underline{m}_{k_s}) \right. \right. \\
    & \quad\quad\quad\quad\quad \left. \left. -p_{\underline{Z}_{\mathcal{W}}|\underline{M}_{\mathcal{K}_w}\underline{M}_{\mathcal{K}_s}}(\underline{z}_{w}|\underline{m}_{k_w},\underline{m}_{k_s}) \right. \right. \\
    & \quad\quad\quad\quad\quad\quad\quad\quad\quad\quad\quad \left. \left. \cdot \tilde{p}_{\underline{M}_{\mathcal{K}_w}|\underline{M}_{\mathcal{K}_s}}(\underline{m}_{k_w}|\underline{m}_{k_s})\right)\right| \\
    & \quad = \sum_{\underline{z}_\mathcal{W}} p_{\underline{Z}_{\mathcal{W}}|\underline{M}_{\mathcal{K}_w}\underline{M}_{\mathcal{K}_s}}(\underline{z}_{w}|\underline{m}_{k_w},\underline{m}_{k_s}) \cdot \\ 
    & \quad\quad\quad\quad \left|\sum_{\underline{m}_{k_w}} \left(p_{\underline{M}_{\mathcal{K}_w}|\underline{M}_{\mathcal{K}_s}}(\underline{m}_{k_w}|\underline{m}_{k_s}) \right. \right. \\
    &  \quad\quad\quad\quad\quad\quad\quad\quad\quad \left. \left. -\tilde{p}_{\underline{M}_{\mathcal{K}_w}|\underline{M}_{\mathcal{K}_s}}(\underline{m}_{k_w}|\underline{m}_{k_s})\right)\right| \\
    & \quad \overset{(a)}{\leq} \sum_{\underline{z}_\mathcal{W}} \sum_{\underline{m}_{k_w}} p_{\underline{Z}_{\mathcal{W}}|\underline{M}_{\mathcal{K}_w}\underline{M}_{\mathcal{K}_s}}(\underline{z}_{w}|\underline{m}_{k_w},\underline{m}_{k_s}) \cdot \\
    & \quad\quad\quad \left|p_{\underline{M}_{\mathcal{K}_w}|\underline{M}_{\mathcal{K}_s}}(\underline{m}_{k_w}|\underline{m}_{k_s})-\tilde{p}_{\underline{M}_{\mathcal{K}_w}|\underline{M}_{\mathcal{K}_s}}(\underline{m}_{k_w}|\underline{m}_{k_s})\right| \\
    & \quad \overset{(b)}{=} \sum_{\underline{m}_{k_w}} \left|p_{\underline{M}_{\mathcal{K}_w}}(\underline{m}_{k_w}) - \tilde{p}_{\underline{M}_{\mathcal{K}_w}}(\underline{m}_{k_w})\right| \\
    & \quad = \mathbb{V}\left(p_{\underline{M}_{\mathcal{K}_w}},p_{U_{\underline{M}_{\mathcal{K}_w}}}\right) \leq \mathbb{V}\left(p_{\underline{M}_{\mathcal{L}}},p_{U_{\underline{M}_{\mathcal{L}}}}\right),
\end{align*}
where (a) holds from the triangle inequality, (b) holds from the independence between messages and $p_{U_{\underline{M}_{\mathcal{K}_w}}}$ is the uniform distribution of the uniform matrix $\underline{\tilde{M}}_{\mathcal{K}_w}$. We continue with bounding \eqref{eq:FullZ}
\begin{multline*}
    \begin{aligned}
    &\mathbb{V}\left(\tilde{p}_{\underline{Z}_{\mathcal{W}}},p_{\underline{Z}_{\mathcal{W}}}\right) \\
       & \quad= \sum_{\underline{z}_{w}}\left|\tilde{p}_{\underline{Z}_{\mathcal{W}}}(\underline{z}_{w}) - p_{\underline{Z}_{\mathcal{W}}}(\underline{z}_{w})\right| \\
       & \quad = \sum_{\underline{z}_{w}}\left|\sum_{\underline{m}_{k_w},\underline{m}_{k_s}}\left(\tilde{p}_{\underline{Z}_{\mathcal{W}} \underline{M}_{\mathcal{K}_w} \underline{M}_{\mathcal{K}_s}}(\underline{z}_{w},\underline{m}_{k_w},\underline{m}_{k_s})  \right. \right. \\ 
       & \quad\quad\quad\quad\quad\quad\quad\quad\quad \left. \left. -p_{\underline{Z}_{\mathcal{W}} \underline{M}_{\mathcal{K}_w} \underline{M}_{\mathcal{K}_s}}(\underline{z}_{w},\underline{m}_{k_w},\underline{m}_{k_s})\right)\right| \\
       & \quad = \sum_{\underline{z}_{w}}\left|\sum_{\underline{m}_{\ell}}\left(p_{\underline{Z}_{\mathcal{W}}|\underline{M}_{\mathcal{L}}}(\underline{z}_{w}|\underline{m}_{\ell})  \cdot p_{U_{\underline{M}_{\mathcal{L}}}}(\underline{m}_{\ell}) \right. \right. \\
       & \quad\quad\quad\quad\quad\quad \left. \left.- p_{\underline{Z}_{\mathcal{W}} | \underline{M}_{\mathcal{L}}}(\underline{z}_{w}|\underline{m}_{\ell}) \cdot p_{\underline{M}_{\mathcal{L}}}(\underline{m}_{\ell})\right)\right| \\
       & \quad \overset{(a)}{\leq} \sum_{\underline{m}_{\ell}}\sum_{\underline{z}_{w}} p_{\underline{Z}_{\mathcal{W}}|\underline{M}_{\mathcal{L}}}(\underline{z}_{w}|\underline{m}_{\ell}) \cdot \left| p_{U_{\underline{M}_{\mathcal{L}}}}(\underline{m}_{\ell}) - p_{M_{\mathcal{L}}}(\underline{m}_{\ell})\right| \\
       & \quad = \sum_{\underline{m}_{\ell}} \left|p_{U_{\underline{M}_{\mathcal{L}}}}(\underline{m}_{\ell}) - p_{\underline{M}_{\mathcal{L}}}(\underline{m}_{\ell})\right|  = \mathbb{V}\left(p_{\underline{M}_{\mathcal{L}}},p_{U_{\underline{M}_{\mathcal{L}}}}\right),
    \end{aligned}
\end{multline*}
where inequality (a) follows from the triangle inequality.

We now bound $\mathbb{D}(p_{\underline{M}_{\mathcal{L}}}||p_{U_{\underline{M}_{\mathcal{L}}}})$\off{. For ease of notation, we denote a column from $\underline{M}_{\mathcal{L}}$ by $M(j)$} and obtain 
\begin{equation*}
\begin{aligned}
    &\mathbb{D}\left(p_{\underline{M}_{\mathcal{L}}}||p_{U_{\underline{M}_{\mathcal{L}}}}\right) \\
    & \quad = \sum_{\underline{m}_{\ell}} p_{\underline{M}_{\mathcal{L}}}(\underline{m}_{\ell}) \cdot \log_{2} \left(\frac{p_{\underline{M}_{\mathcal{L}}}(\underline{m}_{\ell})}{p_{U_{\underline{M}_{\mathcal{L}}}}(\underline{m}_{\ell})}\right) \\
    & \quad = \log_{2} (2^{\ell \cdot \tilde{n}}) - H\left(\underline{M}_{\mathcal{L}}\right) \\
    & \quad = \ell \tilde{n} - H\left(\underline{M}_{\mathcal{L}}\right) \\
    & \quad \overset{(a)}{=} \ell \tilde{n} - \ell \cdot \sum_{j=1}^{\tilde{n}}  H\left(M^{(j)}|M^{j-1}\right) \\
    & \quad \overset{(b)}{\leq} \ell \tilde{n} - \ell  \tilde{n} (1-\delta_n) = \ell  \tilde{n}  \delta_n \leq \ell  \tilde{n} 2^{-n^\beta},
\end{aligned}
\end{equation*}
where (a) holds from the independence between messages, and (b) is given using the source scheme in Appendix~\ref{appendix:src_code}.

Now, we invoke the Pinsker inequality \cite{1053968} to bound the variational distance between the distribution $p_{\underline{M}_{\mathcal{L}}}$ and the uniform distribution $p_{U_{\underline{M}_{\mathcal{L}}}}$. Thus, we have
\begin{equation*}
    \begin{aligned}
        \frac{1}{2}\mathbb{V}^2\left(p_{\underline{M}_{\mathcal{L}}},p_{U_{\underline{M}_{\mathcal{L}}}}\right) \leq  \mathbb{D}\left(p_{\underline{M}_{\mathcal{L}}}||p_{U_{\underline{M}_{\mathcal{L}}}}\right) \leq \ell \tilde{n} 2^{-n^\beta},
    \end{aligned}
\end{equation*}
s.t. we bound the expressions in \eqref{eq:CondSec} and \eqref{eq:FullZ} by
\begin{equation} \label{eq:uniform-bound}
    \begin{aligned}
        \mathbb{V}\left(p_{\underline{M}_{\mathcal{L}}},p_{U_{\underline{M}_{\mathcal{L}}}}\right) \leq \sqrt{2 \ell \tilde{n} 2^{-n^\beta}}.
    \end{aligned}
\end{equation}

\off{To bound \eqref{eq:UnifSec}, we first present the following lemma.
\begin{lemma} \label{Lemma:HelpLemma}
Let $a_i,b_i$, $1 \leq i \leq n$ be a series of non-negative numbers. Then, the following inequality holds 
\begin{gather*}
    \left|\prod_{i=1}^{n}a_i - \prod_{i=1}^{n}b_i\right| \leq \sum_{i=1}^{n}|a_i-b_i|\cdot \prod_{j=1}^{i-1}a_j \cdot \prod_{k=i+1}^{n}b_k.
\end{gather*}
\end{lemma}
Using Lemma \ref{Lemma:HelpLemma}, we have}
We now bound \eqref{eq:UnifSec} and obtain
\begin{equation} \label{eq:HelpLemmaBound}
\begin{aligned}
    &\mathbb{V}\left(\tilde{p}_{\underline{Z}_{\mathcal{W}}|\underline{M}_{\mathcal{K}_s}},\tilde{p}_{\underline{Z}_{\mathcal{W}}}\right) \\
    & \quad = \sum_{\underline{z}_w}\left|\tilde{p}_{\underline{Z}_{\mathcal{W}}|\underline{M}_{\mathcal{K}_s}}(\underline{z}_w|\underline{m}_{k_s})-\tilde{p}_{\underline{Z}_{\mathcal{W}}}(\underline{z}_w)\right| \\
    & \quad \overset{(a)}{=} \sum_{\underline{z}_\mathcal{W}}\left|\prod_{j=1}^{\tilde{n}}\tilde{p}_{\underline{Z}_{\mathcal{W}}^{(j)}|\underline{M}_{\mathcal{K}_s}^{(j)}}(\underline{z}^{(j)}|\underline{m}_{k_s}^{(j)}) - \prod_{j=1}^{\tilde{n}}\tilde{p}_{\underline{Z}_{\mathcal{W}}^{(j)}}(\underline{z}^{(j)})\right| \\
    & \quad \overset{(b)}{\leq} \sum_{\underline{z}_\mathcal{W}} \sum_{j=1}^{\tilde{n}} \left|\tilde{p}_{\underline{Z}_{\mathcal{W}}^{(j)}|\underline{M}_{\mathcal{K}_s}^{(j)}}(\underline{z}^{(j)}|\underline{m}_{k_s}^{(j)})-\tilde{p}_{\underline{Z}_{\mathcal{W}}^{(j)}}(\underline{z}^{(j)})\right| \cdot \\
    & \quad\quad\quad\quad \prod_{q=1}^{j-1}\tilde{p}_{\underline{Z}_{\mathcal{W}}^{(q)}|\underline{M}_{\mathcal{K}_s}^{(q)}}(\underline{z}^{(q)}|\underline{m}_{k_s}^{(q)}) \cdot \prod_{k=j+1}^{\tilde{n}}\tilde{p}_{\underline{Z}_{\mathcal{W}}^{(k)}}(\underline{z}^{(k)}) \\
    & \quad \overset{(c)}{=} \sum_{j=1}^{\tilde{n}} \sum_{\underline{z}^{(j)}}\left|\tilde{p}_{\underline{Z}_{\mathcal{W}}^{(j)}|\underline{M}_{\mathcal{K}_s}^{(j)}}(\underline{z}^{(j)}|\underline{m}_{k_s}^{(j)})-\tilde{p}_{\underline{Z}_{\mathcal{W}}^{(j)}}(\underline{z}^{(j)})\right|\cdot \\
    & \quad\quad\quad \sum_{\underline{\hat{z}}^{(j)}}\prod_{q=1}^{j-1}\tilde{p}_{\underline{Z}_{\mathcal{W}}^{(q)}|\underline{M}_{\mathcal{K}_s}^{(q)}}(\underline{z}^{(q)}|\underline{m}_{k_s}^{(q)}) \cdot \prod_{k=i+1}^{\tilde{n}}\tilde{p}_{\underline{Z}_{\mathcal{W}}^{(k)}}(\underline{z}^{(k)}) \\
    & \quad = \sum_{j=1}^{\tilde{n}} \sum_{\underline{z}^{(j)}}\left|\tilde{p}_{\underline{Z}_{\mathcal{W}}^{(j)}|\underline{M}_{\mathcal{K}_s}^{(j)}}(\underline{z}^{(j)}|\underline{m}_{k_s}^{(j)})-\tilde{p}_{\underline{Z}_{\mathcal{W}}^{(j)}}(\underline{z}^{(j)})\right| \overset{(d)}{\leq} \tilde{n} \ell^{-\frac{t}{2}},
\end{aligned} 
\end{equation}
where (a) holds since $\tilde{p}_{\underline{Z}_{\mathcal{W}}|\underline{M}_{\mathcal{K}_s}}$ and $\tilde{p}_{\underline{Z}_{\mathcal{W}}}$ are induced from a completely uniform distribution, and (b) holds from \cite[Telescoping Expansion]{korada2010polar}. (c) is from denoting $\underline{\hat{z}}^{(j)}=\left(\underline{z}^{(1)},...,\underline{z}^{(j-1)},\underline{z}^{(j+1)},...,\underline{z}^{(\tilde{n})}\right)$. (d) holds according to \cite[Theorem 1]{SMSM}, by choosing $k_s \leq \ell - w - \ell\epsilon$, and $\ell\epsilon = \lceil t \log{\ell}\rceil$ for $t \geq 1$.

Now, we return to \eqref{FullSecrecy}. By substituting \eqref{eq:CondSec}-\eqref{eq:FullZ} with \eqref{eq:uniform-bound}-\eqref{eq:HelpLemmaBound}, we show that
\[
\mathbb{V}\left(p_{\underline{Z}_{\mathcal{W}} | \underline{M}_{\mathcal{K}_s}=\underline{m}_{k_s}},p_{\underline{Z}_{\mathcal{W}}}\right) \leq  \tilde{n} \ell^{-\frac{t}{2}} + 2 \sqrt{2 \ell \tilde{n} 2^{-n^\beta}}.
\]
\off{\begin{multline*}
\begin{aligned}
    &\mathbb{V}\left(p_{\underline{Z}_{\mathcal{W}} | \underline{M}_{K_s}=\underline{m}_{k_s}},p_{\underline{Z}_{\mathcal{W}}}\right) \\
    & \quad \leq \mathbb{V}\left(p_{\underline{Z}_{\mathcal{W}}|\underline{M}_{K_s}},\tilde{p}_{\underline{Z}_{\mathcal{W}}|\underline{M}_{K_s}}\right) +\\
    & \quad\quad\quad  \mathbb{V}\left(\tilde{p}_{\underline{Z}_{\mathcal{W}}|\underline{M}_{K_s}=\underline{m}_{k_s}},\tilde{p}_{\underline{Z}_{\mathcal{W}}}\right) + \mathbb{V}\left(\tilde{p}_{\underline{Z}_{\mathcal{W}}},p_{\underline{Z}_{\mathcal{W}}}\right) \\
    & \quad \leq \tilde{n} \cdot \ell^{-\frac{t}{2}} + 2\cdot \sqrt{2\cdot \ell \cdot \tilde{n} \cdot 2^{-n^\beta}}.
\end{aligned}
\end{multline*}}

We have shown that the channel scheme employed on $\underline{M}_{\mathcal{L}}$ is $k_s$-IS. To conclude the leakage proof against IT-Eve, we show that the scheme remains $k_s$-IS secured even when using the source coding scheme, by bounding $\mathbb{V}\left(p_{\underline{Z}_{\mathcal{W}} | \underline{V}_{\mathcal{K}_s}=\underline{v}_{k_s}},p_{\underline{Z}_{\mathcal{W}}}\right)$, and showing that for any $\underline{v}_{k_s} \in \underline{\mathcal{V}}_{\mathcal{K}_s}$ the
information leakage becomes negligible. Thus, we have
\begin{align*}
    &\mathbb{V}\left(p_{\underline{Z}_{\mathcal{W}} | \underline{V}_{\mathcal{K}_s}=\underline{v}_{k_s}},p_{\underline{Z}_{\mathcal{W}}}\right) \nonumber \\
    & \quad = \sum_{\underline{z}_w} \left|p_{\underline{Z}_{\mathcal{W}}|\underline{V}_{\mathcal{K}_s}}(\underline{z}_{w}|\underline{v}_{k_s})-p_{\underline{Z}_{\mathcal{W}}}(\underline{z}_{w})\right| \nonumber \\
    & \quad = \sum_{\underline{z}_{w}}\left|\sum_{\underline{m}_{k_s}} p_{\underline{Z}_{\mathcal{W}}|\underline{M}_{\mathcal{K}_s}\underline{V}_{\mathcal{K}_s}}(\underline{z}_{w}|\underline{m}_{k_s}\underline{v}_{k_s}) \cdot \right. \nonumber \\
    & \quad\quad\quad\quad\quad\quad\quad\quad\quad\quad \left. p_{\underline{M}_{\mathcal{K}_s}|\underline{V}_{\mathcal{K}_s}}(\underline{m}_{k_s}|\underline{v}_{k_s}) \right. \nonumber \\ 
    & \quad\quad\quad\quad\quad \left. - p_{\underline{Z}_{\mathcal{W}}}(\underline{z}_{w})\cdot \sum_{\underline{m}_{k_s}} p_{\underline{M}_{\mathcal{K}_s}|\underline{V}_{\mathcal{K}_s}}(\underline{m}_{k_s}|\underline{v}_{k_s})\right| \nonumber \\
    & \quad \overset{(a)}{\leq} \sum_{\underline{m}_{k_s}} \sum_{\underline{z}_{w}} p_{\underline{M}_{\mathcal{K}_s}|\underline{V}_{\mathcal{K}_s}}(\underline{m}_{k_s}|\underline{v}_{k_s}) \cdot \nonumber \\
    & \quad\quad\quad\quad\quad\quad \left|p_{\underline{Z}_{\mathcal{W}}|\underline{M}_{\mathcal{K}_s}\underline{V}_{\mathcal{K}_s}}(\underline{z}_{w}|\underline{m}_{k_s}\underline{v}_{k_s}) - p_{\underline{Z}_{\mathcal{W}}}(\underline{z}_{w})\right| \nonumber \\
    & \quad = \sum_{\underline{m}_{k_s}} p_{\underline{M}_{\mathcal{K}_s}|\underline{V}_{\mathcal{K}_s}}(\underline{m}_{k_s}|\underline{v}_{k_s}) \cdot \nonumber \\
    & \quad\quad\quad\quad\quad\quad \mathbb{V}\left(p_{\underline{Z}_{\mathcal{W}} | \underline{V}_{\mathcal{K}_s}=\underline{v}_{k_s}\underline{M}_{\mathcal{K}_s}=\underline{m}_{k_s}},p_{\underline{Z}_{\mathcal{W}}}\right) \nonumber \\
    & \quad \overset{(b)}{=} \sum_{\underline{m}_{\mathcal{K}_s}} p_{\underline{M}_{\mathcal{K}_s}|\underline{V}_{\mathcal{K}_s}}(\underline{m}_{k_s}|\underline{v}_{k_s}) \cdot \mathbb{V}\left(p_{\underline{Z}_{\mathcal{W}} | \underline{M}_{\mathcal{K}_s}=\underline{m}_{k_s}},p_{\underline{Z}_{\mathcal{W}}}\right) \\
    & \quad \leq \tilde{n} \ell^{-\frac{t}{2}} + 2 \sqrt{2 \ell \tilde{n}  2^{-n^\beta}},
\end{align*}
where (a) is from the triangle inequality, and (b) is since $\underline{V}_{\mathcal{K}_s} \rightarrow \underline{M}_{\mathcal{K}_s} \rightarrow \underline{Z}_{\mathcal{W}}$ is a Markov chain.
For the information leakage to be negligible, we give an upper and lower bound to $\ell$ compared to $n$. $\ell$ is lower bounded by $\omega(\tilde{n}^{\frac{2}{t}})$, s.t. the expression $\tilde{n} \ell^{-\frac{t}{2}}$ is negligible. In addition, $\ell$ is upper bounded by $o\left(2^{n^\beta}/\tilde{n}\right)$, s.t. the expression $2 \sqrt{2 \ell \tilde{n} 2^{-n^\beta}}$ is negligible. By choosing $\ell$ upholding both bounds, the information leakage to IT-Eve becomes negligible. This holds for any set of $\mathcal{K}_s \subset \mathcal{L}$  source messages, thus the code is $k_s$-IS.
\off{The lower bound of $\ell$ is required for $\tilde{n} \ell^{-\frac{t}{2}}$ to be negligible. This expression is negligible by taking $l$ to be any function of $n$ s.t. $\ell = \omega(\tilde{n}^{\frac{2}{t}})$. The upper bound on $l$ is required for $2\cdot \sqrt{2 \cdot \ell \cdot \tilde{n} \cdot 2^{-n^\beta}}$ to be negligible. This expression is negligible by taking $l$ to be any function of $\ell$ s.t. $\ell = o\left(2^{n^\beta}/\tilde{n}\right)$. By choosing $\ell$ s.t. those bounds hold, the information leakage to IT-Eve becomes negligible. 
This holds for any set of $\mathcal{K}_s \subset \mathcal{L}$  source messages, thus the code is $k_s$-IS.}

\section{Individual SS-CCA1 Against Crypto-Eve}\label{appendix:individual-ss-cca1}
We give here the full security proof of Theorem~\ref{Individual-SS-CCA1}. We aim to show each column $j \in \{1,...,\tilde{n}\}$ of the message matrix $\underline{M}_{\mathcal{L}}$ is ISS-CCA1. The proof is based on the equivalence between semantic security and indistinguishability \cite{goldwasser2019probabilistic}. We consider the maximal advantage for Crypto-Eve (maximal divination from uniform distribution probability), given the almost uniform messages after the source encoding stage (see Section~\ref{sec:main_results}) and prove that each column of the message matrix $\underline{M}_{\mathcal{L}}$ is IIND-CCA1 as given in \cite[Definition 4]{cohen2022partial}, and thus it is ISS-CCA1.

We assume the message matrix is the output of the source encoder from Appendix~\ref{appendix:src_code} with almost uniform messages. Alice and Crypto-Eve start playing the game as defined in \cite[definition 4]{cohen2022partial}. First, using the security parameter $c$, public and secret keys are created and denoted by $(p_c,s_c)$. We note that the security parameter $c$ is a function of the number of encrypted bits $1 \leq c < \ell$. Crypto-Eve sends a polynomial amount of ciphertexts to Alice and receives back their decryption. At this stage, Crypto-Eve chooses $i^{*} \in \{1,...,k_s\}$ (the case for $i^{*} \in \{k_s+1,...,l\}$ follows analogously), and two possible messages $M_{i^{*},1}$ and $M_{i^{*},2}$.

We add a step to the game that gives Crypto-Eve an additional advantage over Alice to show a stronger statement than in Definition~\ref{def:individuall-SS-CCA1}. Crypto-Eve is given the bits in positions $\{1,...,k_s\} \setminus i^{*}$. We show that still, Crypto-Eve is not able to distinguish between $M_{i^{*},1}$ and $M_{i^{*},2}$. Alice draws the bits in positions $\{k_s+1,...,\ell\}$ from the distribution induced by the source encoder, and chooses $h \in \{1,2\}$ uniformly at random s.t. the bit in position $i^{*}$ is $M_{i^{*},h}$. The message received is denoted by $M^\ell_h(j) \in \mathbb{F}_{2}^{\ell}$.

Alice encrypts $M^\ell_h(j)$ using $Crypt_2$ from \eqref{eq:encryption_crypt2}. First, Alice employs the IS channel code from Appendix~\ref{appendix:msg_encoder} to receive the encoded codeword denoted by $X^\ell_h(j) \in \mathbb{F}_{2}^{\ell}$. Second, Alice encrypts the first $c$ bits from $X^\ell_h(j)$ using $Crypt_1$. For the purpose of this proof, we denote the encrypted ciphertext by $\kappa = Crypt_2(M^\ell_h(j))$.

Upon receiving the ciphertext, $\kappa$, Crypto-Eve tries to guess $h$. First, $w = \ell - c$ of the bits from $\kappa$ are seen by Crypto-Eve as plaintext. Thus, Crypto-Eve can potentially reduce the number of possible codewords in each of the bins. \off{The average number of codewords that remain per bin after Crypto-Eve observes $w$ of the bits is $2^{\ell\epsilon}$ \cite{SMSM}, where $k_s = \ell - w - \ell\epsilon = c - \ell\epsilon$.} We denote the set of the remaining possible codewords from bin 1 and bin 2 by $\mathcal{B}_1$ and $\mathcal{B}_2$, respectively. It was shown in \cite{SMSM} that with high probability, the number of remaining possible codewords per bin deviates between $(1-\epsilon') 2^{\ell\epsilon} \leq |\mathcal{B}_1|,|\mathcal{B}_2| \leq (1+\epsilon') 2^{\ell\epsilon}$
for some $\epsilon' > 0$ and $\ell\epsilon$ s.t. $k_s = \ell - w -\ell\epsilon = c - \ell\epsilon$.

We assume the best possible scenario for Crypt-Eve s.t.: 1) The number of possible codewords remaining in bin 1 is as high as possible where the number of possible codewords remaining in bin 2 is as low as possible, $|\mathcal{B}_1| \geq |\mathcal{B}_2|$, 2) the induced probability of the codewords from bin 1 is as high as possible while the induced probability of the codewords from bin 2 is as low as possible \footnote{In \cite{cohen2022partial} the induced probabilities of each of the codewords were equal since the probability of the messages was uniform.}, and 3) the advantage Crypto-Eve has from $Crypt_1$ is as high as possible.

We start by bounding the maximum and minimum possible probabilities of the codewords in $\mathcal{B}_1$ and $\mathcal{B}_2$. We denote those probabilities by $p_{max}$ and $p_{min}$. First, we consider $j \in \mathcal{H}_V$, i.e. the bits in column $j$ were not padded by the uniform seed, and their entropy is lower bounded by $1 - \delta_n$. The case for $j \in \mathcal{J}_V$ will be addressed later on.

For each column $j$, we denote by $M_{k_w} \in \mathbb{F}_{2}^{k_w}$ the set of possible messages from bits $\{k_s+1,...,\ell\}$. In each column $j$ of the message matrix, the bits are independent since they are obtained from independent source messages. Thus, we have
\begin{align*}
    p(M_{k_w} =m_{k_w}) = \prod_{i=1}^{k_w}p(M_{k_w,i} = m_{k_w,i}),
\end{align*}
From \eqref{eq:unreliable-group}, we conclude that
\begin{equation*} \label{eq:bitEntropy}
    \begin{aligned}
        H(m_{k_w,i}) \geq H\left(m_{k_w,i}|m_{k_w,i}^{j-1}\right) \geq 1 - \delta_n,
    \end{aligned}
\end{equation*}
where $m_{k_w,i}^{j-1} = \left(m_{k_w,i}(1),...,m_{k_w,i}(j-1)\right)$ are the bits in columns $1$ to $j-1$ of the $i$-th row in the message matrix. We denote $\zeta \in (0,\frac{1}{2})$ s.t. $H(\frac{1}{2} - \zeta) = H(\frac{1}{2} + \zeta) = 1 - \delta_n$, and use $\zeta$ to bound $p_{min}$ and $p_{max}$
\begin{equation*}\label{eq:p-bounds}
    \begin{aligned}
        p_{min} \triangleq \left(\frac{1}{2} - \zeta\right)^{k_w} \leq p(\underline{m}_{k_w}) \leq \left(\frac{1}{2} + \zeta\right)^{k_w} \triangleq p_{max}.
    \end{aligned}
\end{equation*}
We bound $\zeta$ using the binary entropy bounds given in \cite{topsoe2001bounds}, $4p(1-p) \leq H_b(p) \leq \left(4p(1-p)\right)^{\ln{4}}$ for $0 < p < 1$. That is, by replacing $p = \frac{1}{2} + \zeta$ we obtain the upper bound by
\begin{equation*} \label{eq:right-side}
    \begin{aligned}
        H_b\left(\frac{1}{2} + \zeta\right) \leq \left(4\left(\frac{1}{2} + \zeta\right)\cdot \left(\frac{1}{2} - \zeta\right)\right)^{\frac{1}{\ln 4}},
    \end{aligned}
\end{equation*}
and the lower bound by
\begin{equation*} \label{eq:left-side}
    \begin{aligned}
        H_b\left(\frac{1}{2} + \zeta\right) \geq 4\left(\frac{1}{2} + \zeta\right)\left(\frac{1}{2} - \zeta\right).
    \end{aligned}
\end{equation*}
From which, we conclude
\begin{align*}
    \frac{1}{2} + \zeta \leq \frac{1}{2} + \frac{1}{2}\cdot \sqrt{1 - (1 - \delta_n)^{\ln 4}},
\end{align*}
and therefore, the upper bound for $\zeta$ is given by
\begin{equation}\label{eq:zeta-bound}
    \begin{aligned}
        \zeta \leq \frac{\sqrt{1 - (1 - \delta_n)^{\ln 4}}}{2}.
    \end{aligned}
\end{equation}

Now, we give the probability for some codeword from the set $\mathcal{B}_1 \cup \mathcal{B}_2$. We denote the codeword by $\alpha \in \mathcal{B}_1 \cup \mathcal{B}_2$. The probability for $\alpha$ is
\begin{equation} \label{eq:one-set}
    \begin{aligned}
        \mathbb{P}[\alpha | \alpha \in \mathcal{B}_1 \cup \mathcal{B}_2] = 
        \begin{cases}
            \frac{p_{max}}{|\mathcal{B}_1| \cdot p_{max} + |\mathcal{B}_2| \cdot p_{min}} & \text{if } \alpha \in \mathcal{B}_1 \\
            \frac{p_{min}}{|\mathcal{B}_1| \cdot p_{max} + |\mathcal{B}_2| \cdot p_{min}} & \text{if } \alpha \in \mathcal{B}_2
        \end{cases}
    \end{aligned}
\end{equation}
From \eqref{eq:one-set}, we conclude that the probability for a codeword $\alpha$ to be from the set $\mathcal{B}_1$ is
\begin{equation} \label{eq:bin1}
    \begin{aligned}
        \mathbb{P}[\alpha \in \mathcal{B}_1 | \alpha \in \mathcal{B}_1 \cup \mathcal{B}_2] = 
            \frac{|\mathcal{B}_1| \cdot p_{max}}{|\mathcal{B}_1| \cdot p_{max} + |\mathcal{B}_2| \cdot p_{min}}.
    \end{aligned}
\end{equation}

Since $Crypt_1$ is SS-CCA1, Crypto-Eve can obtain some negligible information about the original codeword or some function of the codeword. We consider the function $f: \{\mathcal{B}_1 \cup \mathcal{B}_2\} \rightarrow \{1,2\}$, i.e. the function that takes a codeword from the set $\{\mathcal{B}_1 \cup \mathcal{B}_2\}$ and outputs whether it belongs to $\mathcal{B}_1$ or $\mathcal{B}_2$
\begin{equation*} \label{eq:f(c*)}
    \begin{aligned}
        f(\alpha \in \{\mathcal{B}_1 \cup \mathcal{B}_2\}) = 
        \begin{cases}
            1 & \text{if }  \alpha \in \mathcal{B}_1\\
            2 & \text{if }  \alpha \in \mathcal{B}_2
        \end{cases}
    \end{aligned}
\end{equation*}
From the definition of SS-CCA1, we conclude
\begin{equation} \label{eq:P(c*)}
    \begin{aligned}
        \mathbb{P}[M_{i^{*},1}] =  \mathbb{P}[\alpha \in \mathcal{B}_1] + \epsilon_{ss-cca1},
    \end{aligned}
\end{equation}
where $\epsilon_{ss-cca1}$ is a negligible function in $c$ s.t. $\epsilon_{ss-cca1} \leq \frac{1}{c^{d}}$. Thus, by substituting \eqref{eq:bin1} into \eqref{eq:P(c*)}, we have
\begin{align}
    \mathbb{P}[M_{i^{*},1}] - \frac{1}{2} &=  \mathbb{P}[\alpha \in \mathcal{B}_1] + \epsilon_{ss-cca1} - \frac{1}{2} \nonumber \\
        \off{&= \frac{|\mathcal{B}|_1 \cdot p_{max}}{|\mathcal{B}_1| \cdot p_{max} + |\mathcal{B}_2| \cdot p_{min}} - \frac{1}{2} + \epsilon_{ss-cca1} \nonumber \\}
        &= \frac{|\mathcal{B}_1| \cdot p_{max} - |\mathcal{B}_2| \cdot p_{min}}{|\mathcal{B}_1| \cdot p_{max} + |\mathcal{B}_2| \cdot p_{min}} + \frac{1}{c^{d}} .\label{eq:adv-M}
\end{align}
We focus on bounding the left term of \eqref{eq:adv-M}
\begin{align}
    &\frac{|\mathcal{B}_1| \cdot p_{max} - |\mathcal{B}_2| \cdot p_{min}}{|\mathcal{B}_1| \cdot p_{max} + |\mathcal{B}_2| \cdot p_{min}} \nonumber \\
        &\overset{(a)}{=} \frac{(1+\epsilon') 2^{\ell\epsilon} p_{max} - (1-\epsilon') 2^{\ell\epsilon} p_{min}}{(1+\epsilon') 2^{\ell\epsilon} p_{max} + (1-\epsilon') 2^{\ell\epsilon} p_{min}} \nonumber \\
        &= \frac{1}{\frac{p_{max}+p_{min}}{p_{max}-p_{min}} + \epsilon'} + \frac{1}{\frac{p_{max}-p_{min}}{p_{max}+p_{min}} + \frac{1}{\epsilon'}}, \label{eq:p_max-p_min}
\end{align}
where (a) is from the assumption that $|\mathcal{B}_1| = (1+\epsilon')\cdot 2^{\ell\epsilon}$ and $|\mathcal{B}_2| = (1-\epsilon')\cdot 2^{\ell\epsilon}$. \off{By choosing $\epsilon' = \ell^{-t}$ s.t. $\ell\epsilon = \lceil t \log{\ell} \rceil$, and $t \geq 1$ \cite{SMSM},} We bound the term $\frac{p_{max}-p_{min}}{p_{max}+p_{min}}$ from \eqref{eq:p_max-p_min} by
\begin{align}
    \frac{p_{max}-p_{min}}{p_{max}+p_{min}} &= \frac{(\frac{1}{2}+\zeta)^{k_w}-(\frac{1}{2}-\zeta)^{k_w}}{(\frac{1}{2}+\zeta)^{k_w}+(\frac{1}{2}-\zeta)^{k_w}} \nonumber \\
    & = \frac{\sum_{i=0}^{k_w}\binom{k_w}{i}(2\zeta)^i - \sum_{i=0}^{k_w}\binom{k_w}{i}(-2\zeta)^i}{\sum_{i=0}^{k_w}\binom{k_w}{i}(2\zeta)^i + \sum_{i=0}^{k_w}\binom{k_w}{i}(-2\zeta)^i} \nonumber \\
    & = \frac{2\sum_{i=0}^{\lfloor \frac{k_w-1}{2} \rfloor}\binom{k_w}{2i+1}(2\zeta)^{2i+1}}{2\sum_{i=0}^{\lfloor \frac{k_w}{2} \rfloor}\binom{k_w}{2i}(2\zeta)^{2i}} \nonumber \\
    & \overset{(a)}{<} 2\zeta k_w \overset{(b)}{\leq} 2^{\frac{3}{2}}2^{\frac{-n^{\beta}}{2}}k_w, \label{eq:pmax-pmin-bound}
\end{align}
where (a) follows from the upper bound  $\frac{\sum_{i=0}^{\lfloor (k_w-1) / 2 \rfloor}\binom{k_w}{2i+1}(2\zeta)^{2i}}{\sum_{i=0}^{\lfloor k_w / 2 \rfloor}\binom{k_w}{2i}(2\zeta)^{2i}} < k_w$, which can be easily proved, and (b) is from \eqref{eq:zeta-bound}. By choosing $l \leq 2^{\frac{n^\beta}{2(t+1)}}$ s.t. $\ell\epsilon = \lceil t \log{\ell} \rceil$, for any $t \geq 1$, we have that $2^{\frac{3}{2}}2^{\frac{-n^{\beta}}{2}}k_w < 2^{\frac{3}{2}} \ell^{-t}$.

By substituting \eqref{eq:pmax-pmin-bound} into \eqref{eq:p_max-p_min} and taking $\epsilon' = \ell^{-t}$ \cite{SMSM}, it can be shown that for every $d'$ Crypto-Eve's advantage can be made smaller than $\frac{1}{c^{d'}}$ by choosing an appropriate $d$, large enough $\ell$ s.t $\ell \leq 2^{\frac{n^\beta}{2(t+1)}}$ and 
taking $t=\log{\ell}$. Thus, we showed that column $j \in \mathcal{H}_V$ is ISS-CCA1.

For the case of a column $j$ s.t. $j \in \mathcal{J}_V$, ISS-CCA1 follows directly from the SS-CCA1 of $Crypt_1$, considering Crypto-Eve has the maximal advantage in guessing the seed from its ciphertext. Thus, we can conclude the cryptosystem is IIND-CCA1 and thus ISS-CCA1 as well \cite{goldwasser2019probabilistic}.

\section{Seed Length}\label{appendix:seed_length}
The optimal size of the seed was defined in \cite{NegligbleCost,chou2013data}. For a message of size $n$ bits, the optimal size of the seed is 1) $d_{opt} = O(k_n\sqrt{n})$, $\forall k_n$ s.t. $\lim\limits_{n \rightarrow \infty} k_n \rightarrow \infty$, and 2) $d_{opt} = \Omega(\sqrt{n})$. However, it was shown by Chou et al. in \cite{NegligbleCost} that the seed size for the polar source coding scheme from Appendix~\ref{appendix:src_code} is sub-linear, i.e. $d_{opt} = o(n)$.\off{ and is considered sub-optimal.}
\off{However, the size of the seed for the polar codes source encoder/decoder from Appendix~\ref{appendix:src_code} is sub-optimal.}

The length of the seed is directly affected by the polarization rate which determines the portion of the bits that remain unpolarized after the polarization transform, i.e. the size of the group $\mathcal{J}_v$. This size is often referred to as the gap to capacity \cite{hassani2010scaling,wang2023sub} since it determines how close the channel code gets to the optimal capacity of the communication channel. Polar codes for source and channel coding are directly linked, \cite{cronie2010lossless}, thus the gap to capacity is equal to the gap of the polar codes for source coding to the optimal compression rate \cite[Corollary 3.16]{wang2021complexity}. In \cite{hassani2010scaling} it was shown that the number of unpolarized bits can be bounded by
\vspace{-0.15cm}
\begin{equation*} \label{SeedBound}
    \begin{aligned}
        n^{0.7214} \leq d_{J} \leq n^{0.7331}.
    \end{aligned}
\end{equation*}
This bound is sub-optimal since for $k_n = n^{0.2}$, we have that $d_J = \Omega(n^{0.7})$. Yet, this overhead for the proposed NU-HUNCC scheme with non-uniform messages, for sufficiently large $n$, is still negligible compared to the compressed message size, as illustrated in Fig. \ref{fig:SeedSize}.
\vspace{-0.3cm}
\begin{figure}[htbp]
    \centering
    \includegraphics[width=0.95\linewidth]{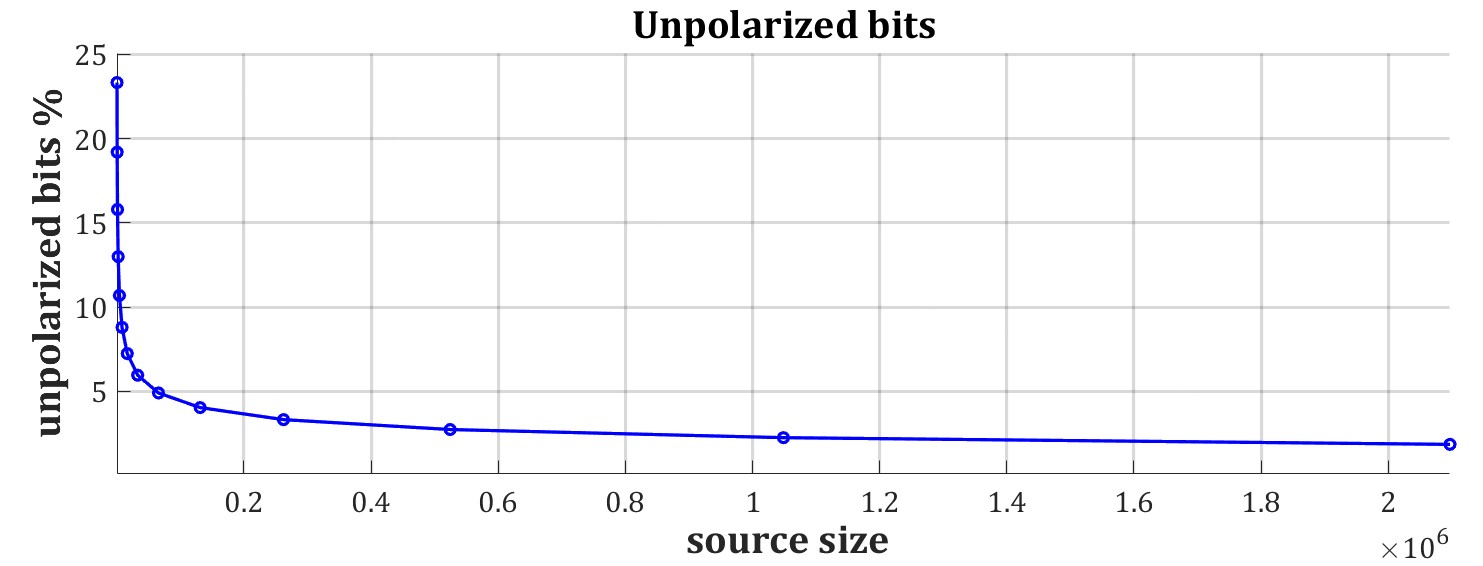}
    \caption{Numerical simulation of the seed size for a source $(\mathcal{V},p_V)$ with entropy $H(V) = 0.9$. For messages with a size greater than $2^{18}$ bits, the seed size already decreases to about $2.2\%$ of the compressed message size.}
    \label{fig:SeedSize}
\end{figure}
\off{
This bound is sub-optimal since for $k_n = n^{0.2}$, we have that $d_J = \Omega(n^{0.7})$. However, numeric simulations show that for $n=2^{19}$ bits, the length of the seed is already $2\%$ of the message size and $2.2\%$ of the compressed message size considering the entropy of the source is $H(V) = 0.9$.}
\off{
The seed length is directly affected by the rate of the polarization of the bits by the source encoder. The rate of polarization of polar codes was studied mainly for channel codes where it was called the capacity gap. i.e. the gap between the actual capacity and the achieved capacity due to the unpolarized bits. Since there is a direct link between polar codes for source coding and for channel coding \cite{cronie2010lossless}, it was shown in \cite[Corollary 3.16]{wang2021complexity} that the capacity gap can also be applied for source coding. The number of unpolarized bits depends on the blocklength, $n = 2^m$. In \cite{hassani2010scaling} it was shown that the number of unpolarized bits can be bounded by
\begin{equation*} \label{SeedBound}
    \begin{aligned}
        n^{0.7214} \leq d_{J,n} \leq n^{0.7331}
    \end{aligned}
\end{equation*}

\off{This bound is better than $d_n = o(n)$ but is still not optimal as was defined in \cite{NegligbleCost}, \cite{chou2013data}. We get that $d_n = \Omega(\sqrt{n})$, but there are some $k_n$ s.t. $\lim \limits_{n \rightarrow \infty} k_n = \infty$ for them $k_nd_{J,n} = \omega(k_n\sqrt{n})$, thus the second condition does not hold. However, by numeric simulations on the different blocklengths sizes, it can be shown that even for $n = 2^{19}$ the length of the seed is already $2\%$ of the source message size. For a source $(\mathcal{V},p_V)$ with entropy $H(V) = 0.9$ the seed length is $2.2\%$ of the compressed message.}
The bounding provided in \cite{hassani2010scaling} sufficiently serves the objectives of this study, with further exploration of the subject being beyond its scope. Nevertheless, further exploring the seed size within the polar code uniform compression scheme could present an intriguing avenue for future research.}}
\end{document}